%% file: main-OOPSLA25-ExtendedVersion.tex
\documentclass[acmsmall]{acmart}
\input{main-version-switching}
\input{headers}
\ToggleExtendedVersion
\begin{document}
\acmDOI{10.1145/3763142}
\acmYear{2025}
\acmJournal{PACMPL}
\acmVolume{9}
\acmNumber{OOPSLA2}
\acmArticle{364}
\acmMonth{10}
\received{2025-03-26}
\received[accepted]{2025-08-12}
\input{main.oopsla}
\end{document}

%% file: main-version-switching.tex
\usepackage{ifthen}
\newboolean{extendedVersionFlag}

\newcommand{\ToggleExtendedVersion}{
    \setboolean{extendedVersionFlag}{true}
}

\newcommand{\IfCameraReady}[2]{\ifthenelse{\boolean{extendedVersionFlag}}{#2}{#1}}

\newcommand{\TestExtendedVersion}
{\ifthenelse{\boolean{extendedVersionFlag}}{
    \textbf{\color{red}Is extended version}
}{
    \textbf{\color{red}Is not extended-version}
}}

%% file: headers.tex
\AtBeginDocument{%
  }
\setcopyright{none}
\AtBeginDocument{%
  }
\usepackage[T1]{fontenc}

\usepackage{graphicx}
\usepackage{amsmath}
\usepackage{amsfonts}
\usepackage{flexisym}
\usepackage{ebproof}
\usepackage{breqn}
\usepackage{tikz}
\usepackage{pgfplots}
\usepackage{xspace}
\usepackage{enumitem}
\usepackage{float}
\usepackage{placeins}
\usepackage{rotating}
\usepackage{mdframed}
\usepackage{tabularx} 
\usepackage{stmaryrd}
\usepackage{hyperref}
\usepackage{wrapfig}
\usepackage{xcolor}
\usepackage{pifont}
\usepackage{booktabs}
\usepackage{extarrows}

\usepackage[nameinlink]{cleveref}
\usepackage{comment}

\newtheorem{problem}{Problem}
\newtheorem{lemma}{Lemma}
\newtheorem{theorem}{Theorem}
\newtheorem{definition}{Definition}
\newtheorem{example}{Example}

\usepackage{colortbl}
\AtBeginDocument{%
  }
\usepackage{centernot}
\usepackage[noend]{algpseudocode}
\usepackage[plain,figure]{algorithm}
\algnewcommand\Input{\textbf{input: }}
\algnewcommand\Output{\textbf{output: }}
\algblockdefx[Match]{Match}{EndMatch}[1][]{\textbf{match} #1 \textbf{with}}{\vspace{-\baselineskip}}
\algdef{SE}[SUBALG]{Indent}{EndIndent}{}{\algorithmicend\ }%
\algdef{SE}[DOWHILE]{Do}{doWhile}{\algorithmicdo}[1]{\algorithmicwhile\ #1}%
\algtext*{Indent}
\algtext*{EndIndent}

\algnewcommand\algorithmicswitch{\textbf{match}}
\algnewcommand\algorithmiccase{\textbf{case}}
\algnewcommand\algorithmicassert{\texttt{assert}}
\algnewcommand\Assert[1]{\State \algorithmicassert(#1)}%

\algdef{SE}[SWITCH]{Switch}{EndSwitch}[1]{\algorithmicswitch\ #1\ \algorithmicdo}{\algorithmicend\ \algorithmicswitch}%
\algdef{SE}[CASE]{Case}{EndCase}[1]{\algorithmiccase\ #1 \algorithmicdo}{\algorithmicend\ \algorithmiccase}%
\algtext*{EndSwitch}%
\algtext*{EndCase}%

\usepackage{array}
\usepackage{booktabs}

\newcommand{\src}{\mathsf{src}}
\newcommand{\dst}{\mathsf{dst}}

\renewcommand{\S}{\mathbf{SF}}
\renewcommand{\P}{\mathcal{P}}

\newcommand{\post}{\text{post}}
\newcommand{\sem}[1]{\llbracket #1 \rrbracket}
\newcommand{\TF}[1]{\textbf{TF}(#1)}
\newcommand{\SF}[1]{\textbf{SF}(#1)}
\newcommand{\Paths}[3]{\textit{Paths}_{#1}(#2,#3)}
\newcommand{\defeq}{\triangleq}
\newcommand{\PathExp}[3]{\textit{PathExp}_{#1}(#2,#3)}

\newcommand{\zak}[1]{{\color{orange}[Zak: #1]}}
\newcommand{\ruijie}[1]{{\color{purple}[Ruijie: #1]}}
\newcommand{\twr}[1]{{\color{magenta}{{T: }{#1}}}}
\newcommand{\twrchanged}[1]{{\color{blue}{#1}}}
\newcommand{\zakchanged}[1]{{\color{orange}{#1}}}
\newcommand{\rfchanged}[1]{{\color{purple}{#1}}}
\newcommand{\spost}[2]{\textit{post}({#1},{#2})}
\newcommand{\tfrel}[3]{#1 \xrightarrow{#2} #3}
\newcommand{\set}[1]{\{#1\}}
\newcommand{\tblch}[1]{{\color{red}#1}}
\renewcommand{\tblch}[1]{{#1}}
\renewcommand{\twrchanged}[1]{{#1}}
\renewcommand{\zakchanged}[1]{{#1}}
\renewcommand{\rfchanged}[1]{{#1}}
\renewcommand{\zak}[1]{}
\renewcommand{\twr}[1]{}
\renewcommand{\ruijie}[1]{}

\input{macros}

\pgfplotsset{compat=1.18} 

\algrenewcommand\algorithmicindent{0.5em}%

%% file: macros.tex
\usepackage{enumitem}
\setlist{nosep,leftmargin=*}
\usepackage{tikz}
\newcommand\encircle[1]{%
  {\tiny\tikz[baseline=(X.base)] 
    \node (X) [draw, shape=circle, inner sep=0, fill=black, text=white] {\strut  #1};}%
}
\newcommand\encircleR[1]{%
  {\tiny \tikz[baseline=(X.base)] 
    \node (X) [draw, shape=circle, inner sep=0, fill=red, text=white] {\strut  #1};}%
}

\newcommand\encircleB[1]{%
  {\tiny \tikz[baseline=(X.base)] 
    \node (X) [draw, shape=circle, inner sep=0, fill=blue, text=white] {\strut  #1};}%
}

\newcommand\encircleO[1]{%
  {\tiny \tikz[baseline=(X.base)] 
    \node (X) [draw, shape=circle, inner sep=0, fill=orange, text=white] {\strut \tiny #1};}%
}

\newcommand{\bfpara}[1]{\vspace{5pt}\noindent\emph{\textbf{#1}}}

\definecolor{Bittersweet}{rgb}{1.0, 0.44, 0.37}

\definecolor{highlight}{rgb}{.8,.8,.8}

\newcommand{\minVar}{\texttt{min}}
\newcommand{\N}{\texttt{N}}
\newcommand{\I}{\texttt{i}}

\newcommand{\CFG}{G}
\newcommand{\false}{\bot}
\newcommand{\true}{\top}
\newcommand{\Omit}[1]{}
\usepackage{soul}

\newcommand{\tuple}[1]{{({#1})}}

\newcommand{\CheckSat}[1]{\mathsf{Check}^{\textbf{+}}(#1)}

\renewcommand{\Check}[1]{\mathsf{Check}(#1)}
\newcommand{\xmark}{\ding{55}\xspace}

\newcommand{\GPSlite}{{GPSLite}\xspace}
\newcommand{\GPSLite}{\GPSlite}

\newcommand{\GPSnogas}{{GPS-nogas}\xspace}
\newcommand{\GPSnocra}{{GPS-nogas-nosum}\xspace}

\renewcommand{\defeq}{\triangleq}

\newcommand{\pdst}[1]{\textit{dst}(#1)}
\newcommand{\psrc}[1]{\textit{src}(#1)}
\newcommand{\pparent}[1]{\textit{parent}(#1)}
\newcommand{\TreeNodes}[1]{T}
\newcommand{\TreeRoot}[1]{\epsilon}
\newcommand{\TreeVtx}[2]{\pdst{#2}}
\newcommand{\TreeLbl}[2]{L(#2)}

\newcommand{\TreeCovers}[1]{\mathsf{Cov}}
\newcommand{\TreeFrontiers}[1]{\mathsf{frontierQueue}}

\newcommand{\prefixord}[0]{\sqsubseteq}
\usepackage{cleveref}
\crefname{problem}{problem}{problems}
\crefname{lemma}{lemma}{lemmas}
\crefname{claim}{claim}{claims}

\newcommand{\TreePrefixTwo}[2]{\textit{prefix}_{#1}(#2)}

\newcommand{\TreeSuffixTwo}[2]{\textit{suffix}_{#1}(#2)}

\newcommand{\barrier}{\FloatBarrier} 

\usepackage{listings}
\lstdefinestyle{base}{
  language=C,
  emptylines=1,
  breaklines=true,
  basicstyle=\color{black}\fontfamily{pzc}\selectfont\tt,
  commentstyle=\color{gray}\rm\itshape,
  keywordstyle=\rm\bfseries,
  identifierstyle=\rm\itshape,
  escapechar=@,
  morekeywords=[1]{then,do},
  morekeywords=[2]{assert},
  morekeywords=[3]{requires,ensures},
  keywordstyle	= [2]\color{red}\bf,
  keywordstyle	= [3]\bf\color{gray},
  numberstyle=\footnotesize\color{gray},
  moredelim=**[is][\bf\color{green}]{@!}{!@},
  moredelim=**[is][\bf\color{red}]{@?}{?@},
  literate=
    {<=}{{$\leq$}}1
    {>=}{{$\geq$}}1
    {||}{{$\lor$}}1
    {&&}{{$\land$}}1
    {->}{{$\rightarrow$}}1
    {_1}{$_{1}$}2
    {_2}{$_{2}$}2
    {_3}{$_{3}$}2
}
\newcommand{\cinline}[1]{\mbox{\lstinline[style=base,mathescape]{#1}}}
\usepackage{subcaption}

\tikzset{pathtree/.style={grow'=right,edge from parent/.style={->,draw},v/.style={draw,circle,fill=blue,text=white},
l/.style={draw,circle,fill=orange,text=white},
d/.style={draw,circle,fill=black,text=white},node distance=1cm,thick,>=stealth},label/.style={fill=black,fill opacity=0.1,text opacity=1}}
\newcommand{\artlabel}[2]{
  \node[label,below of=#1,node distance=0.5cm] (L#1) {\scriptsize $#2$};
}
\newcommand{\exthree}{
    \node [v] (A) {A}
    child {
      node [v] (AB) {B}
        child {
          node[v] (ABB) {B}
          child { node[l] (ABBB) {B} }
          child {
            node[v] (ABBC) {C}
            child {  node[l] (ABBCC) {C} }
            child {
              node[v] (ABBCD) {D}
              child { node[l] (ABBCDE) {E} }
            }
          }
        }
        child { node[l] (ABC) {C} }
    };
}
\newcommand{\exthreepl}{
 \exthree{}
     \node[right of=ABC, node distance=0.6cm] (pi1) {$\pi_1$};
     \node[right of=ABBB, node distance=0.6cm] (pi2) {$\pi_2$};
     \node[right of=ABBCC, node distance=0.6cm] (pi3) {$\pi_3$};
     \node[right of=ABBCDE, node distance=0.6cm] (pi4) {$\pi_4$};
}
\usepackage{wrapfig}

%% file: main.oopsla.tex
\sloppy

\IfCameraReady{
\title{Software Model Checking via Summary-Guided Search}
}{
\title{Software Model Checking via Summary-Guided Search (Extended Version)}
\titlenote{This is the extended version of our paper at OOPSLA 2025.}}

\author{Ruijie Fang}
\orcid{0000-0001-5348-5468}
\affiliation{%
  \institution{University of Texas at Austin}
  \city{Austin, TX}
  \country{USA}
}
\email{ruijief@cs.utexas.edu}

\author{Zachary Kincaid}
\orcid{0000-0002-7294-9165}
\affiliation{
  \institution{Princeton University}
  \city{Princeton, NJ}
  \country{USA}
}
\email{zkincaid@cs.princeton.edu}

\author{Thomas Reps}
\orcid{0000-0002-5676-9949}
\affiliation{
    \institution{University of Wisconsin---Madison}
    \city{Madison, WI}
    \country{USA}
}
\email{reps@cs.wisc.edu}

\input{abstract}


\maketitle              

\input{zak_introduction.tex}
\input{overview}
\input{background}

\input{sgt}
\input{intraproc}
\input{apa}
\input{evaluation_intraproc}

\input{related-work}

\input{conclusion.tex}

\section*{Data-Availability Statement}
{
The artifact related to this paper (including benchmarks, an implementation of GPS, CRA, as well as other tools we compared against in the evaluations) is packaged as a Docker container on Docker Hub: \url{https://hub.docker.com/r/ruijiefang/gps-oopsla25-ae/}. Instructions for using the Docker container, as well as all benchmarks used, are available at the following Zenodo Link: \url{https://doi.org/10.5281/zenodo.16914159}.
}

\begin{acks}
This work was supported in part by the NSF under grant number 1942537. Opinions, findings, conclusions, or recommendations expressed herein are those
of the authors and do not necessarily reflect the views of the sponsoring agencies.
\end{acks}

\bibliographystyle{ACM-Reference-Format}
\bibliography{merged}
\IfCameraReady{
}{
\appendix 
\input{appendix}
}

%% file: abstract.tex
\begin{abstract}
In this work, we describe a new software model-checking algorithm {called} GPS.
GPS treats the task of model checking a program as a directed search of the program states, guided by a compositional, summary-based static analysis. The summaries produced by static analysis are used both to prune away infeasible paths and to drive test generation to reach new, unexplored program states.
GPS can find both 
proofs of safety and counter-examples to safety (i.e., inputs that trigger bugs), {and features a novel two-layered search strategy that renders it particularly efficient at finding bugs in programs featuring long, input-dependent error paths. }
{To make GPS} refutationally complete (in the sense that it will find an error if one exists, if it is allotted enough time), {we introduce an instrumentation technique and show that it helps GPS achieve refutation-completeness without sacrificing overall performance.} We benchmarked GPS on a diverse suite of benchmarks {including} programs from the Software Verification Competition (SV-COMP), from prior literature, as well as synthetic programs based on
examples in this paper. We found that our implementation of GPS outperforms state-of-the-art software model checkers (including the top performers in SV-COMP ReachSafety-Loops category), both in terms of the number of benchmarks solved
and {in terms of} running time.
\keywords{Software Model Checking \and Static Analysis \and Algebraic Program Analysis}
\end{abstract}

%% file: zak_introduction.tex
\section{Introduction}
\label{Se:Introduction}

Three prominent approaches to reasoning about safety properties of programs are static analysis, automated testing, and software model checking.  Static analysis is typically concerned with computing conservative over-approximations of program behavior---static analyzers can prove safety, but not refute it.  
Automated testing derives inputs on which to run programs, aiming to maximize the amount of code exercised by the tests---testing can be used to refute safety (that is, find bugs), but not verify it.  Finally, software model-checking techniques take both a program and property of interest, and attempt to determine whether it holds.  Software model checkers are capable of both verification and refutation, but are typically slow, and sometimes fail to terminate.


One may understand the failure modes of software model checkers by viewing model checking as a search problem: \textit{given a control-flow graph that represents the program and a designated error location, find a path from entry to error that is feasible (in the sense that it corresponds to a program execution)}.  Assuming that the error location is \textit{unreachable}, a model checker may fail to discover a proof because it proves
the safety of individual paths
one path at a time, failing to find a correctness argument that generalizes to all paths.
Assuming that the error location is \textit{reachable}, a  model checker may fail to find the error path because it exhausts its computational resources searching paths that are either infeasible or do not reach the error location.

This paper proposes a software model-checking algorithm, GPS (\underline{G}uided by \underline{P}ath \underline{S}ummaries), that uses static analysis and testing techniques to mitigate  both of these failure modes.  
In particular, GPS makes use of a recent line of work on program summarization via algebraic program analysis
\cite{DBLP:conf/pldi/KincaidBBR17,cav:apa}, a static-analysis technique {that uses SMT solving} to derive a \emph{path summary}, which is a transition formula that over-approximates the input/output relation of a given set of paths.  Summarization is immediately applicable to the \textit{unreachable} failure mode: rather than prove safety of paths one-by-one, we may compute a summary $F$ and prove that $F$ is unsatisfiable using an SMT solver.  Our key insight is if $F$ is instead \textit{satisfiable}, then a model of $F$ is an input/output pair of states and the \textit{input} state corresponds to a test that may reach the error location. We can thus address the \textit{reachable} failure mode by using this kind of ``directed testing'' to guide exploration of paths through the program, ensuring that the paths searched by GPS are feasible and at least
\textit{might}
lead to the error location.

Of course, safety checking is an undecidable problem, and GPS has its own failure modes.  The summary $F$ for a set of paths is over-approximate, and so a test input generated from a model of $F$ may fail to reach the error location either because 
(1) the error location is unreachable {under any input} (but the summary $F$ was not precise enough to prove it), or 
(2) the error location {\textit{is reachable via some program path}, but cannot be reached under the given test input.} 
In an attempt to determine which, GPS explores different paths through the program (similarly to symbolic execution), and for each path $\pi$, {GPS} tries to generate a {new} test {that satisfies two criteria:}  (a) {the new test} \textit{extends} $\pi$ (ensuring a novel test because it traverses a previously-unexplored path) and (b) {the new test} \textit{may} reach the error
(where ``may'' is judged according to
a path summary for all continuations of $\pi$ to the error location).  
Successful test generation results in more paths being explored, making progress towards refuting safety. Failure to find a test case identifies a ``dead end''---a program path that cannot be extended to reach the error location.  Dead-end detection can help find bugs by leading the search toward more productive areas of the search space, and 
can prove safety by establishing that all paths lead to dead ends.  Furthermore, by using Craig interpolation from failed test generation, GPS can synthesize candidate invariants for a proof of safety.  GPS may still fail to terminate on some programs, but it is \textit{refutationally complete} in the sense that GPS will find an error if one exists, if provided enough computational resources ({similar} to bounded model checking).
\vspace*{-0.1in}
\paragraph{Contributions.}
We show how to {combine synergistically} the strengths of a summary-producing static-analysis method,
a test generator, and a software model checker, leveraging
the following mechanisms
from prior work:
\begin{enumerate}
    \item  
    Directed test generation \`{a} la \textsc{Dart},  \textsc{Synergy}, \textsc{Dash}, and \textsc{Smash} \cite{DART, synergy,Dash,Smash};
    \item Path summaries \`{a} la algebraic program analysis 
    \cite{cav:apa,fmcad:fk15};
    \item Craig interpolation and abstract reachability trees for {invariant synthesis} from the lazy abstraction for interpolants {(\textsc{Impact})} algorithm \cite{cav:mcmillan06}.
\end{enumerate}
The main novelties of 
our
work are as follows:
\begin{enumerate}
  \item GPS provides a generic method of using over-approximate summaries for \emph{directed testing},
    where the goal is
    to find a test that reaches an error location rather than to maximize code coverage. 
  {Aided by summaries, GPS is particularly effective at finding non-deterministic error paths: over-approximate summaries can guide GPS's test generator to effectively resolve non-determinism.}
  \item GPS makes use of \textit{failed} attempts 
    at generating
    directed tests to generate candidate invariants that we call \textit{dead-end interpolants}. 
    {Dead-end interpolants} are high quality in the sense that they are suitable for proving infeasibility of a possibly infinite (regular) set of paths---namely, all possible extensions of the dead-end path.
    \item {GPS employs a novel, two-layered search strategy that interleaves a depth-first algorithm for executing tests and a breadth-first algorithm for generating new tests. 
    } {To ensure \textit{refutation-completeness} of this strategy (that is, if a bug exists, GPS will eventually find it), we require that the test executor does not get stuck executing an infinite-length test.  We achieve this  {goal} by}
    {instrumenting the program with a special variable called \texttt{gas}, along with additional assumptions that ensure (a) the \texttt{gas} variable is bounded from below by zero at all times, and (b) every infinite path through the control-flow contains infinitely many decrements to \texttt{gas}.
    } Rather than searching for iteratively higher amounts of gas, GPS uses summaries to infer values of gas {in each test it generates, which, when combined with GPS's search strategy,} allows for a 
    \emph{non-uniform}
    search that is capable of finding long counter-examples.
     {We further show that this} instrumentation technique {can help} the algorithm converge on certain classes of benchmarks featuring highly input-dependent loops (\Cref{ssec:gas}).

    \item Almost every step of the GPS algorithm requires a \emph{single-target} path-summary query to the static-analysis engine. We show that such queries can be efficiently precomputed offline by using Tarjan's algorithm for finding single-source/multi-target path expressions \cite{tarjan} in dual mode, so that it instead computes single-target/multi-source path expressions (\Cref{Se:SingleTargetSummaries}). 
\end{enumerate}
\noindent  We implemented GPS 
 using a variant of compositional recurrence analysis \cite{fmcad:fk15} to compute summaries, and evaluated GPS on a rich suite of benchmarks, including intraprocedural programs from the Software Verification Competition;
 from prior literature;
 as well as synthetic programs based on examples in this paper.
 Our experiments show that \emph{GPS outperforms several state-of-the-art software model checkers}
 (\Cref{tab:rq13-benchmark-stats})
 {while being especially effective at bug-finding for a class of programs featuring input-dependent loops, which we dub \emph{``Lock \& Key" problems} (defined in \Cref{sec:overview}).}

\begin{wrapfigure}{r}{5cm} 
\begin{lstlisting}[style=base] 
int main() {
  int N = havoc();
  int i = 0, min = 0;
  while (i < N) {
    min++; i++;
  }
  if (min < 1000) return 0
  else assert(0)
}\end{lstlisting}
\caption{Example EX-1 in \Cref{sec:overview}. }
\label{fig:ex1}
\vspace*{-0.2in}
\end{wrapfigure}

The remainder of the paper is structured as follows.
\Cref{sec:overview} illustrates how GPS can prove and refute safety using a series of examples.  \Cref{sec:background} reviews necessary background.  To facilitate the presentation of our technique, we develop it in two steps: \GPSLite and GPS.
\GPSLite (\Cref{sec:gpslite_algorithm}) is a simplified variant of GPS that illustrates how program summaries can be used for both verification and refutation.  GPS (\Cref{sec:algorithm}) augments \GPSLite with the facility to generate invariants from dead-end interpolants.  \Cref{sec:apa} discusses how algebraic program analysis can be used to compute path summaries, and gives an efficient algorithm for computing the single-target/multi-source path summaries that are required for GPS. Experimental results appear in~\Cref{sec:eval}, and a discussion of related {work} {is given} in~\Cref{sec:relwork}.
\IfCameraReady{Additional statistics about our experiments, as well as proofs about theorems in this paper may be found in \rfchanged{the extended version of this paper \cite{GPSExtendedVersion}}.}{Additional statistics about our experiments, as well as proofs about theorems in this paper may be found in~\Cref{appendix:proofs}.}




%% file: overview.tex
\section{Overview}
\label{sec:overview}


In this section, we highlight key insights behind the GPS algorithm via three motivating examples. {The first example illustrates how over-approximate path summaries produced by static analysis can aid in finding an assertion violation.
The second and third examples illustrate how GPS uses over-approximate (but not necessarily precise) summaries in verifying safety. }

\subsection{Summary-Directed Test Generation}

We consider a bug-hunting task adapted from a real-world web-server bug, discussed in \cite{ISSTA:SPMS09} and presented in~\Cref{fig:ex1}.
EX-1 is a buggy program with a long error path---the loop must be executed $N$ times---and moreover, for a tool to exercise the bug, it must find one of the values for $N$ that trigger the bug---in this case, at least 1000.
This combination {of a long error path and input-dependent control flow} is challenging for existing tools because they must find an appropriate value for a nondeterministic expression (in this case, $N = \mathsf{havoc()}$) {\emph{and} {efficiently} traverse a long path to trigger the error}.
One could imagine another kind of buggy program {that takes in an arbitrary integer value as input, but} an assertion violation could only be triggered when one guesses the specific input value of 37. Collectively, we refer to these kinds of programs as \emph{``Lock and Key" (L\&K) problems}: \emph{unlocking} the assertion violation is contingent on a good \emph{key} (i.e., correct values for non-determinstic {expressions}).
Algorithms that search for errors along paths of increasing length
(e.g., bounded model checking or IC3 \cite{ICThree}) may exhaust their resources before reaching the necessary length bound, while testing techniques may fail to find the correct value of $N$ to reach the error (e.g., Saxena et al.\ \cite{ISSTA:SPMS09} identified EX-1 as a challenging problem for concolic-execution engines).

The GPS algorithm begins by computing a summary $F$ that over-approximates all paths from entry to the assertion location {(using an off-the-shelf static analyzer)}:
\[
  F \defeq \exists k. k \geq 0 \land (k \geq 1 \Rightarrow 0 < N \land i' \leq N') \land \textit{min}' = i' = k \land N = N' \land \textit{min}' \geq 1000
\]
In the formula $F$, program variables correspond to their values at the beginning of the program, while primed variables correspond to their values at the assertion location. {GPS then uses an SMT solver to determine the satisfiability of $F$.}  If this formula is \textit{unsatisfiable}, then the assertion is safe; however in this instance it is satisfiable---for example, we have the model
\[ M \defeq \{ N \mapsto 1000, i \mapsto 0, \textit{min} \mapsto 0, N' \mapsto 1000, i' \mapsto 1000, \textit{min} \mapsto 1000 \}\]
Because summaries are over-approximate, satisfiability of the summary $F$ does not imply that the assertion can fail.  {However, in this case, $M$ corresponds to an actual counter-example input that witnesses the assertion violation. } GPS is {thus} able to confirm that the assertion can be violated by simulating executing from the 
initial state $\{ N \mapsto 1000, i \mapsto 0, \textit{min} \mapsto 0 \}$ (corresponding to the values of the \textit{unprimed} variables in the model $M$). {Here, the key insight is that} \emph{if a summary-based static analysis fails to prove safety of an assertion, the failure provides an opportunity to refute the assertion by finding a test case using the summary}.

\smallskip
\begin{mdframed}
    \textbf{Insight \#1: } If a given path summary is satisfiable, we can use a model recovered from
    a
    satisfiable query as a new test.
\end{mdframed}

\subsection{Verification through Dead-End Detection}
\label{sec:VerificationThroughDeadEndDetection}

EX-1 demonstrates how path summaries can be used to guide test generation toward an error.  However, path summaries are over-approximate in nature, and so a test generated using a path summary may not actually reach the error location.  In such cases (where the generated test is spurious), we must manage the search space for errors to allow \rfchanged{GPS} an opportunity to select another potential error path, or to exhaust the space and prove that the error is unreachable.  This section illustrates how summaries can be used for \textit{dead-end} detection to prove safety in the context of the simplified algorithm \GPSLite (which omits the invariant-synthesis capability of GPS, and will be illustrated in
\Cref{sec:InvariantSynthesisWithDeadEndInterpolation}).

\begin{figure}
\begin{subfigure}{0.45\textwidth}
\begin{lstlisting}[style=base,numbers=left]
void main(int N) {
  int r = 0;
  while (N > 0) {
    r = r + N--;
  }
  assert(r != 2);
}
\end{lstlisting}
\subcaption{Pseudo-code for program EX-2.}
\end{subfigure}
\begin{subfigure}{0.45\textwidth}
\small
\scalebox{0.8}{
\begin{tikzpicture}[v/.style={draw,circle,fill=black,text=white},
e/.style={draw,circle,fill=red,text=white}, node distance=2cm,thick,>=stealth]
  \node [v] (A) {A}; 
  \node [v,right of=A] (B) {B}; 
  \node [v,right of=B] (C) {C}; 
  \node [e,right of=C] (D) {D}; 
  \draw (A) edge[->] node[above]{$r' = 0$} (B);
  \draw (B) edge[->] node[above]{$\begin{array}{l@{}l}&N \leq 0\\ \land& N' = N\\ \land& r' = r\end{array}$} (C);
  \draw (C) edge[->] node[above]{$r = 2$} (D);
  \draw (B) edge[loop below,->] node[below]{$\begin{array}{l@{}l}&N > 0\\ \land& N' = N - 1\\ \land& r' = r + N\end{array}$} (B);
\end{tikzpicture}
}
\subcaption{A control-flow graph for (a).}
\end{subfigure}
\begin{subfigure}{0.48\textwidth}
\footnotesize
\scalebox{0.8}{
\begin{tikzpicture}[style=pathtree]
  \node [v] (A) {A}
    child {
      node [v] {B}
        child {
          node[v] {B}
          child { node[l] {B} }
          child {
            node[v] {C}
              child { node[l] {D} }
          }
        }
        child { node[l] {C} }
    };
\end{tikzpicture}
}
\subcaption{A \textit{path tree} produced by the test $\set{r \mapsto 0, N \mapsto 1}$ \label{fig:ex2-pathtree}}
\end{subfigure}
\caption{EX-2: An illustration of the use of dead-end detection to verify safety.}
\label{fig:ex2}
\end{figure}
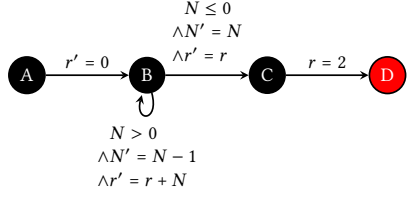
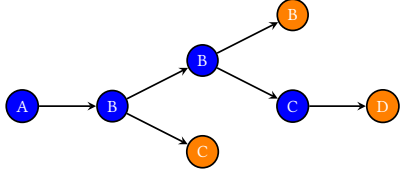

Consider the example EX-2 shown in \Cref{fig:ex2}. EX-2 contains a safe assertion; we wish to verify it. \GPSLite begins by computing a summary that over-approximates the set of paths from the entry vertex \encircle{A} to the error vertex \encircleR{D}.  This summary can be understood as the sequential composition of the formula $F_0 \defeq r' = 0 \land N' = N$, which labels the edge from \encircle{A} to \encircle{B},  with the formula
\begin{align*}
F \defeq \exists k . ((k = 0 \land r' = r) \lor (k \geq 1 \land 0 < N \land N' \geq 0)) \land N' = N-k \land r+k \leq r' \land r' = 2,
\end{align*}
where $F$ summarizes \emph{all} paths from $\encircle{B}$ to $\encircleR{D}$ (including any number of iterations of the $\encircle{B}$ loop).
The formula $F$ is over-approximate:
it represents the fact that $r$ must increase by \textit{at least} $k$ after $k$ iterations of the $\encircle{B}$ loop, but does not represent its exact closed form.  

The over-approximation
in turn makes the formula $F_0 \circ F$ \emph{satisfiable};\footnote{As defined in~\Cref{sec:background}, here $\circ$ denotes relational composition of two transition formulas.} 
thus, \GPSLite generates a test input
$M \defeq \set{N \mapsto 1, r \mapsto 0}$ from a model of $F_0 \circ F$.  In contrast to EX-1, simulating execution from $M$ does \textit{not} lead to an assertion violation.  Instead, the path traversed by this test is used to unwind the control-flow graph of EX-2, producing the \textit{path tree} shown in~\Cref{fig:ex2-pathtree}.
In the path tree, blue nodes represent the path traversed by the test, while the leaves of the path tree are endpoints of unexplored paths, shown in orange.\footnote{Concretely, the path tree includes all prefix paths of $\encircle{A}\to\encircle{B}\to\encircle{B}\to\encircle{C}$.}

\GPSLite continues its search by selecting a path to one of the orange vertices, say $\encircleB{A} \rightarrow \encircleB{B} \rightarrow \encircleO{C}$, and checks whether the composition of the transition formulas along this new path ($N \leq 0 \land N'=N \land r'=0$)
and the summary of all paths from \encircle{C} to \encircle{D} ($r = 2$) is satisfiable.
The formula obtained in this way is
not satisfiable---meaning that \emph{no feasible path} with prefix $\encircle{A}\to \encircle{B}\to \encircle{C}$ can arrive at the error vertex \encircleR{D}---so \GPSLite declares the path $\encircle{A} \rightarrow \encircle{B} \rightarrow \encircle{C}$ to be a dead end.

\GPSLite proceeds to consider a path represented by the next leaf in the path tree,
$\pi = \encircleB{A} \rightarrow \encircleB{B} \rightarrow \encircleB{B} \rightarrow \encircleO{B}$, and again checks whether the composition of the transition formulas along $\pi$ ($r' = 2N - 1 \land N' = N-2 \land N>1$)
and the summary of paths from $\encircle{B}$ to $\encircle{D}$ (formula $F$ above) is satisfiable.  Again, the answer is no:
$r$ must be at least 3 after executing $\pi$, whereas
the path-to-error summary $F$ indicates that $r$ is increasing and bounded above by 2;
thus, $\pi$ is a dead end as well.

Finally, \GPSLite
considers the {last} unexplored path $\encircle{A} \rightarrow \encircle{B} \rightarrow \encircle{B}  \rightarrow \encircle{C} \rightarrow \encircle{D}$, finds that this path
too is a dead end, and concludes that the program is safe.
The behavior of the full algorithm, GPS, follows the same recipe outlined above. 
This example illustrates how GPS(Lite) can use path summaries to prove safety, even in cases where the path summary for the whole program is not sufficiently precise to do so.

\begin{mdframed}
    \textbf{Insight \#2: }
    Over-approximate summaries can help establish that a control-state is a dead end, thereby eliminating a whole class of paths from consideration.
\end{mdframed}

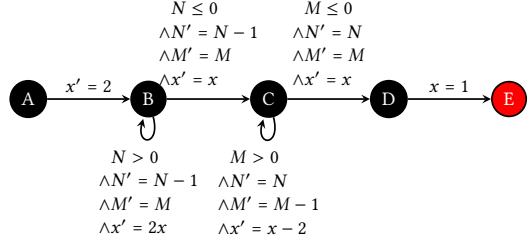
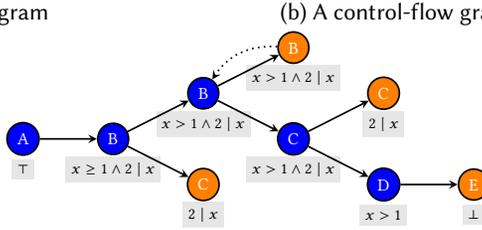
\begin{figure}
\begin{subfigure}{0.35\textwidth}
\begin{lstlisting}[style=base,numbers=left]
void main(int N, int M) {
  int x = 2;
  while (N-- > 0)
    x = 2 * x;
  while (M-- > 0)
    x = x - 2;
  assert(x != 1);
}
\end{lstlisting}

\subcaption{An example program}
\end{subfigure}
\begin{subfigure}{0.6\textwidth}\small
\scalebox{0.8}{
\begin{tikzpicture}[v/.style={draw,circle,fill=black,text=white},
e/.style={draw,circle,fill=red,text=white}, node distance=2cm,thick,>=stealth]
  \node [v] (A) {A}; 
  \node [v,right of=A] (B) {B}; 
  \node [v,right of=B] (C) {C}; 
  \node [v,right of=C] (D) {D};
    \node [e,right of=D] (E) {E}; 
  \draw (A) edge[->] node[above]{$x' = 2$} (B);
  \draw (B) edge[->] node[above]{$\begin{array}{l@{}l}&N \leq 0\\ \land& N' = N-1\\\land&M'=M\\ \land& x' = x\end{array}$} (C);
    \draw (C) edge[->] node[above]{$\begin{array}{l@{}l}&M \leq 0\\ \land& N' = N\\ \land&M'=M\\ \land& x' = x\end{array}$} (D);
  \draw (D) edge[->] node[above]{$x = 1$} (E);
  \draw (B) edge[loop below,->] node[below]{$\begin{array}{l@{}l}&N > 0\\ \land& N' = N - 1\\ \land&M'=M\\ \land& x' = 2x\end{array}$} (B);
    \draw (C) edge[loop below,->] node[below]{$\begin{array}{l@{}l}&M > 0\\ \land& N' = N\\ \land & M' = M-1 \\\land& x' = x-2\end{array}$} (C);
\end{tikzpicture}
}
\subcaption{A control-flow graph for (a).}
\end{subfigure}
\begin{subfigure}{\textwidth}
\footnotesize
\begin{center}
\scalebox{0.8}{
\begin{tikzpicture}[style=pathtree,label/.style={fill=black,fill opacity=0.1,text opacity=1}]
    \exthree{}
    \draw (ABBB) edge[->,dotted, bend right] (ABB);

    \artlabel{A}{\true}
    \artlabel{AB}{x \geq 1 \land 2 \mid x}
    \artlabel{ABB}{x > 1 \land 2 \mid x}
    \artlabel{ABC}{2 \mid x}
    \artlabel{ABBC}{x > 1 \land 2 \mid x}
    \artlabel{ABBB}{x > 1 \land 2 \mid x}
    \artlabel{ABBCD}{x > 1}
    \artlabel{ABBCDE}{\false}
    \artlabel{ABBCC}{2 \mid x}
\end{tikzpicture}
}
\end{center}
\subcaption{An \textit{abstract reachability tree} (ART)---a path tree with labels and covering edges (shown in dashes)---proving safety of EX-3.}
 \label{fig:ex3-art}
\end{subfigure}
\caption{EX-3: An illustration of the use of dead-end interpolation to verify safety.}
\label{fig:ex3}
\end{figure}

\subsection{Invariant Synthesis with Dead-End Interpolation}
\label{sec:InvariantSynthesisWithDeadEndInterpolation}

As previously described, \GPSLite is essentially a variant of concolic testing \cite{FSE:SMA05,MajumdarSenICSE07}, but augmented with the use of static analysis-generated summaries to guide test generation and detect dead ends.  In some cases (such as EX-2), dead-end detection is sufficient to prove safety; in other cases---especially programs with complex loops---it is necessary to synthesize an invariant that separates the reachable states of the program from the error. Here, we give an example of how the full GPS algorithm can synthesize such invariants by combining dead-end detection---the idea
illustrated in \Cref{sec:VerificationThroughDeadEndDetection}---with
Craig interpolation and abstract reachability trees from the {\sc Impact} family of software-model-checking algorithms \cite{cav:mcmillan06,ufo,fmcad:bkw10}.

\paragraph{Example EX-3.} Consider another program with a safe assertion, shown in~\Cref{fig:ex3}. Suppose that one were to run \GPSLite on EX-3. 
 We begin by computing a summary that over-approximates all paths from the entry vertex \encircle{A} to the error vertex \encircleR{E}.  
Conceptually, we can understand this summary as the sequential composition of the 
two summaries 
\begin{align*}
F_1 &\defeq \exists k_1. N' = N - k_1 \land M' = M \land x' \geq 2 + k_1 \land (k_1 \geq 1 \Rightarrow N' \geq 0) \land N' \leq 0\\
F_2 &\defeq \exists k_2. N' = N \land M' = M - k_2 \land x' = x - 2k_2 \land (k_2 \geq 1 \Rightarrow M' \geq 0) \land M' \leq 0 \land x' = 1
\end{align*}
where $F_1$ summarizes paths from \encircle{A} to \encircle{C} (excluding looping paths at \encircle{C}), and $F_2$ summarizes paths from \encircle{C} to \encircleR{E}.  Note that $F_2$ is precise, while $F_1$ \emph{is not} (in particular, it does not represent the fact that $x' = 2^{k_1+1}$,
and hence is over-approximate).

\GPSLite determines that the composition of $F_1$ and $F_2$ is satisfiable, and generates a test input
{$\{ N \mapsto 1, x \mapsto 2, M \mapsto 0 \}$}
from a model.
Similar to the previous example, simulating execution
with 
{this input} does not lead to an assertion violation, and instead produces a path tree, {depicted} in Fig.~\ref{fig:ex3-art}
(for now, ignore 
the gray boxes and dashed edges). 
\GPSLite is able to determine that
$\encircle{A} \rightarrow \encircle{B} \rightarrow \encircle{C}$ is a dead end, but 
(because $F_1$ over-approximates the behaviors along the paths from \encircle{A} to \encircle{C})
cannot establish that $\encircle{A} (\rightarrow \encircle{B})^i$ is a dead end for all $i \in \mathbb{N}$;
as a result, \GPSLite diverges for EX-3, producing infinitely many tests, none of which reach the error location
(because the error location is unreachable).

Our full algorithm, GPS, is able to prove safety of EX-3 by exhibiting a \textit{complete and well-labeled abstract reachability tree (ART)}, shown in Figure~\ref{fig:ex3-art}.  Intuitively, we can think of an ART as a Floyd/Hoare proof that every path in EX-3 is a dead end.
More concretely, an ART is a path tree in which (a) each {path} 
is labeled with a state formula 
(depicted
in gray)---which {describes} {an \emph{over-approximation} of the states reachable along} {the} path{, and} {functions as an invariant candidate}---and (b)
there is
an additional set of \textit{covering edges} (dashed lines) between identical control locations on a path.
{The presence of a covering edge means that an {inductive} invariant has been found for a loop}. 
Observe that each leaf in the pictured ART (with the exception of the \encircleO{B} leaf, which has an outgoing covering edge) is labeled with a condition {that is inconsistent with the path-to-error summary for that node, thus indicating that it is a ``dead end''.}
For instance, each $\encircleO{C}$ leaf is labeled with the condition ``$2 \mid x$'' (i.e., $x$ is even), which is inconsistent with $F_2$, and the $\encircleO{E}$ leaf is labeled with ``$\false$''
{
($\false$ denotes $\textit{false}$,
}
which is inconsistent with the identity transition formula).  The label of the $\encircle{B}$ leaf is not a dead-end condition, however it is an inductive invariant of the first loop in EX-3---and indeed, this is indicated by the dashed covering edge between $\encircleO{B}$ and the second $\encircleB{B}$ node.

To derive the ART shown in~\Cref{fig:ex3-art}---and thus arrive at a safety
{
proof---GPS  generates an initial test, attempts to explore the path
$\encircle{A} \rightarrow \encircle{B} \rightarrow \encircle{C}$, and discovers that it is a dead end.
}
GPS uses the unsatisfiability proof from dead-end detection to
generate a sequence of three state formulas
$\top$, $2\mid x$, $2 \mid x$, which constitute a Floyd/Hoare proof that the path $\encircle{A} \rightarrow \encircle{B} \rightarrow \encircle{C}$ is a dead end: we call these formulas \textit{dead-end interpolants} for the path $\encircle{A} \rightarrow \encircle{B} \rightarrow \encircle{C}$.  Similarly, GPS finds dead-end interpolants for the paths
$\encircle{A} \rightarrow \encircle{B} \rightarrow \encircle{B} \rightarrow \encircle{C}$ ($\top$, $2 \mid x$, $2 \mid x$, $2 \mid x$, $2 \mid x$) and
$\encircle{A} \rightarrow \encircle{B} \rightarrow \encircle {B} \rightarrow \encircle{C} \rightarrow \encircle{D} \rightarrow \encircle{E}$ ($\top$, $x \geq 1$, $x > 1$, $x > 1$, $x > 1$, $\bot$), and conjoins the dead-end interpolants across these three paths to arrive at the labels pictured in Figure~\ref{fig:ex3-art}.  Finally, GPS checks that the label of the path $\encircle{A} \rightarrow \encircle{B} \rightarrow \encircle{B}$ can be extended to $\encircle{A} \rightarrow \encircle{B} \rightarrow \encircle{B} \rightarrow \encircle{B}$ (i.e., $x > 1 \land 2 \mid x$ is an inductive invariant of the first loop), thus completing the proof.

Intuitively, dead-end interpolants are good candidates for inductive invariants, because they prove that \textit{all} extensions of a given
{
path---of which there are typically infinitely many---are unsatisfiable.
}

\smallskip
\begin{mdframed}
    \textbf{Insight \#3: }
    Interpolants generated from summary-based dead-end detection can be used to derive high-quality candidate invariants.
\end{mdframed}

\smallskip
{
There is mutually beneficial synergy between Insight \#2 and Insight \#3:
each step of the GPS algorithm typically considers a non-error leaf of the ART, and thus the path-to-error summary for the leaf describes a (possibly infinite) set of paths.
If a dead end is detected in that step, interpolation is forced to generalize with respect to that set of paths.
When the set of paths includes a loop, such generalization steps {produce} good candidates for inductive invariants.
}

\barrier %

%% file: background.tex
\section{Background} \label{sec:background}
Fix a set of variable symbols $X$.  We use $\SF{X}$ denote the set of \textit{state formulas} over $X$ (whose free variables are drawn from $X$).  A \textbf{transition formula} over $X$ is a formula whose free variables range over $X \cup X'$, where $X'$ denotes the set of ``primed'' variants of the variables in $X$.  We use $\TF{X}$ to denote the set of all transition formulas over $X$.  For transition formulas $F$ and $G$ in $\TF{X}$, we use $F \circ G \defeq \exists X''. F[X' \mapsto X''] \land G[X \mapsto X'']$ to denote the relational composition of $F$ and $G$.
{For states $M$ and $M'$ and a transition formula $F$, we use
$\tfrel{M}{F}{M'}$ to denote that $M$ can transition to $M'$ along $F$.
For a transition formula $F$ and a state formula $\psi$, we define $\spost{\psi}{F} = (\exists X. \psi \land F)[X' \mapsto X]$ to be the strongest post-condition of $\psi$ along $F$.}

{We represent a program as a weighted graph.} A \textbf{weighted graph} 
 is a tuple $G = \tuple{V,E,w,s,t}$ where $V$ is a finite set of vertices, $E \subseteq V \times V$ is a set of directed edges, $w : E \rightarrow \TF{X}$ is a weight function associating each
edge
with a transition formula, and $s,t \in V$ are designated \textit{source} and \textit{terminal} vertices, such that $s$ is the unique vertex of in-degree 0 in $G$ and $t$ is the unique vertex of out-degree 0 in $G$. A \textbf{path} in $G$ is a sequence of 
edges $e_1 e_2 \dots e_n$ such that the destination of each $e_i$ is the source of $e_{i+1}$.   For vertices $u,v \in V$, we use
$\Paths{G}{u}{v}$ to denote the set of paths in $G$ that begin at $u$ and end at $v$.  

Let  $\pi = e_1 e_2 \dotsi e_n$ be  a path. We use $\psrc{\pi}$ to denote the source vertex of $e_1$, $\pdst{\pi}$ to denote the destination vertex of $e_n$; if $\pi$ is the empty path, define $\psrc{\pi} = \pdst{\pi} = s$. We extend the weight function $w$ to paths by defining $w(\pi) = w(e_1)\circ w(e_2)\circ ...\circ w(e_n)$ to be the composition of edge weights along $\pi$. 
For paths $\pi$ and $\pi'$, we use $\pi \prefixord \pi'$ to denote that $\pi$ is a prefix of $\pi'$; we call this the \emph{prefix order}.  {For two paths $\pi$ and $\sigma$ with $\pdst{\pi}=\psrc{\sigma}$, we write $\pi\sigma$ to denote the concatenation of $\pi$ with $\sigma$.} For any prefix $\tau = e_1 \dots e_i$ of $\pi$, let $\tau^{-1}\pi$ denote the suffix $e_{i+1}\dots e_n${; note that in this notation, $\tau (\tau^{-1}\pi) = \tau e_{i+1},...,e_n = \pi$.} 

Let $G = \tuple{V,E,w,s,t}$ be a weighted graph.  We say that $G$ is \textbf{unsafe} if there exists a path $\pi \in \Paths{G}{s}{t}$ from $s$ to $t$ such that $w(\pi)$ is satisfiable; if so, we say that such a path $\pi$ is a \textbf{feasible} $s$-$t$ path.
If no such feasible $s$-$t$ path exists, we say that $G$ is safe. The GPS algorithm solves the following problem:

\begin{problem}[Control-state Reachability]
    Given a weighted graph $G$, decide whether $G$ is safe.
\end{problem}

For presenting the GPS algorithm, we assume access to a \textit{path summary oracle} $\mathsf{Sum}(G,u,v)$, which given a weighted graph $G$ and two vertices $u$ and $v$, returns a transition formula that over-approximates all paths from $u$ to $v$ in $G$ (i.e., $w(\pi) \models \mathsf{Sum}(G, u, v)$ for all $\pi \in \textit{Paths}_G(u,v)$).  In ~\Cref{sec:apa}, we review how to use algebraic program analysis \cite{cav:apa} to produce such an oracle, and give an algorithms that is particularly well-suited to GPS.  However, the GPS algorithm is agnostic as to how path summaries are obtained.

%% file: sgt.tex
\section{\GPSlite: Summaries for Test Generation and Dead-End Detection}
\label{sec:gpslite_algorithm}

In this section, we describe \GPSlite, a simplified variant of the GPS algorithm that we use to stage the exposition of the GPS algorithm---and highlight the role of path summaries in generating tests and detecting dead ends.  In Section~\ref{sec:algorithm}, we describe how \GPSlite can be combined with Craig interpolation and abstract reachability trees \`{a} la lazy abstraction with interpolants ({\sc Impact}) \cite{cav:mcmillan06} to perform invariant synthesis, thus yielding the GPS algorithm.

\GPSlite can be understood as a combination of path summaries with testing and symbolic execution, similarly to concolic testing (as in {\sc Dart} and {\sc Cute}).  In contrast to pure concolic testing, \GPSlite uses summaries to generate test inputs that are ``likely'' to reach an error location (rather than generating inputs randomly, or to maximize code coverage).

\GPSlite uses a \textbf{path tree} to represent its search space.
A path tree for a weighted graph $G = \tuple{V,E,w,s,t}$ is a finite collection of paths of $G$ emanating from the source vertex $s \in G$.  Intuitively, the non-maximal paths of a path tree $T$ represent paths that have already been explored (and proven to be feasible), while the maximal paths of $T$ (the \textit{leaves} of the path tree) represent paths that are either dead ends (cannot be extended to an error trace) or remain to be explored.  

\begin{definition}[Path Tree]
\label{def:path_tree}
A \textbf{path tree} for 
a weighted graph $G = \tuple{V,E,w,s,t}$ 
is a set of paths $T \subseteq \bigcup_{u \in V} \text{Paths}_G(s, u)$ satisfying the following conditions:
\begin{itemize}
\item \textbf{$T$ is prefix-closed:} For any path $\pi \in T$ and any prefix $\tau$ of $\pi$, we have $\tau \in T$.
\item \textbf{$T$ is full:} For any path $\pi \in T$, if there is some edge $e \in E$ such that $\pi e \in T$, then $\pi e' \in T$ for any edges $e' \in E$ with $\dst(\pi) = \src(e')$.  
\end{itemize}
Note that these two conditions ensure that for any path $\pi$ beginning at $s$, there is a leaf of $T$ that is a prefix of $\pi$---in this sense, the leaves of a path tree represent all possible opportunities for expanding the tree. 
\end{definition}

The \GPSlite algorithm is shown in Figure~\ref{fig:SummaryGuidedTesting}. \GPSlite takes as input a weighted graph $G = \tuple{V,E,w,s,t}$, and determines whether $G$ is safe.  It maintains two data structures: a path tree $T$, and a queue $\TreeFrontiers{T}$, with the invariant that each maximal path $\pi \in T$ is either a dead end or belongs to $\TreeFrontiers{T}$.
It terminates when it either finds a feasible $s$-$t$ path witnessing that $G$ is unsafe with respect to $\psi$, or it exhausts \textsf{frontierQueue} (proving that $G$ is safe with respect to $\psi$ by virtue of establishing each leaf of $T$ as a dead end).

In each iteration of the main loop, \GPSlite selects a path $\pi$ from $\TreeFrontiers{T}$ and attempts to generate a \textit{directed} test that extends $\pi$---that is, a state $M$ that
(1) \textit{is reachable} along the path $\pi$
and (2) \textit{may reach} the target location $t$ along the path summary 
$\mathsf{Sum}(\CFG,\dst(\pi),t)$.
This step is accomplished using the following primitive (which can be implemented via a query to an SMT solver):

\vspace*{0.1in}
\noindent
\begin{minipage}{0.61\textwidth}
\begin{mdframed}
\small
    \textbf{Primitive }$\CheckSat{F, G}$: 
    
    \begin{itemize}
        \item \textbf{Input: } A pair of transition formulas $F$ and $G$
        \item \textbf{Output: } Returns either:
            \begin{enumerate}
        \item $(\mathsf{SAT}, M)$ if {$F \circ G$} is satisfiable, where
        $M$ is a state such that 
        $L \xrightarrow{F} M \xrightarrow{G} N$
        for some $L$ and $N$  (i.e., $M$ is a witness for the midpoint of the composition of $F$ and $G$).
        \item Or $\mathsf{UNSAT}$, if $F \circ G$ is unsatisfiable.
    \end{enumerate}
    \end{itemize}
\end{mdframed}
\end{minipage}
\hfill
\begin{minipage}{0.38\textwidth}
\scriptsize\centering
        \includegraphics[width=\linewidth]{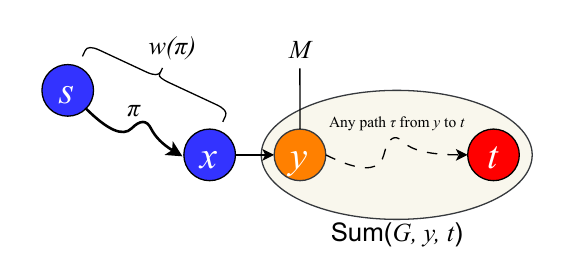} \\
        Illustration of $\CheckSat{w(\pi), \mathsf{Sum}(G, \dst(\pi), t)}$ where $y = \dst(\pi)$ and $x$ is predecessor of $y$.
\end{minipage}
\vspace*{0.1in}

If $\CheckSat{w(\pi), \mathsf{Sum}(G,\dst(\pi),t)}$ fails to find a model $M$, then $\pi$ is a dead end: for any
path $\tau$ from $\dst(\pi)$ to $t$, we must have
that $w(\tau)$ entails $ \mathsf{Sum}(G,\dst(\pi),t)$, so if $w(\pi) \circ \mathsf{Sum}(G,\dst(\pi),t)$ is unsatisfiable, so is
$w(\pi) \circ w(\tau) = w(\pi\tau)$.  On the other hand, if $\CheckSat{w(\pi), \mathsf{Sum}(G,\dst(\pi),t)}$ finds a model $M$, then we have established that $\pi$ is feasible, and executing from $M$ 
\emph{might}
to lead to the error because $M$ lies in the {domain} of $\mathsf{Sum}(G, \dst(\pi), t)$; thus, we can simulate execution from $M$ to find yet more feasible paths (extending $\pi$), with the goal of reaching $t$.

\begin{figure}[tb!]

    \begin{algorithmic}[1]
\small
\Procedure{GPSLite}{$G, \psi$}
    \Statex \Input{Weighted graph $G=(V, E, w, s, t)$ }
    \Statex \Output{Either $\textsf{Safe}$, if there exists no feasible $s$-$t$ path in $G$, or $(\textsf{Unsafe}, \pi)$ where $\pi$ is a feasible $s$-$t$ path.}
    \State $T \gets \set{ \epsilon }$;
    \State $\textsf{frontierQueue} \gets [\epsilon]$
    \While{\textsf{frontierQueue} is not empty}
    \State $\pi \gets \textsf{frontierQueue}.\textsf{dequeue()}$\;
    \Switch{$\mathsf{Check}^+(w(\pi), \textsf{Sum}(G,\dst(\pi),t))$}
    \Case{Sat(M)}
    \Switch{\textsc{Explore}($G,T,\textsf{frontierQueue},\pi,M$)}
    \Case{Safe} \textbf{continue}
    \EndCase
    \Case{$(\text{Unsafe},\tau)$}
    \Return{$(\text{Unsafe},\tau)$}
    \EndCase
    \EndSwitch
    \EndCase
    \Case{Unsat} \textbf{continue}\;
    \EndCase
    \EndSwitch
    \EndWhile
    \State \Return{Safe}
\EndProcedure

\Procedure{Explore}{$G, T, \textsf{frontierQueue}, \pi, M$}
    \Statex \Input{Weighted graph $G = (V, E, w, s, t)$, path tree $T$, path $\pi \in T$,  model $M$}
    \Statex \Output{$\text{Safe}$ when {$t$} is not reached; otherwise $(\text{Unsafe}, \tau)$, where $\tau$ is a feasible $s$-$t$ path}
        \If{$\pdst{\pi} = t$} 
            \Return $(\text{Unsafe}, \pi)$; \Comment{Terminal location reached; return}
        \EndIf 
        \For{$e\in E$ with $\pdst{\pi}$ = $\psrc{e}$}\label{line:explore:expand}
        \State $T.\text{add}(\pi e)$;
            \State $\textit{Next} \gets w(e)[X \mapsto M(X)]$ \Comment{{Set of states $M'$ such that $\tfrel{M}{w(e)}{M'}$}}
            \Switch{$\mathsf{Check}^+(\textit{Next},
            \mathsf{Sum}(G, \pdst{e}, t))$}\label{line:Explore_guided} 
                \Case{$\text{SAT}(M')$} \Comment{$M$ has successor $M'$ along $e$. }
                \Switch{\textsc{Explore}($G,T,\textsf{frontierQueue}, \pi e, M'$)}
                \Case{$(\text{Unsafe}, \tau)$)}
                \Return{$(\text{Unsafe}, \tau)$}
                \EndCase
                \Case{Safe}
                \textbf{continue}\label{line:GPSLite:unsat}
                \EndCase
                \EndSwitch
                \EndCase
                \Case{$\text{UNSAT}$} \Comment{$M$ has no successor along $e$. }
                    \State $\textsf{frontierQueue}.\text{enqueue}(\pi e);$  \Comment{Halt simulation at $\pi e$ and add to frontier.}
                \EndCase
            \EndSwitch
        \EndFor
    \State \Return \text{Safe};
\EndProcedure

\end{algorithmic}
    \caption{The \GPSlite algorithm.}
\label{fig:SummaryGuidedTesting}
\end{figure}

\GPSlite uses the  
\textsc{Explore} sub-routine (\Cref{fig:SummaryGuidedTesting}) to simulate executing a test.
Given a weighted graph $\CFG$, a path tree $T$, a frontier queue $\TreeFrontiers{T}$, a (feasible) path $\pi$, and model $M$ (reachable along $\pi$), $\textsc{Explore}$ simulates execution of $G$ beginning at location $\TreeVtx{\Lambda}{\pi}$ with state $M$.
Should $\dst(\pi) = t$, then $\pi$ is a feasible $s$-$t$ path, and so we may halt execution (and the \GPSlite algorithm).
Otherwise, we enter the loop at line~\ref{line:explore:expand} of
$\textsc{Explore}(-)$ and simulate execution along each edge $e$ emanating from $\pdst{\pi}$.
For each edge $e$ along which execution may progress---we have $\tfrel{M}{w(e)}{M'}$ for some state $M'$---we may recursively explore the (feasible) path $\pi e$ starting at state $M'$.  For each edge $e$ for which no such $M'$ exists, $\pi e$ is added to $\TreeFrontiers{T}$, to be revisited in the main loop of \GPSlite.

Should an edge $e$ correspond to a deterministic program command, we may find the state $M'$ such that $M \xrightarrow{w(e)} M'$ (should it exist) simply by executing it.  However, our program model permits non-deterministic commands: in general there may be infinitely many models $M'$ such that $\tfrel{M}{w(e)}{M'}$.  One may determine whether such an $M'$ exists by checking satisfiability of the formula
$\textit{Next} \defeq w(e)[X \mapsto M(X)]$ (replacing each pre-state variable with its corresponding value in $M$, thus producing formula whose models correspond exactly to the set $\set{ M' : \tfrel{M}{w(e)}{M'}}$).  The \textsc{Explore}(-) procedure refines this approach by further constraining the successor state $M'$ to be one that may reach the target location
(according to the path summary $\textsf{Sum}(G,\pdst{e},t)$).  By doing so, we reap two benefits: (1) this
choice may do a better job of directing
execution toward the target location, and (2) we can terminate execution early if 
\textsc{Explore}(-) cannot reach the error location using $M'$.

{
\bfpara{Discussion.} The main insight behind \GPSlite is two-fold: first, (over-approximate) summaries can effectively guide test generation; second, the summary-guided tests can be used to guide exploration of program states in a two-layered search: a breadth-first exploration algorithm can be used to generate targeted tests along frontiers of the path tree, and a depth-first search can be used to execute each generated test. Intuitively, this search strategy can be particularly performant when the quality of generated tests are good (which, as we observe in \Cref{sec:eval}, is often true) but additional care must be taken to ensure that the test generator does not get stuck executing an infinite-length test; we discuss in detail how to address this issue in \Cref{ssec:gas}. 
}

%% file: intraproc.tex
\section{GPS: Combining \GPSlite with Abstraction Refinement}
\label{sec:algorithm}

\GPSlite has no facility for invariant generation---it is capable of proving safety of a program only when path summaries are sufficiently strong to show that all paths are dead-ends.  This section presents GPS, a software-model-checking algorithm that integrates \GPSlite with Craig interpolation and abstract reachability trees from the lazy-abstraction-with-interpolants algorithm \cite{cav:mcmillan06}.  In particular, {we develop \textit{dead-end interpolation}, which uses summaries to generate high-quality candidate invariant assertions that prove 
safety of a regular set of paths}
simultaneously.


\subsection{From Path Trees to Abstract Reachability Trees (ARTs)} 
The main ingredient that bridges GPSLite (described in \Cref{sec:gpslite_algorithm}) and GPS is an \emph{abstract reachability tree} (ART).  Informally speaking, an ART is an extension of a path tree (\Cref{def:path_tree}), where every path inside the path tree is also paired with a 
{
\emph{label}---a state formula that over-approximates the post-state of the path---and some pairs of paths are connected by
}
\emph{covering edges}, which describe entailment relationships between paths and are used to check for inductiveness of loop labels.
An example of an ART for the program EX-3 in~\Cref{sec:overview} appears in Figure~\ref{fig:ex3-coverings}.  

{
\begin{definition}[Abstract Reachability Trees]
    \label{def:art}
    An abstract reachability tree (ART) for a weighted graph $G = (V, E, w, s, t)$  is a 3-tuple $\Lambda = (T, L, \TreeCovers{\Lambda})$, {with each element of the 3-tuple defined as follows:} 
    \begin{enumerate}
        \item \textbf{Path tree}. $T$ is a path tree of  $G$ (Cf.~\Cref{def:path_tree}), 
        \item \textbf{Label function}. $L : T \to \SF{X}$ maps each path  $\pi \in T$ to a state formula in $\SF{X}$.
        \item \textbf{Covering relation}.  $\TreeCovers{\Lambda} \subset T \times T$ is called the {covering relation}, such that for every $(\pi, \pi') \in \TreeCovers{\Lambda}$, $\pi'$ is a prefix of $\pi$ and $\dst(\pi) = \dst(\pi')$
    \end{enumerate}
\end{definition}

{For a path {$\pi = e_1e_2 \dotsi e_n$,} 
if $\pi$ is non-empty, we use $\pparent{\pi}$ to denote its immediate prefix $e_1e_2 \dotsi e_{n-1}$ and say that $\pi$ is a \textbf{child} of $\pparent{\pi}$ (corresp. $\pparent{\pi}$ is a \textbf{parent} of $\pi$). Furthermore, recall from \Cref{sec:background} that $\prefixord$ denotes the prefix order over paths.} 
Consider an ART $\Lambda = (T, L, \TreeCovers{\Lambda})$. 
For a path $\pi \in T$, we refer to $\pi$ as a \textbf{leaf path} if it {maximal} w.r.t.\ $\prefixord$. We say that a path $\pi \in \TreeNodes{\Lambda}$ is \textbf{pruned} if $L(\pi) \wedge \mathsf{Sum}(G, \dst(\pi), t)$ is unsatisfiable (where $t$ is the terminal vertex of the weighted graph associated with ART $\Lambda$). 
One may think of a pruned path $\pi$ as {a path} 
for which the
ART represents a proof that $\pi$ is a dead-end---every pruned path is a dead end, but not every dead end is necessarily pruned.
For a path $\pi \in T$, we say that $\pi$ is \textbf{covered} if there is a pair $(\pi, \pi') \in \TreeCovers{\Lambda}$. We say a leaf path $\pi$ is a \textbf{frontier} 
if it is neither pruned nor covered. Intuitively, $\pi$ is a frontier path if it \emph{might} be the prefix of a previously unexplored, feasible $s$-$t$ path in $G$. We define $\Lambda$ to be \textbf{complete} when there are no frontier paths in $T$ (i.e., \emph{every} {maximal} path $\pi \in T$ is either pruned or covered). 

For any $\pi \in \TreeNodes{\Lambda}$, we can associate $\pi$ with a set of paths $\textit{Paths}_{\Lambda}(\pi)$ in $G$ by treating $\Lambda$ as a finite-state automaton with start state $\TreeRoot{\Lambda}$, final state $\pi$, {and with an $e$-labeled transition from $\pi$ to $\pi e$ for each $\pi e \in \TreeNodes{\Lambda}$, and an $\epsilon$-transition from $\pi$ to $\pi'$ for each $(\pi,\pi') \in \TreeCovers{\Lambda}$}.  
{For instance, the set of paths associated with $\encircle{A} \rightarrow \encircle{B} \rightarrow \encircle{B}$ in Figure~\ref{fig:ex3-art} is 
$\encircle{A} \rightarrow \encircle{B} (\rightarrow \encircle{B})^+$.}
Formally, $\textit{Paths}_\Lambda$ is the least function such that $\textit{Paths}_{\Lambda}(\TreeRoot{\Lambda})$ contains $\TreeRoot{\Lambda}$,
$\textit{Paths}_{\Lambda}(\pi e)$ contains $\set { \tau e : \tau \in \textit{Paths}_{\Lambda}(\pi) }$, and
$\textit{Paths}_{\Lambda}(\pi')$ contains $\textit{Paths}_{\Lambda}(\pi)$ for each $(\pi,\pi') \in \TreeCovers{\Lambda}$.

Let $\Lambda$ be an ART for a weighted graph $G$.  The following \textit{well-labeledness} condition ensures that
for any $\pi \in \TreeNodes{\Lambda}$ and any
$\tau \in \textit{Paths}_{\Lambda}(\pi)$, we have $\spost{\top}{w(\tau)} \models \TreeLbl{\Lambda}{\pi}$.
 \begin{definition}[Well-labeledness]
 \label{def:well_labeled_ness}
 Given an ART $\Lambda$ of a weighted graph $G$, we call $\Lambda$ well-labeled (w.r.t.\ $G$) if the following conditions are satisfied:
    \begin{enumerate}
        \item \textbf{Initiation}. {The formula labeling the root ($\TreeLbl{\Lambda}{\epsilon}$) 
is $\true$.}
    \item \textbf{Consecution}. For any path $\pi e \in \TreeNodes{\Lambda}$, 
$\spost{\TreeLbl{\Lambda}{\pi}}{e} \models \TreeLbl{\Lambda}{\pi e}$
\item \textbf{Well-covered}. For any $(\pi,\pi') \in \TreeCovers{\Lambda}$, we have $\TreeLbl{\Lambda}{\pi} \models \TreeLbl{\Lambda}{\pi'}$
    \end{enumerate}
 \end{definition}

The following lemma states that the safety of a weighted graph $\CFG$ can be witnessed by a complete and well-labeled ART $\Lambda$:

\begin{lemma}
\label{thm:art_cfg_safe}
Let $G = \tuple{V,E,w,s,t}$ be a weighted graph, and let $\Lambda$ be a well-labeled ART for $G$. If $\Lambda$ is complete and well-labeled, then $G$ is safe.
\end{lemma}
}

\input{fig-intraproccheck}

\input{fig-runtestandrefine}

\subsection{From GPSLite to GPS}
\label{sec:FromGPSLiteToGPS}
\bfpara{At a glance.} Recall that as discussed in~\Cref{sec:gpslite_algorithm}, the basic idea of the \GPSlite algorithm is as follows: in each iteration of the main loop of~\Cref{fig:SummaryGuidedTesting}, we select a path $\pi$ from $\TreeFrontiers{\Lambda}$---which is a queue storing paths that are potentially frontiers. Then, \GPSlite tries to generate a longer path with $\pi$ as a prefix, in an effort to make progress in a search for a feasible source-sink path in a given weighted graph $G$. Such an attempt is not always guaranteed to be successful; notably, the query to $\CheckSat{w(\pi), \mathsf{Sum}(G, \dst(\pi), t)}$ can return UNSAT (See line~\ref{line:GPSLite:unsat} in ~\Cref{fig:SummaryGuidedTesting}), showing that $\pi$ leads to a dead-end. The key idea here---borrowed from lazy abstraction with interpolants \cite{cav:mcmillan06}--- is that in the case of UNSAT, one can extract a sequence of intermediate state formulas that can form candidate invariants constituting a proof of infeasibility from the unsatisfiable input formula---and such intermediate state formulas can be maintained as \emph{labels}, or invariant candidates, in an ART. This in turn enables us to synthesize invariants from UNSAT proofs by maintaining coverings between ART paths.

Similar to  \GPSlite, the GPS algorithm  maintains a queue $\TreeFrontiers{\Lambda}$ of leaf paths in $\Lambda$, maintaining the invariants that (1) all \emph{frontier paths of $\Lambda$} are in $\TreeFrontiers{\Lambda}$, and (2) every path in $\TreeFrontiers{\Lambda}$ is either a frontier path or a pruned path. {We call the paths in $\TreeFrontiers{\Lambda}$ \emph{possible frontiers.}}  Recall that in \GPSlite, {possible frontier paths are selected} in breadth-first order from the queue $\TreeFrontiers{\Lambda}${; for each selected path $\pi$} \GPSlite makes $\pi$ a \emph{non}-frontier by either generating a test that uses $\pi$ as a proper prefix, or proves that $\pi$ is a dead-end. The full GPS algorithm differs in two main ways: (1) GPS extracts information from dead-end detection --- in form of interpolants --- and uses them to refine candidate invariants; (using the \textsc{Refine} procedure in \Cref{fig:Explorerefine}) (2) GPS uses the \textsc{TryCover} procedure shown in \Cref{fig:Explorerefine} to check whether the candidate invariants are in fact inductive. The \textsc{Refine} and \textsc{TryCover} procedures in GPS {build} upon the procedures of the same name in the \textsc{Impact} algorithm \cite{cav:mcmillan06}. The main difference is that our \textsc{Refine} procedure incorporates information from path summaries (discussed below), and our \textsc{TryCover} procedure only introduces covering edges between descendants and ancestors.\footnote{We chose to only introduce coverings between descendants and ancestors to simplify presentation. In practice, one could still implement an \textsc{Impact}-like covering strategy of allowing coverings between paths that aren't ancestors and descendants.}

In what follows, we will first explain how the covering step and the summary-based dead-end refinement works, and then reinforce our explanation through a concrete example.

\bfpara{Synthesis of candidate invariants via dead-end interpolation.}
{We assume access to a primitive $\mathsf{Check}(-)$, extending the $\CheckSat{-}$ primitive from the previous section.
It
}
 takes as input a sequence of transition formulas
$F_1, \dots, F_n$, 
 determines whether the path
$F_1, \dots, F_n$ is feasible, and produces either a witness sequence of states demonstrating feasibility, or a sequence of intermediate state formulas constituting a proof of infeasibility. 
The intermediate assertions can be computed, for instance, using Craig interpolation
\cite{mcmillanInterpolantSummary}.

\vspace*{0.1in}
\begin{mdframed}
\small
    \textbf{Primitive }$\mathsf{Check}(F_1, ..., F_n)$: 
    
    \begin{itemize}
        \item \textbf{Input: } A sequence of transition formulas $F_1, \dots, F_n$
        \item \textbf{Output: } Returns either:
            \begin{enumerate}
        \item $(\mathsf{SAT}, [M_0, \dots, M_n, M_{n+1}]$ if {$(F_1 \circ \dots \circ F_n)$ is satisfiable, where $M_0, \dots, M_{n+1}$ is a sequence of states such that
        $\tfrel{M_i}{F_i}{M_{i+1}}$ for all $i \in \set{0,\dots,n}$}
        
        \item $(\mathsf{UNSAT}, [\phi_0, \dots, \phi_n, \phi_{n+1}])$ if {$(F_1 \circ \dots \circ F_n)$ is unsatisfiable, where
        $\phi_0,\dots,\phi_{n+1}$ is a sequence of state formulas such that
        $\phi_0 = \true$,
        $\spost{\phi_i}{F_i} \models \phi_{i+1}$
        for all $i \in \set{0,\dots,n}$,
        and $\phi_{n+1}$ is $\false$.}
    \end{enumerate}
    \end{itemize}
\end{mdframed}
\vspace*{0.1in}
{Thus, the $\Check{-}$ primitive generalizes $\CheckSat{-}$ in two ways: (1) it checks satisfiability of the composition of a sequence of transition formulas rather than two, and (2) when the result is UNSAT, $\Check{-}$ produces a proof of unsatisfiability.
}

{GPS uses the $\Check{-}$ primitive analogously to how \GPSlite uses $\CheckSat{-}$---to generate a test from a path $\pi$ drawn from the frontier queue or to detect that it is a dead end.  In addition, when GPS detects a dead end, it uses the resulting sequence of state formulas (\textit{dead-end interpolants}) to strengthen the labels of $\pi$ and its prefixes with the help of the auxiliary {\sc Refine} routine (Figure~\ref{fig:Explorerefine}) so that $\pi$ becomes a pruned path in the resulting ART.
}
Given an ART $\Lambda = (T, L, \TreeCovers{\Lambda})$, a path $\pi e_1 \dots e_n$, and a sequence of state formulas $\phi_1 \dots \phi_n$, $\textsc{Refine}$ {(due to \cite{cav:mcmillan06})} adds 
$\phi_i$ as a conjunct to the state-formula label of $\pi e_1 \dots e_i$ in $\Lambda$. 
Strengthening the label at a path $\tau$ may violate the \textit{well-covered} condition (Definition~\ref{def:well_labeled_ness}) for covering edges to $\tau$;
thus, $\textsc{Refine}$ checks if the condition remains satisfied (line~\ref{line:refine:check-covering}) and removes
covering edges that now violate the condition.

The novelty of GPS is not in the $\Check{-}$ primitive, but the way in which it is used.  GPS derives interpolants from a sequence of $n+1$ transition formulas, where the first $n$ correspond to a path $\pi$ and the last to a summary $\mathsf{Sum}(G, \dst(\pi), t)$ of a set of paths.

\begin{figure}
\centering
\begin{subfigure}{0.3\textwidth}
\centering
\includegraphics[width=\linewidth]{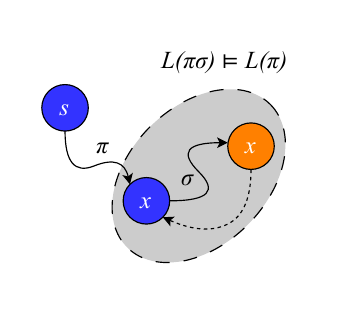}
\subcaption{The covering condition.}
\label{fig:covering-motivation}
\end{subfigure}
\hfill
\begin{subfigure}{0.6\textwidth}
    \centering
   \includegraphics[width=0.8\linewidth]{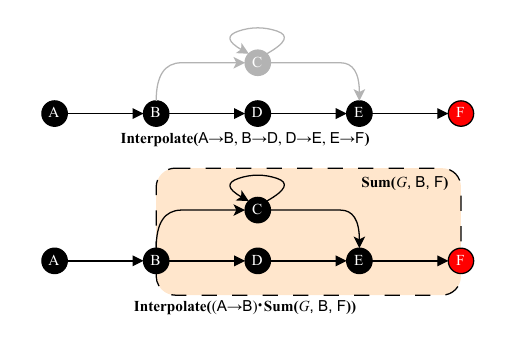}
    \caption{Two styles of interpolation ($G$ denotes the weighted graph).}
    \label{fig:interpolation-styles}
\end{subfigure}
\caption{(a): Illustration of the covering condition {in an ART}: If taking path $\sigma$ from $\pi$ (with $x = \dst(\pi)$) ends in the same vertex $x$, and $L(\pi\sigma) \vDash L(\pi)$, then $\pi\sigma$ and $\pi$ essentially end in the same states, and $L(\pi\sigma)$ is an inductive invariant of the loop at $\sigma$. (b): Two styles of interpolation explained via a weighted graph with $A$, ..., $F$ being vertices: (top) ``One path-to-error at a time'' interpolation as in {the \textsc{Impact} algorithm}; versus (bottom) GPS-style dead-end interpolation, which {interpolates a concrete prefix path (from A to B) and an (over-approximate) summary-to-error (shown in orange; representing all paths that start at B and end at F).}}
\end{figure}

{\Cref{fig:interpolation-styles} illustrates the main difference between refinement in GPS and refinement in the \textsc{Impact} algorithm \cite{cav:mcmillan06}. In this figure, vertices \encircle{A} through \encircleR{F} denote weighted-graph vertices, with \encircle{A} being the source and \encircleR{F} being the sink (which represents the error vertex).
In \textsc{Impact}, interpolation-based refinement is performed only for each source-to-error path, one path at a time (shown
{in}
the top half of the figure).
}
{In contrast, in GPS interpolation-based refinement is performed for a \emph{regular set of paths} $\set{ \pi\tau : \tau \in \textit{Paths}_G(\dst(\pi),\mathit{error})}$:
$\pi$ is a particular path-prefix that GPS chooses (of the form $s$-to-$\dst(\pi)$), and $\textit{Paths}_G(\dst(\pi),\mathit{error})$ is the set of all suffixes to $\mathit{error}$.
This approach increases the generality of the resulting interpolants (shown
{in}
the bottom half of the figure).
}

\bfpara{Using coverings to test for inductive invariants.}
Consider the situation illustrated in \Cref{fig:covering-motivation}.
Suppose that we have a well-labeled ART $\Lambda = (T, L, \TreeCovers{\Lambda})$ for a weighted graph $G= (V, E, w, s, t)$, and that a path $\pi \in T$ is a prefix of another path $\pi' \in T$ such that $\dst(\pi) = \dst(\pi')$. Let $\sigma$ be a suffix path such that $\pi' = \pi \sigma$. Then $\sigma$ is necessarily a \emph{cycle} in $G$---as $\sigma$ starts at $\dst(\pi)$ and ends at the same vertex $\dst(\pi') = \dst(\pi)$. By the well-labeledness condition of $\Lambda$, we have that $\post(L(\pi), w(\sigma))\vDash L(\pi')$. Now, note that if $L(\pi') \vDash L(\pi)$, then we may conclude that $L(\pi')$ is a loop invariant for the cycle $\sigma$. {In other words, one can view the labels of the ART as \textit{candidate} loop invariants, and checking whether a given candidate is an invariant is simply an implication check.} 

{
{Invariant-checking is  implemented by the $\textsc{TryCover}(\Lambda, v)$ subroutine (\Cref{fig:Explorerefine}) of GPS (again essentially due to \cite{cav:mcmillan06}).} This routine seeks to test whether the label of a path $\pi \in T$ can be strengthened to permit $\pi$ to be covered by one of its ancestors $\tau$ w.r.t.\ the prefix order (establishing the label at $\tau$ to be an inductive invariant of the cycle $\tau^{-1}\pi$).
  For each candidate ancestor $\tau$ 
 (with $\pdst{\tau} = \pdst{\pi}$), the \textsc{TryCover} routine
 forms an inductiveness query that checks whether there is an execution of $\tau^{-1}\pi$
 that begins in a state satisfying  $\TreeLbl{\Lambda}{\tau}$
 and ends in a state satisfying 
 $\lnot \TreeLbl{\Lambda}{\tau}$ (line~\ref{line:trycover:check}).  If such an execution exists, the candidate $\tau$ is discarded; if it does not, then \textsc{TryCover} uses sequence interpolants derived from the inductiveness check to refine the labels of prefix paths on the path from $\tau$ to $\pi$ (ensuring that $\TreeLbl{\Lambda}{\pi} \models \TreeLbl{\Lambda}{\tau}$ and  $\Lambda$ remains well-labeled), and adds $(\pi,\tau)$ to  $\TreeCovers{\Lambda}$.
}

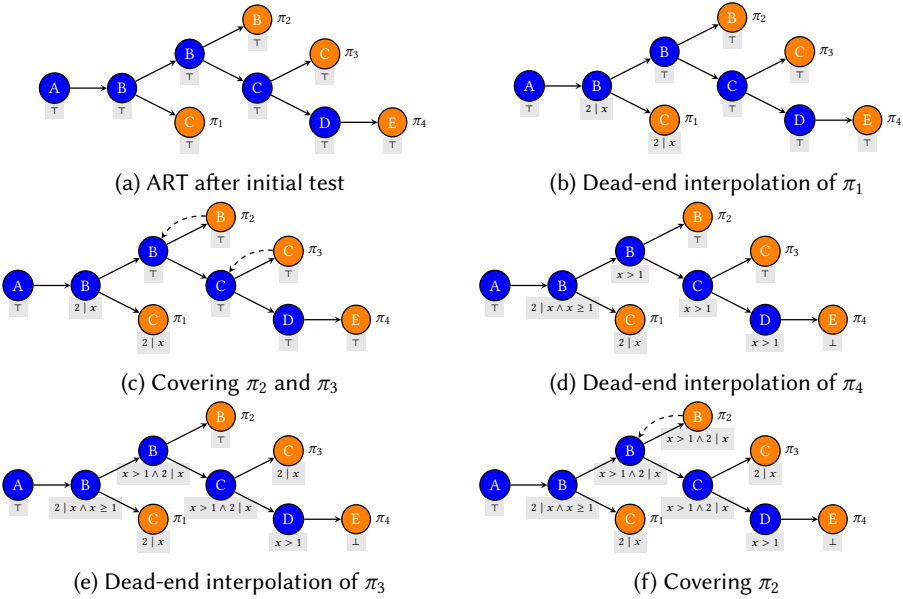
\begin{figure}[tb!]
\begin{subfigure}{0.45\textwidth}\centering
\scalebox{0.6}{
\begin{tikzpicture}[style=pathtree]
    \exthreepl{}

    \artlabel{A}{\true}
    \artlabel{AB}{\true}
    \artlabel{ABB}{\true}
    \artlabel{ABC}{\true}
    \artlabel{ABBC}{\true}
    \artlabel{ABBB}{\true}
    \artlabel{ABBCD}{\true}
    \artlabel{ABBCDE}{\true}
    \artlabel{ABBCC}{\true}

\end{tikzpicture}}
\caption{ART after initial test \label{fig:ex3-1}}
\end{subfigure}
\begin{subfigure}{0.45\textwidth}\centering
\scalebox{0.6}{
\begin{tikzpicture}[style=pathtree]
    \exthreepl{}

    \artlabel{A}{\true}
    \artlabel{AB}{2 \mid x}
    \artlabel{ABB}{\true}
    \artlabel{ABC}{2 \mid x}
    \artlabel{ABBC}{\true}
    \artlabel{ABBB}{\true}
    \artlabel{ABBCD}{\true}
    \artlabel{ABBCDE}{\true}
    \artlabel{ABBCC}{\true}
\end{tikzpicture}
}
\caption{Dead-end interpolation of $\pi_1$ \label{fig:ex3-2}}
\end{subfigure}
\begin{subfigure}{0.45\textwidth}
\scalebox{0.6}{
\begin{tikzpicture}[style=pathtree]
    \exthreepl{}
    
    \draw (ABBB) edge[->,dashed,bend right] (ABB);
    \draw (ABBCC) edge[->,dashed,bend right] (ABBC);
    \artlabel{A}{\true}
    \artlabel{AB}{2 \mid x}
    \artlabel{ABB}{\true}
    \artlabel{ABC}{2 \mid x}
    \artlabel{ABBC}{\true}
    \artlabel{ABBB}{\true}
    \artlabel{ABBCD}{\true}
    \artlabel{ABBCDE}{\true}
    \artlabel{ABBCC}{\true}
\end{tikzpicture}
}
\caption{Covering $\pi_2$ and $\pi_3$ \label{fig:ex3-3}}
\end{subfigure}
\begin{subfigure}{0.45\textwidth}
\scalebox{0.6}{
\begin{tikzpicture}[style=pathtree]
    \exthreepl{}
    \artlabel{A}{\true}
    \artlabel{AB}{2 \mid x \land x \geq 1}
    \artlabel{ABB}{x > 1}
    \artlabel{ABC}{2 \mid x}
    \artlabel{ABBC}{x > 1}
    \artlabel{ABBB}{\true}
    \artlabel{ABBCD}{x > 1}
    \artlabel{ABBCDE}{\false}
    \artlabel{ABBCC}{\true}
\end{tikzpicture}
}
\caption{Dead-end interpolation of $\pi_4$ \label{fig:ex3-4}}

\end{subfigure}
\begin{subfigure}{0.45\textwidth}
\scalebox{0.6}{
\begin{tikzpicture}[style=pathtree]
    \exthreepl{}
    \artlabel{A}{\true}
    \artlabel{AB}{2 \mid x \land x \geq 1}
    \artlabel{ABB}{x > 1 \land 2 \mid x}
    \artlabel{ABC}{2 \mid x}
    \artlabel{ABBC}{x > 1 \land 2 \mid x}
    \artlabel{ABBB}{\true}
    \artlabel{ABBCD}{x > 1}
    \artlabel{ABBCDE}{\false}
    \artlabel{ABBCC}{2 \mid x}
\end{tikzpicture}
}
\caption{Dead-end interpolation of $\pi_3$ \label{fig:ex3-5}}
\end{subfigure}
\begin{subfigure}{0.45\textwidth}
\scalebox{0.6}{
\begin{tikzpicture}[style=pathtree]
    \exthreepl{}
     \draw (ABBB) edge[->,dashed,bend right] (ABB);
    \artlabel{A}{\true}
    \artlabel{AB}{2 \mid x \land x \geq 1}
    \artlabel{ABB}{x > 1 \land 2 \mid x}
    \artlabel{ABC}{2 \mid x}
    \artlabel{ABBC}{x > 1 \land 2 \mid x}
    \artlabel{ABBB}{x > 1 \land 2 \mid x}
    \artlabel{ABBCD}{x > 1}
    \artlabel{ABBCDE}{\false}
    \artlabel{ABBCC}{2 \mid x}
\end{tikzpicture}
}
\caption{Covering $\pi_2$ \label{fig:ex3-6}}
\end{subfigure}
\caption{Illustration of GPS on the example EX-3}
\label{fig:ex3-coverings}
\end{figure}

\bfpara{Putting everything together: how GPS works on example EX-3 from \Cref{sec:overview}.} We illustrate these steps by revisiting example EX-3 from \Cref{sec:overview}, with the CFG $G = (V, E, w, s, t)$ and ART $\Lambda = (T, L, \TreeCovers{\Lambda})$ given in Figures~\ref{fig:ex3}, and the full execution trace of GPS on EX-3 (shown as a sequence of ART operations) shown in~\Cref{fig:ex3-coverings}.

As discussed in Section~\ref{sec:overview}, GPS first executes a test $\{N \mapsto 1, x \mapsto 2, M \mapsto 0 \}$ to obtain the (incomplete) ART shown in Figure~\ref{fig:ex3-1}. After this step, no refinements have been performed, so the label for each path in the ART is simply $\top$. Next, GPS then detects that $\pi_1$ is a dead end via interpolating $w(\pi_1)$ composed with $\mathsf{Sum}(\dst(G, \pi_1), t)$, producing Figure~\ref{fig:ex3-2} where the labels of the vertices on $\pi_1$ have been refined with dead-end interpolants.

Since $\pi_2$ and $\pi_3$ both have the same label as their parent ($\true$), both can be covered during a \textsc{TryCover}(-) call to produce Figure~\ref{fig:ex3-3}. 
 These covering operations indicate that $\true$ is an inductive invariant of the $\encircle{B}$ and $\encircle{C}$ (though not necssarily an  invariant that separates the initial states from the error states--GPS will determine this later and uncover these paths).
  GPS then detects that $\pi_4$ is a dead end, producing Figure~\ref{fig:ex3-4}---note that $\pi_2$ and $\pi_3$ become \textit{uncovered}, because the label of their parent has changed and so they are no longer considered redundant.  GPS detects that $\pi_3$ is a dead end to produce Figure~\ref{fig:ex3-5}, and finally the covering edges from $\pi_2$ to its parent is reinstated to arrive at the complete and well-labeled ART in Figure~\ref{fig:ex3-6}, and GPS reports that the assertion is safe.

\begin{theorem}[Soundness of GPS]
\label{thm:gps_sound}
{Let $G$ be a weighted graph. If $\textsc{GPS}(G)$ returns $\mathsf{Safe}$, then $G$ is safe; if it returns $(\mathsf{Unsafe}, \pi)$, then $\pi$ is a feasible path from the source of $G$ to its target.}
\end{theorem}

\subsection{Refutation Completeness}
\label{ssec:gas}

The GPS procedure described in \Cref{sec:FromGPSLiteToGPS} is not complete for refutation:
if the program has a non-terminating execution, it is possible for the $\textsc{Explore}$ procedure in Figure~\ref{fig:Explorerefine} to diverge. In fact, the \emph{only} obstacle to refutation completeness is when the $\textsc{Explore}(-)$ procedure fails to terminate by executing a test input that leads to a trace of infinite length.  To address this issue, this section describes a transformation that makes GPS complete by bounding the depth of test executions.
We also formally state and prove a refutation-completeness result. \IfCameraReady{\rfchanged{The detailed proof may be found in the extended version of this paper \cite{GPSExtendedVersion}.}}{
(The detailed proof is in Appendix~\ref{sec:Proofs}.)}


\paragraph{Instrumentation with the global variable $\mathtt{gas}$.} GPS instruments the entire program using a global variable named $\mathtt{gas}$. The purpose of the instrumentation is to ensure the termination of \textsc{Explore} by using $\mathtt{gas}$ to represent a model-checking {bound}.
Using the $\mathsf{havoc}(-)$ primitive, we treat $\mathtt{gas}$ as being initialized to a non-deterministic, positive integer value.
At each loop-header vertex, we add an {assumption} that $\mathtt{gas} \geq 0$ and a statement to decrement $\mathtt{gas}$ by 1. {In a nutshell, the gas-bound instrumentation ensures termination of (calls to) the $\textsc{Explore}(-)$ procedure (\Cref{fig:SummaryGuidedTesting}). Without gas-instrumentation, GPS might generate a test that is unbounded in length (e.g., by entering a non-terminating loop), in which case the $\textsc{Explore}(-)$ procedure {would get} stuck executing this infinite test, therefore breaking refutation completeness. Intuitively, to guarantee refutation completeness we need to make sure that the algorithm fairly explores all traces of the program-under-test. This property is in turn used in the proof of refutation completeness.
\IfCameraReady{(See the proof of \rfchanged{Lemma 9} in the extended version of this paper \cite{GPSExtendedVersion})}{(See the proof of \Cref{lem:trichotomy} in \Cref{appendix:proofs}.)
}}


\begin{figure}
\begin{subfigure}{0.48\textwidth}
\scriptsize
\begin{lstlisting}[style=base]%numbers=left

void main() {
  int s0, s1, s2, c;
  int state = 0;
  while (1) {
    c = *; s0 = *; s1 = *; s2 = *;
    /* state machine */
    if (c == 68 && state == 0) state = 1;
    if (c == 79 && state == 1) state = 2;
    if (c == 71 && state == 2) state = 3;
    if (state == 3 && s0 == 67 && s1 == 65 && s2 == 84)
        error();
  }
}\end{lstlisting}
\subcaption{An example program implementing a state-machine that loops indefinitely until the user enters a correct input of ASCII-encoded characters  \texttt{C, A, T} at once (via $s0-s2$) and \texttt{D, O, G} in succession (via $c$).}
\end{subfigure}
\hfill
\begin{subfigure}{0.48\textwidth}
\scriptsize
\begin{lstlisting}[style=base]

void main() {
  int s0, s1, s2, c;
  int state = 0;
  int gas = *; // instrumented
  while (1) {
    if (gas-- < 0) break; // instrumented
    c = *; s0 = *; s1 = *; s2 = *;
    /* state machine */
    if (c == 68 && state == 0) state = 1;
    if (c == 79 && state == 1) state = 2;
    if (c == 71 && state == 2) state = 3;
    if (state == 3 && s0 == 67 && s1 == 65 && s2 == 84)
        error();
    }
}\end{lstlisting}
\subcaption{Instrumented version of the program on the left.}
\end{subfigure}
    
    \caption{An example illustrating a class of loop patterns on which the refutation-completeness transformation for GPS also speeds up convergence.}
    \label{fig:gas_really_helps}
\end{figure}

\begin{example}
\ruijie{TODO: change the text below}
 Consider again the programs {EX-1} and {EX-2} from \Cref{fig:motivatingexamples}.
 The code to instrument them with $\mathtt{gas}$ is shown in grey.
\end{example}

For each choice of the \texttt{gas} value, $\textsc{Explore}(-)$ generates a finite  sub-tree of a given ART, which represents the set of all executions bounded by the particular \texttt{gas} value. Thus, the above instrumentation guarantees that  $\textsc{Explore}(-)$ always terminates.
\begin{theorem}[Refutation completeness]
    \label{thm:intraproc_refutation_complete}
Let $G$ be a weighted graph, instrumented with gas as above.
If $P$ is unsafe, then $\textsc{GPS}(P)$  terminates with the result ``Unsafe.''
\end{theorem}

\bfpara{Discussion.}
The purpose of \textsf{gas}-instrumentation is to obtain refutation-completeness as a  theoretical guarantee of GPS. Its effect on \textit{practical} performance is another matter: on the one hand, \textsf{gas} may be helpful by terminating exploration of paths that do not lead to the error location, but on the other it may be harmful by terminating exploration of paths that
would (if provided with enough gas).
We will return to this question in our experimental evaluation (Section~\ref{sec:eval}).  
For the moment, we remark that the potential negative effects of the transformation are mitigated by the use of summaries in GPS: by performing \textsf{gas}-instrumentation before computing summaries, we obtain summaries that (can) provide useful information about how long an error path must be.  In the remainder of this section, we discuss such an example.

Consider the program in Figure~\ref{fig:gas_really_helps}. This program implements a small state machine accepting user inputs via the variables $c$, $s1$, $s2$, and $s3$. To reach the error location, GPS's test generator has to \emph{both} guess the value for $c$ correctly across three successions of the loop---as well as guessing correct values for variables $s0$, $s1$, and $s2$ when the state machine is in its accepting state. 

Note that this example also falls within the category of ``Lock \& Key'' programs we discussed in~\Cref{sec:overview}---here, the key to unlock a feasible path-to-error is precisely the sequence of inputs required to arrive at the accepting state of the state machine. Without performing \textsf{gas}-instrumentation, GPS {times out when running its first test.  The essential problem is that the {\sc Explore} algorithm uses the path-to-error summary to maintain the invariant that the state \textit{may} reach the error location, but there is no inherent bias for an SMT solver to choose values for the $c, s1, s2,$ and $s3$ variables that make \textit{progress} towards the error. Because every execution can be extended to reach the error location, {\sc Explore} may simply choose the value 0 for all non-deterministic expressions and enter an infinite loop.
}
In contrast, once the instrumentation is performed (the result is shown in \Cref{fig:gas_really_helps}(b)), a summary-generating static analysis---such as compositional recurrence analysis \cite{fmcad:fk15}---is able to generate a loop summary {of the following form}:
\[
\exists k. k \geq 0 \land \mathit{gas}' = \mathit{gas} - k \wedge \mathit{state} \leq \mathit{state}' \leq \mathit{state} + k,
\]
which, in essence, forces the test generator to select paths in which variable $\mathit{state}$ changes between loop iterations, because the \textit{possible} change in \cinline{state} is bounded by the decrease in \cinline{gas} (e.g., from a state with $\set{ \textit{gas} \mapsto 2; \textit{state} \mapsto 0}$, GPS \textit{must} select $c=68$, because if variable \textit{state} fails to advance then the error state cannot be reached while consuming $\leq 2$ units of \textit{gas}). As such, the \textsf{gas}-instrumentation technique is potentially beneficial in \emph{helping GPS generate higher-quality tests} in loops where non-deterministically-valued expressions can lead to non-terminating executions.

\barrier 

%% file: fig-intraproccheck.tex
\begin{figure}[tb!]
        \begin{algorithm}[H]
\begin{algorithmic}[1]
\small
\Procedure{GPS}{$G$}
\Statex \Input{Weighted graph $G = (V, E, w, s, t)$ }
\Statex \Output{Safe if $G$ is safe; 
otherwise $(\text{Unsafe}, \pi)$, where $\pi$ is a feasible $s$-$t$ path in $G$.}
\label{line:intraproc:create-weighted-graph}
\State $\Lambda \gets \text{CreateART}(G);$\label{line:intraproc:create-art}\Comment{{Initialize ART containing only the empty path $\epsilon$}}
\State $\TreeFrontiers{\Lambda}.\text{enqueue}(\TreeRoot{\Lambda});$
    \While{$\TreeFrontiers{\Lambda} \neq []$} \label{line:GPS:mainloop}
        \State $\pi \gets \text{dequeue}(\TreeFrontiers{\Lambda});$
        \If{$\textsc{TryCover}(\Lambda, \pi)$} \Comment{Attempt to cover $\pi$ by an ancestor}\label{line:GPS:trycover}
            \State \textbf{continue;}\Comment{$\pi$ is covered in $\Lambda$, no need to explore further}
        \Else
            \State $\textit{Fs} \gets \mathsf{map}(w, \pi) :: \mathsf{Sum}(\CFG,\TreeVtx{\Lambda}{\pi}, t)$;\Comment{Sequence of $|\pi|+1$ transition formulas}\label{line:interproccheck_guideditps}
            \Switch{$\mathsf{Check}(\textit{Fs})$}
                \Case{$\text{UNSAT}(\phi_0, \phi_1,...,\phi_{n})$}\label{line:intraproc:refine}
                     $\textsc{Refine}( \Lambda, [\phi_0,...,\phi_{n-1}], \pi);$\label{line:check-weighted-graph:prune}
    
                \EndCase
                \Case{$\text{SAT}(M_0,...,M_n,M_{n+1})$}\label{line:intraproc:testgen}
                    \Switch{$\textsc{Explore}(\CFG, \Lambda, M_n, \pi)$}\label{line:check-weighted-graph:newtest}
                        \Case{$\text{Safe}$}
                     \textbf{continue};
                        \EndCase 
                        \Case{$(\text{Unsafe}, \tau)$}
                             \Return $(\text{Unsafe}, \tau)$;
                        \EndCase 
                        \EndSwitch 
                \EndCase
            \EndSwitch
            \EndIf
    \EndWhile 
    \State \Return $\text{Safe}$; \Comment{$\Lambda$ is complete and the procedure is proven safe.}
\EndProcedure 
\end{algorithmic}

    \end{algorithm}

                     \caption{Main procedure for the full GPS algorithm.}
    \label{fig:gps-intraproc}
\end{figure}

%% file: fig-runtestandrefine.tex
\begin{figure}[tb!]
    \begin{minipage}[t]{0.5\textwidth}
\begin{algorithmic}[1]
\small
\Procedure{TryCover}{$\Lambda = (T, L, \TreeCovers{\Lambda}), \pi$}

    \For{each $\tau \sqsubset \pi$ with $\pdst{\tau} = \pdst{\pi}$}
    \State $I \gets \TreeLbl{\Lambda}{\tau}$
        \Switch{$\text{Check}(I, \mathsf{map}(w,\tau^{-1}\pi) :: \lnot I)$}\label{line:trycover:check}
            \Case{$\text{UNSAT}(T_0,...,T_n)$}
                \State $\textsc{Refine}(\Lambda, \pi, [T_0,...,T_n])$;
                \State Add $(\pi,\tau)$ to $\TreeCovers{\Lambda}$;
                \State \Return $\top$;
            \EndCase
            \Case{$\text{SAT}(M_0,...,M_n)$}
                 \Return $\bot$;
            \EndCase
        \EndSwitch
    \EndFor
\EndProcedure
    
\end{algorithmic}    

\end{minipage}\hfill 
    \begin{minipage}[t]{0.48\textwidth}
    
            \begin{algorithmic}[1]
\small
        \Procedure{Refine}{$\Lambda, \pi e_1 \dots e_n, \phi_1 \dots \phi_n$}
    \State $(T, L, \TreeCovers{\Lambda}) \gets \Lambda$;
    \For{$i = 1$ to $n$}
        \State $\pi' \gets \pi e_1 \dots e_i$
        \For{each $\tau$ with $(\tau,\pi') \in \TreeCovers{\Lambda}$}\label{line:refine:check-covering}
            \If{$\TreeLbl{\Lambda}{\tau} \not\models \phi_i$}
                \State Remove $(\tau,\pi')$ from $\TreeCovers{\Lambda}$;
                \State $\TreeFrontiers{\Lambda}.\text{enqueue}(\tau)$;
            \EndIf
        \EndFor
        \State $\TreeLbl{\Lambda}{\pi'} \gets \TreeLbl{\Lambda}{\pi'} \land \phi_i$;
    \EndFor
    \EndProcedure
            \end{algorithmic}
    \end{minipage}

    \vspace*{0.1in}
    \caption{Auxilliary procedures for the GPS algorithm: \textsc{Refine}(-) and \textsc{TryCover}(-). {We use $a :: t$ to denote concatenating a head element $a$ with a list $t$, and use $\mathsf{map(f, L)}$ to denote the result of applying a function $f$ to every element of list $L$ (as 
    {is}
    standard in functional programming languages).}
    }
    \label{fig:Explorerefine}
    \end{figure}

%% file: apa.tex
\section{Summary Computation} \label{sec:apa}

The GPS algorithm heavily relies on the use of path summaries: the test generator uses summaries to generate tests; the test executor uses summaries to help guide path selection; summaries are also instrumental in helping prove that some paths are dead ends. In this section, we present background information on how \emph{path summaries} may be computed via static analysis---in particular, algebraic program analysis \cite{cav:apa}.

Algebraic program analysis is a program-analysis framework in which one analyzes a program, represented by a control-flow graph (CFG), by (1) computing a regular expression recognizing a set of paths of interest, and then (2) interpreting that regular expression over an algebra corresponding to the analysis task at hand.
Typically, step (1) is accomplished with an algorithm for the \emph{single-source path-expression problem}, which computes, for some fixed graph $G$ and vertex $u$, a path expression $\PathExp{G}{u}{v}$ recognizing the set of paths $\Paths{G}{u}{v}$ for each vertex $v \in V$.
In this paper, we are interested in using algebraic program analysis to compute transition formulas that summarize sets of paths in a CFG, represented by a weighted graph $G = \tuple{V,E,w,s,t}$.
There are several program analyses that can be used for this task, which operate in the same fashion \cite{fmcad:fk15,CAV:SK2019,POPL:KCBR2017,POPL:CK2024}.
The idea is to instantiate the interpretation step (step (2)) of algebraic program analysis to evaluate regular expressions to a transition formula by interpreting each edge $e \in E$ as its weight $w(e)$, $0$ as $\bot$, $1$ as $\bigwedge_{x \in X} x' = x$, $+$ as $\lor$, $\cdot$ as $\circ$, and $(-)^*$ as some over-approximate reflexive-transitive-closure operator.
The analyses referred to above differ essentially in their choice of the operator used to interpret $(-)^*$.  In our implementation of GPS, we use a variant of \emph{Compositional Recurrence Analysis} (CRA) \cite{fmcad:fk15}, which, given a formula $F$ representing a loop body, computes $F^*$ by extracting recurrence relations from $F$ and computing their closed forms.

\subsection{Efficient Offline Computation of Single-Target Summaries}
\label{Se:SingleTargetSummaries}


As previously discussed, in line~\ref{line:interproccheck_guideditps} of the \textsc{Explore}(-) procedure (Figure~\ref{fig:gps-intraproc}), GPS relies on path-summary queries from an arbitrary CFG vertex $u$ to the
terminal
vertex,
and such queries are made on almost every step of executing a test.
Thus, relying on a path-summary query via Tarjan's algorithm \cite{tarjan} would incur a somewhat expensive single-source path-expression query with a different vertex at almost every step of the GPS algorithm.
This section describes a key optimization that pre-computes \emph{all} static queries required by GPS offline, using a single call to Tarjan's algorithm, thus lowering pre-computation costs by a linear factor and saving the work of re-analyzing loops.

{In particular,} we can \emph{dualize} Tarjan's algorithm to solve the \emph{single-target path-expression} problem: i.e., 
given graph $G$ and vertex $v$, for each vertex $u \in V$ compute (in near-linear time) a path expression $\PathExp{G}{u}{v}$ whose language is the set of paths $\Paths{G}{u}{v}$.
This step is accomplished via the following three-step process:

\newcommand{\reverse}[1]{#1^{R}}
\begin{itemize}
    \item {\bf Step 1: Reverse weighted graph $G$.} 
    Given $G = (V,E,w,s,t)$, its reversal $\reverse{G} = (V, \reverse{E}, \reverse{w}, t, s)$ reverses the direction of all edges and swaps the $s$ and $t$ vertices.  Formally,
    $\reverse{E} \defeq \set{\tuple{v,u} : \tuple{u,v} \in E}$ and $\reverse{w}(u,v) \defeq w(v,u)$. 
    \item {\bf Step 2: Define a new sequential-composition operator $\reverse{\circ}$.}
    Define the \emph{reversed} sequential-composition operator $F \reverse{\circ} G \triangleq G \circ F$, which reverses the direction of composition of two transition formulas $F , G \in \TF{X}$ to compensate for the fact that all edge directions in $\reverse{G}$ have been reversed.
    Note that $F_n \reverse{\circ} \dots \reverse{\circ} F_1 = F_1 \circ \dots \circ F_n$.  

    \item {\bf Step 3: Run the usual analysis on $\reverse{G}$, using $\reverse{\circ}$ to interpret $(\cdot)$.} Given the results of the two steps above, one can run {algebraic program analysis} as usual, using Tarjan's algorithm to compute path expressions over the reversed weighted graph $\reverse{G}$ with $\reverse{\circ}$ as the new sequential-composition operator,
    producing a transition formula $F(v)$ for each vertex $v$.
    We have that $F(v)$ over-approximates all paths in
     $\textit{Paths}(G,v,t)$.
\end{itemize}

Via dualization, one can
efficiently compute all single-target path expressions off-line before invoking the GPS algorithm. As such, the queries discharged by the test-execution procedure $\textsc{Explore}(-)$ can be answered efficiently.

%% file: evaluation_intraproc.tex
\section{Implementation and Evaluation}
\label{sec:eval}

We {prototyped} a software model checker for C implementing the GPS algorithm, using a variant of compositional recurrence analysis  (CRA) \cite{fmcad:fk15} for path summarization, and a variant of Newton refinement \cite{DBLP:conf/sigsoft/DietschHMNP17}
for interpolation.
The implementation of Newton refinement, and thus the GPS algorithm, is limited to programs that can be expressed in linear integer arithmetic (LIA; in particular, our implementation cannot handle programs with pointers or arrays).  We disabled non-linear invariant generation in CRA to ensure that summaries are expressed in LIA.
These limitations are features of our implementation and not intrinsic to GPS {\it per se}.
As described here, GPS is an intra-procedural algorithm;
our implementation also makes a best-effort attempt to in-line procedure calls, and then analyzes the resulting call-free program.

\bfpara{Research Questions.} 
GPS combines components from testing, static analysis, and software model checking.  We start with two research questions aimed towards understanding how each of these components contribute to the overall performance of GPS: 
\begin{itemize}
    \item {\bf RQ \#1 (Effectiveness of different GPS components):} Which features of GPS are crucial in its overall performance?
    \item  {\bf RQ \#2 (GPS vs. baseline algorithms):} GPS builds upon both CRA (its underlying summarization algorithm) and {\sc Impact} \cite{cav:mcmillan06,ufo,fmcad:bcgks09} (the basis of its invariant-synthesis mechanism).  How does GPS compare with each?
\end{itemize}



\noindent
Our last research question compares GPS with the state of the art:
\begin{itemize}
    \item {\bf RQ \#3 (Overall effectiveness of the GPS algorithm):} How does GPS compare against state-of-the-art tools, especially the top-performers in the ReachSafety-Loops category of SV-COMP?
\end{itemize}


\begin{table}[tb!]
    \scriptsize
    \centering
    \begin{tabular}{c|c|c|c}
         \textbf{Benchmark} & \textbf{Type} & \textbf{Description} & \textbf{Source} \\
         \hline
        \texttt{ccbse-refute1.c} & Parametrizable & Example from Fig. 4 of Ma et. al & \cite{directedSymbExec} \\ 
        \texttt{ccbse-refute2.c} & Parametrizable & Example from Fig. 6 of Ma et. al & \cite{directedSymbExec} \\ 
        \texttt{godefroid-issta-1a.c} & Parametrizable & Example from Fig. 1a of Godefroid et. al & \cite{ISSTA:GL11} \\ 
        \texttt{godefroid-issta-1b.c} & Parametrizable & Example from Fig. 1b of Godefroid et. al & \cite{ISSTA:GL11} \\ 
        \texttt{godefroid-issta-2.c} & Non-parametrizable & Example from Fig. 2 of Godefroid et. al & \cite{ISSTA:GL11} \\ 
        \texttt{lese.c} & Parametrizable & Example from Saxena et. al & \cite{ISSTA:SPMS09}  \\ 
        \texttt{majumdar-icse07.c} & State machine & 10-state, 7-input string-processing state machine & \cite{MajumdarSenICSE07} \\ 
        \texttt{godefroid-ndss09.c}  & State machine & 5-state, 4-input string-processing state machine & \cite{ndss:glm08} \\ 
        \texttt{refcomplete-ex.c} & State machine & 4-state, 4-input state machine example from~\Cref{ssec:gas} & Synthetic \\ 
        \texttt{cadabra.c} & State machine & 8-state state machine accepting 'cadabra' & Synthetic \\ 
        \texttt{abracadabra.c} & State machine & 12-state state machine accepting 'abracadabra' & Synthetic \\ 
        \texttt{abracadabraabra.c} & State machine & 16-state state machine accepting 'abracadabraabra' & Synthetic \\
        \rfchanged{\texttt{abracadabraabracadabra.c}} & State machine & 23-state state machine accepting 'abracadabraabracadabra' & Synthetic \\ 
    \end{tabular}
    \caption{Details about the L\&K programs benchmark suite. 10 benchmarks of different parametrized values in range $[10, 10000]$ are created for each parametrizable benchmark.}
    \label{tab:l_and_k_details}
\end{table}

\bfpara{Benchmark Selection.} 
We performed the comparisons on three suites of benchmarks, with \rfchanged{297} programs in total.
\begin{itemize}
    \item {\bf Suite \#1: 230 intraprocedural benchmarks from SV-COMP.} We sub-setted 230 benchmarks from the ReachSafety-Loops category of the Software Verification Competition (SV-COMP) 2024 benchmark suite \cite{svcomp2024}, filtering out \emph{only} programs containing non-LIA syntax. 
    \item {\bf Suite \#2: 9 Safety-verification examples from this paper.} We created a benchmark suite (of safe programs) consisting of examples \textbf{EX-2} and \textbf{EX-3} in \Cref{sec:overview}, as well as 7 variants of these examples that exhibit similar loop patterns; this suite includes a total of 9 programs.
    \item {\bf Suite \#3: \rfchanged{58} Lock \& Key-style unsafe benchmarks.} We created a suite of ``L\&K''-style benchmarks similar to \textbf{EX-1} discussed in \Cref{sec:overview}, as well as the state-machine example discussed in the section on \texttt{gas}-instrumentation (\Cref{ssec:gas}). This suite of benchmarks consists of the following benchmarks (for details, see \Cref{tab:l_and_k_details}):
    \begin{enumerate}
        \item 6 programs with ``L\&K''-style, input-dependent loops from prior literature \cite{directedSymbExec,ISSTA:GL11,ISSTA:SPMS09,MajumdarSenICSE07}---each program in this collection features paths to errors that can only be discovered by successfully guessing the ``key''---i.e., 
        there are
        non-deterministically-valued expressions inside the program. 
        \item We further observed that 5 of the 6 L\&K examples are \emph{parametrizable}---that is, the length of the shortest feasible path-to-error can be prolonged by enlarging a constant inside the program, i.e., changing the ``lock.''
        To
        test
        the robustness of GPS in handling these examples, we benchmarked the tools in our comparison on 10 different values of the parameter (in the range $[10, 10000]$) for each of the parametrizable programs. 
        \item The remaining \rfchanged{7} benchmarks are loops implementing string-processing state machines where an assertion violation is triggered when the state machine arrives at an accepting state. As shown in \Cref{tab:l_and_k_details}, this suite includes two examples from prior work \cite{MajumdarSenICSE07, ndss:glm08}, the state-machine example from \Cref{ssec:gas}, and \rfchanged{four} synthetic state-machine examples.
        \end{enumerate}
    In total, this benchmark suite consists of \rfchanged{58} unsafe programs (i.e., 1 unparametrizable benchmark, 50 parametrized benchmarks, and \rfchanged{7} state machine benchmarks).
\end{itemize}

\input{figures/table-gps-vs-impact-and-cra}

\input{figures/table-rq13}

\bfpara{Evaluation Setup.} We ran all experiments on a Dell XPS 15 laptop running {Ubuntu 24.04 LTS} with an Intel i7-13700H CPU and 16 GiB of RAM. We used the \texttt{benchexec} script standard for evaluating tools in SV-COMP \cite{benchexec}.
The CPU that we used has {6} high-performance cores and {8} efficient cores.
To reduce variance in benchmark timings, we configured \texttt{benchexec} to only use the 6 high-performance cores. For all benchmarks, we used \texttt{benchexec} to enforce a timeout threshold of 600 seconds and a memory limit of 15GB.

We now address each research question in turn.

\begin{figure}
    \centering
    \begin{tabular}{@{\hspace{0ex}}c@{\hspace{0ex}}c@{\hspace{0ex}}c@{\hspace{0ex}}@{\hspace{0ex}}}
    \includegraphics[width=.28\linewidth]{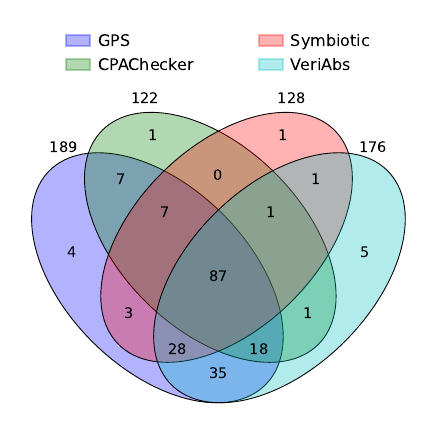}
    &
    \includegraphics[width=.28\linewidth]{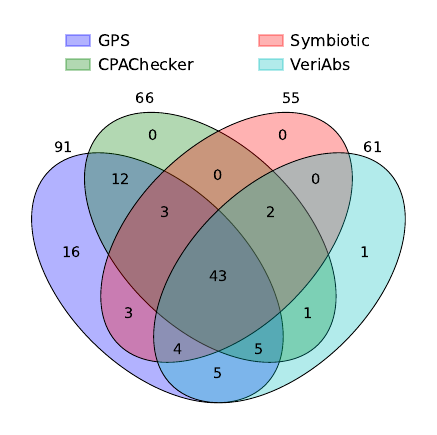}
    &
    \resizebox{.48\textwidth}{!}{
        \input{figures/loops/new-scatter.cameraready}
    }
    \end{tabular}

    \caption{Benchmark results for RQ \#3. From left to right: (a): Venn diagram of safe benchmark results, across all three benchmark suites, for RQ \#3 (200 benchmarks total); (b): Venn diagram of unsafe benchmark results, across all three benchmark suites, for RQ \#3 (96 benchmarks total); (c) Cactus plot of results for RQs \#3. }
    \label{fig:rq3-benchmark-results}
\end{figure}

\begin{mdframed}
         {\bf RQ \#1:} Which features of GPS are crucial in its overall performance?
\end{mdframed}
\vspace*{0.1in}

There are several key ingredients to GPS: (1) the path summaries generated by a static-analysis tool (in our case, CRA); (2) the test-generation and refinement procedures inside the GPS model-checking algorithm; (3) the use of the instrumentation variable \texttt{gas} to make GPS refutation-complete. To understand  the impact of the various components of GPS, we performed an ablation study with several variation of GPS, the results of which are reported in Table~\ref{tab:ablation-benchmark-stats}.  Below, we describe each variation and the result of the comparison.
\begin{itemize}
    \item {\bf (\GPSnocra).} Instead of using CRA as the summary oracle, we use a trivial oracle ($\textsf{Sum}(G,u,v) = \top$ for all vertices $u$ and $v$), and additionally we do not perform \texttt{gas}-instrumentation
    because some of the other tools that we wanted to compare with also did not perform \texttt{gas}-instrumentation.
    This variation essentially omits the main novel features of GPS: with trivial summaries, summary-directed test generation coincides with {\sc Dart}-style directed test generation,
    dead-end detection always fails, and
    dead-end interpolation coincides with {\sc Impact}-style infeasible-path interpolation.  The resulting algorithm essentially follows the tradition of {\sc Synergy} \cite{synergy}, {\sc Dash} \cite{Dash}, and {\sc McVeto} \cite{CAV:TLLBDEAR10} in that it combines concolic testing with abstraction refinement.
    
    We find that \GPSnocra performs significantly worse than GPS on both safe and unsafe benchmarks.  The impact of using summaries to generate high-quality tests is
    apparent when one compares GPS with
    \GPSnocra on the unsafe SV-COMP suite (on which \GPSnocra does well, because that suite is dominated by deterministic benchmarks where summary-guidance is not needed) with the L\&K suite (on which it does poorly due to absence of this guidance).

    \item {\bf (GPSLite)} We measure the impact of invariant synthesis through dead-end interpolation (Section~\ref{sec:algorithm})
    by comparing GPS with GPSLite (Section~\ref{sec:gpslite_algorithm}, instantiated with gas-instrumentation), which omits this feature.  We find that GPSLite is somewhat faster than GPS on unsafe benchmarks (because it avoids the overhead of computing interpolants), but GPS dominates GPSlite on safe benchmarks.  
    \item {\bf (\GPSnogas).}  We study the repercussions of the
    refutation-completeness transformation described in~\Cref{ssec:gas} by comparing
    GPS with \GPSnogas, a variant of GPS without the transformation.
    We find that the theoretical guarantee of refutation-completeness translates to positive effect on the practical performance of GPS.
\end{itemize}

\vspace*{0.1in}
\begin{mdframed}
        \item {\bf RQ \#2:} How does GPS compare with baseline techniques it builds upon?
\end{mdframed}
\vspace*{0.1in}

To address RQ \#2, we benchmarked GPS against a re-implementation of the {\sc Impact} algorithm inside CPAChecker \cite{tool-cpachecker,fmcad:bcgks09}.
We configured the latest release\footnote{Version 4.0 at the time of writing.} of CPAChecker to use the option ``\texttt{-predicateAnalysis-ImpactRefiner-SBE.properties}'', which tells CPAChecker to use the {\sc Impact} algorithm under the hood.
{
The results for {\sc Impact} are given in \Cref{tab:ablation-benchmark-stats}.
}

{
\bfpara{Findings.}
One way to answer RQ \#2 is by thinking of the starting point as {\sc Impact} (see \Cref{tab:ablation-benchmark-stats}, column 6), which uses Craig interpolation and abstract reachability trees, to which we have added (i) a different search strategy combining breadth-first search and concolic testing, and (ii) the novel features of summary-directed testing, dead-end detection, and dead-end interpolation.
Here, to control for {\sc Impact} not using gas instrumentation, we use \GPSnogas as the right end-point:
\[
  \text{{\sc Impact}}~\left(\frac{136}{\rfchanged{297}}\right) \xrightarrow[\text{search strategy}]{+35}
  \mbox{\GPSnocra}~\left(\frac{171}{\rfchanged{297}}\right) \xrightarrow[\text{novel features}]{+\rfchanged{102}}
  \mbox{\GPSnogas}~\left(\frac{\rfchanged{273}}{\rfchanged{297}}\right)
\]
While the different search strategy gives a substantial improvement in the number of benchmarks solved, the novel features of GPS provide a much larger one.
The overall improvement is almost a factor of 2x over {\sc Impact}.
(The full GPS algorithm did even better: \rfchanged{280/297}.)
}

{
Another way to answer RQ \#2 is by thinking of the starting point as CRA (see \Cref{tab:ablation-benchmark-stats}, column 7), with its abilities to create summaries, and to look at the improvement in verification capability---the number of safe benchmarks solved---obtained as we add dead-end detection and then dead-end interpolation to the CRA baseline.
We see the following improvement, with roughly equal contributions by each:
\[
  \textrm{CRA}~\left(\frac{156}{200}\right) \xrightarrow[\text{dead-end detection}]{+\rfchanged{16}}
  \mbox{\GPSLite}~\left(\frac{\rfchanged{172}}{200}\right) \xrightarrow[\text{dead-end interpolation}]{+\rfchanged{17}}
  \textrm{GPS}~\left(\frac{\rfchanged{189}}{200}\right)
\]
While the overall improvement in the number of benchmarks solved, compared to CRA, is a respectable \rfchanged{21.2}\%, GPS also adds the ability to find counter-examples (and is refutation-complete).
}

\vspace*{0.1in}
\begin{mdframed}
        \item {\bf RQ \#3:} How does GPS compare against state-of-the-art tools?
\end{mdframed}
\vspace*{0.1in}

To address RQ \#3, we compared our prototype implementation of GPS against three portfolio-based model checkers that implement a diverse range of techniques.

\bfpara{Baseline Comparisons for RQ \#3.}
We compared GPS against three other model checkers:
\begin{itemize}
    \item {{\bf VeriAbs} \cite{tool-veriabs}, a portfolio of model checkers, which (along with its variant VeriAbsL) ranked first (and second place, for VeriAbsL) in the SV-COMP 2024 ReachSafety-Loops category. }
    \item {\bf Symbiotic} \cite{tool-symbiotic}, a portfolio-based symbolic-execution tool, which was the third-best performer of the SV-COMP 2024 ReachSafety-Loops category.
    \item {\bf CPAChecker} \cite{tool-cpachecker}, a portfolio-based model checker implementing a suite of different strategies;
    although CPAChecker was not a top scorer in the ReachSafety-Loops category of SV-COMP 2024,
{
    the motivation for including it was to compare GPS with
}
    CPAChecker's portfolio mode, because RQ \#2 compared GPS against the \textsc{Impact} implementation inside CPAChecker. 
\end{itemize}
{For each tool involved in this comparison, we used the version and configuration provided for its SV-COMP 2024 contest entry. In particular, in this experiment, both Symbiotic and CPAChecker were used in portfolio modes as specified in their competition entries \cite{tool-symbiotic,veriabs-svcomp24}. }

\bfpara{Findings.} Key statistics about this comparison are shown in~\Cref{tab:rq13-benchmark-stats}.
We observe that GPS has the best performance (both in terms of the number of successful benchmarks and running time) for all benchmark suites.  A cactus plot depicting a more fine-grained view of success rate vs.\ running time appears in the right of \Cref{fig:rq3-benchmark-results}.

Our result confirm that GPS has excellent performance on L\&K examples, owing to the fact that it uses summaries to generate a high-quality test case and then efficiently simulates execution on that test case.   Other tools are more sensitive to the length of the shortest counter-example, which\IfCameraReady{ we illustrate in \rfchanged{the extended version of this paper \cite{GPSExtendedVersion}.}}{ we illustrate in detailed line plots for each parametrized benchmark in~\Cref{appendix:info_about_l_and_k_suite}  (\Cref{fig:parametric_benchmarks_for_rq2_3}). } \rfchanged{GPS solves all except one L\&K benchmarks, with the one unsolved benchmark being a 23-state state machine benchmark on which the underlying static analysis (CRA) times out when generating the initial path summary.}

Venn diagrams for the sets of benchmarks on which each tool is successful can be found in \Cref{fig:rq3-benchmark-results}, from which we conclude that GPS can be profitably incorporated into portfolio solvers.

%% file: figures/table-gps-vs-impact-and-cra.tex

\begin{table}[tb!]
\small
    \centering
\resizebox{\columnwidth}{!}{%

\begin{tabular}{@{}lc|c@{}r|c@{}r|c@{}r|c@{}r|c@{}r|c@{}r}
\toprule
 & & \multicolumn{2}{c|}{GPS} & \multicolumn{2}{c|}{GPSLite}  & \multicolumn{2}{c|}{\GPSnogas} & \multicolumn{2}{c|}{\GPSnocra} &  \multicolumn{2}{c|}{{\sc Impact}} & \multicolumn{2}{c}{CRA}\\
 & \#tasks & \#correct &\quad time & \#correct &\quad time & \#correct &\quad time & \#correct &\quad time   & \#correct &\quad time  & \#correct &\quad time\\\midrule

\textbf{Suite \#1 (SV-COMP)} & 230 & \textbf{\tblch{214}}& \tblch{7389.1}& 200 & \tblch{10302.9}& \tblch{210}& \tblch{7845.7}& 168 & 37969.2   & 119 &68616.9 & 155 &752.3\\
\hline 
 \multicolumn{1}{r}{Safe} & 191 & \textbf{\tblch{180}}& \tblch{4852.7}& \tblch{166}& \tblch{8301.5}& \tblch{176}& \tblch{5482.0}& 136 & 33948.5   & 95 &59559.3 & 155 &746.4\\ 
 
 \multicolumn{1}{r}{Unsafe} & 39 & \textbf{\tblch{34}}& \tblch{2536.4}& \tblch{\textbf{34}}& \tblch{2001.4}& \textbf{34}& \tblch{2363.7}& 32 & 4020.7   & 24 &9057.5 & 0 &5.9\\ 
\hline 
\textbf{Suite \#2 (Safe Exs)} & 9 & \textbf{9} & \tblch{6.0}& \tblch{6}& \tblch{837.5}& \textbf{\tblch{9}}& \tblch{6.4}& 1 & 2780.7   & 2 &4209.4  & 1 &0.7\\
\hline 
\textbf{Suite \#3 (L\&K)} & 57 & \textbf{57}  & \tblch{623.3}& \textbf{57}  & \tblch{620.9}& 54  & \tblch{1749.5}& 2 & \tblch{27718.5}& 15 &\tblch{27808.9}& 0 &22.6\\
\hline 
  \multicolumn{1}{r}{Unparametrized} & 1 & \textbf{1}  & 0.2 & \textbf{1}  & 0.2 & \textbf{1}  & 0.2 & \textbf{1} & 0.2   & \textbf{1} &2.3 & 0 &0.2\\  
  \multicolumn{1}{r}{Parametrized} & 50 & \textbf{50}  & 10.8& \textbf{50}  & \tblch{8.5}& \textbf{50}  & \tblch{8.9}& 1 & 23511.6   & 11 &25370.3  & 0 &7.9\\
  \multicolumn{1}{r}{State Machines} & \tblch{7}& \textbf{\tblch{6}}& \tblch{612.3}& \textbf{6}  & \tblch{612.2}& 3& \tblch{1740.4}& 0 & \tblch{4206.7}& 3 &\tblch{2436.3}& 0 &14.5\\

\midrule
Total & \tblch{297}& \textbf{\tblch{280}}& \tblch{8018.4}& \tblch{263}& \tblch{11761.3}& \tblch{273}& \tblch{9601.6}& 171 & \tblch{68468.4}& 136 &\tblch{100635.2}& 156 &775.6 \\
Failures &  & \multicolumn{2}{c|}{\tblch{8/9/0}} & \multicolumn{2}{c|}{\tblch{13/21/0}} & \multicolumn{2}{c|}{\tblch{9/15/0}} & \multicolumn{2}{c|}{\tblch{110/16/0}} &  \multicolumn{2}{c|}{\tblch{160/0/1}} & \multicolumn{2}{c}{\tblch{2/0/139}}\\
\bottomrule

\end{tabular}
}
\vspace{1.0ex}
      \caption{ 
    Overall benchmark statistics for RQ \#1 and \#2. The Failures row contains triples $T/M/U$ indicating the number of benchmarks for which the tool exceeded the Time bound, exceeded the Memory bound, or reported Unknown, {respectively}. ``GPS'' refers to unablated version of GPS, ``\GPSlite'' refers to the version of GPS without abstraction refinement presented in~\Cref{sec:gpslite_algorithm}, ``\GPSnogas'' refers to ablated GPS without \texttt{gas}-instrumentation, ``\GPSnocra'' refers to ablated GPS without CRA-generated summaries,  ``CRA'' refers to only using CRA to prove safety, and {\sc Impact} refers to the lazy abstraction with interpolants implementation in CPAChecker. 
    }
    \label{tab:ablation-benchmark-stats}

\end{table}

%% file: figures/table-rq13.tex

\begin{table}[tb!]
\small
    \centering
\resizebox{\columnwidth}{!}{%
\begin{tabular}{@{}lc|c@{}r|c@{}r|c@{}r|c@{}r}
\toprule
 & & \multicolumn{2}{c|}{GPS} & \multicolumn{2}{c|}{VeriAbs} & \multicolumn{2}{c|}{Symbiotic} & \multicolumn{2}{c}{CPAChecker (Portfolio)} \\
 & \#tasks & \#correct &\quad time & \#correct &\quad time & \#correct &\quad time & \#correct &\quad  time \\\midrule

\textbf{Suite \#1 (SV-COMP)} & 230 & \textbf{\tblch{214}}& \tblch{7389.1}& 201 & 12362.6 & 158 & 62130.2 & 146 & 64674.9 \\
\hline
 \multicolumn{1}{r}{Safe} & 191 & \textbf{\tblch{180}}& \tblch{4852.7}& 169  & 9223.0 & 126  & 56825.7 & 120 & 56785.3 \\
 
 \multicolumn{1}{r}{Unsafe} & 39 & \textbf{\tblch{34}}& \tblch{2536.4}& 32  & 3139.6 & 32 & 5304.5 & 26 & 7889.6 \\
\hline
\textbf{Suite \#2 (Safe Exs)} & 9 & \textbf{9} & \tblch{6.0}& 7 & 752.7 & 2 & 4667.6 & 2 & 4617.7 \\
\hline 
\textbf{Suite \#3 (L\&K)} & 57 & \textbf{57}  & \tblch{623.3}& \tblch{29}& \tblch{14853.2}& 23  & \tblch{20700.5} & \tblch{40} & \tblch{14751.1} \\
\hline 
  \multicolumn{1}{r}{Unparametrizable} & 1 & \textbf{1}  & 0.2 & 0 & 392.7 & \textbf{1} & 0.7 & \textbf{1} & 3.3 \\  
  \multicolumn{1}{r}{Parametrizable} & 50 & \textbf{50}  & 10.8& 22  & 14393.0 & 17  & 20425.6 & 32 & 14054.4 \\
  \multicolumn{1}{r}{State Machines} & \tblch{7}& {\tblch{6}}& \tblch{612.3}& \tblch{\textbf{7}}& \tblch{67.5}& 5 & \tblch{274.2} & \tblch{\textbf{7}} & \tblch{693.4} \\ 

\midrule
Total & \tblch{297}& \textbf{\tblch{280}}& \tblch{8018.4}& \tblch{237}& \tblch{27968.5}& 183 & \tblch{87498.3} & \tblch{188} & \tblch{84043.7} \\
Failures &  & \multicolumn{2}{c|}{\tblch{8/9/0}} & \multicolumn{2}{c|}{{9/9/42}} & \multicolumn{2}{c|}{{96/\tblch{18}/0}} & \multicolumn{2}{c}{{108/0/1}} \\
\bottomrule
\end{tabular}}\\

\vspace{1.0ex}
    \caption{Overall benchmark statistics for RQ \#3. 
    Times are in seconds.
    The Failures row contains triples $T/M/U$ indicating the number of benchmarks for which the tool exceeded the Time bound, exceeded the Memory bound, or reported Unknown, {respectively}. 
    }
    \label{tab:rq13-benchmark-stats}
    \vspace*{-0.2in}
\end{table}

%% file: figures/loops/new-scatter.cameraready.tex
\begin{tikzpicture}
            \begin{axis}[
            xlabel={Number of Solved Benchmarks},
            ylabel={Running time (s)},
            title=Benchmarks from Suites \#1 and \#2,
            grid=both,
            tick align = outside,
            yticklabel style={/pgf/number format/fixed},     
            scaled x ticks = false,
            xticklabel style={/pgf/number format/fixed},
            legend style={at={(1.05,1)}, anchor=north west}
            ]
            
\addplot+[] coordinates {
(0, 1.1692625429986947)
(1, 1.2074129909997282)
(2, 1.2120662680099485)
(3, 1.2126574469984917)
(4, 1.2264945600036299)
(5, 1.2325118190001376)
(6, 1.2350106869998854)
(7, 1.2710695050045615)
(8, 1.2802168579992212)
(9, 1.2929100540004583)
(10, 1.2965021449999767)
(11, 1.2976406300003873)
(12, 1.3039816939999582)
(13, 1.3075336309993872)
(14, 1.310254972000621)
(15, 1.313283676005085)
(16, 1.3160203709994676)
(17, 1.3163996430012048)
(18, 1.3229445760007366)
(19, 1.3273399999998219)
(20, 1.3342176689911867)
(21, 1.335501696999927)
(22, 1.3388060919969575)
(23, 1.3389683709974634)
(24, 1.3422585270018317)
(25, 1.3470684220083058)
(26, 1.347192389999691)
(27, 1.3486991339996166)
(28, 1.348775463000493)
(29, 1.3536136390011961)
(30, 1.358908892000727)
(31, 1.3605173240002841)
(32, 1.3618238700000802)
(33, 1.3621003799999016)
(34, 1.3656542420012556)
(35, 1.3676373359994614)
(36, 1.3698894879998988)
(37, 1.3844142710004235)
(38, 1.3866915280013927)
(39, 1.3963518969976576)
(40, 1.3971148729997367)
(41, 1.406108946997847)
(42, 1.406127438999647)
(43, 1.4069505160005065)
(44, 1.4071297389891697)
(45, 1.4089786069998809)
(46, 1.4108583739871392)
(47, 1.4203733730009844)
(48, 1.450723170000856)
(49, 1.4560657329984679)
(50, 1.4749391570003354)
(51, 1.600854497999535)
(52, 1.6188544349997755)
(53, 1.6415919129976828)
(54, 1.6470179030002328)
(55, 1.6703389610011072)
(56, 1.6721149810000497)
(57, 1.6875332409999828)
(58, 1.8211982829961926)
(59, 1.9293573119994107)
(60, 2.327879030999611)
(61, 2.529408342001261)
(62, 2.591897174999758)
(63, 3.2216396479998366)
(64, 4.8455828729929635)
(65, 5.4886978790018475)
(66, 5.546739950001211)
(67, 6.351450108006247)
(68, 7.135878136999963)
(69, 22.113656285000616)
(70, 33.49411749900173)
(71, 131.29849154601106)
(72, 132.51882896200186)
(73, 132.91823055699933)
(74, 133.13233325599958)
(75, 133.7754480149997)
(76, 133.90704050200293)
(77, 134.26046258600036)
(78, 134.6905190010002)
(79, 135.47089929399954)
(80, 135.84769284200047)
(81, 135.9805661800001)
(82, 136.15062600300007)
(83, 137.13254734200018)
(84, 137.26485500700073)
(85, 137.4224389049923)
(86, 137.49469679199683)
(87, 138.055620627998)
(88, 138.7702212879958)
(89, 138.9473677699998)
(90, 138.99776600900805)
(91, 139.00068559100328)
(92, 139.01032394099457)
(93, 139.03506932399978)
(94, 139.0821180849889)
(95, 139.1509274880009)
(96, 139.2397971140017)
(97, 139.51209001800453)
(98, 139.6631225149904)
(99, 140.05687792299432)
(100, 140.21840716600127)
(101, 140.27176603499902)
(102, 140.43488724399867)
(103, 140.55511091799417)
(104, 142.56878429599965)
(105, 142.65471291099675)
(106, 142.67220949500188)
(107, 142.93865517400263)
(108, 149.00241300000198)
(109, 149.327668412996)
(110, 151.77793579499848)
(111, 156.38585790499928)
(112, 159.4049892760013)
(113, 159.980390429002)
(114, 161.41905812799996)
(115, 162.92790874099956)
(116, 163.26814681700125)
(117, 163.2823259780016)
(118, 164.47771622501023)
(119, 164.71021940099672)
(120, 165.84582515699913)
(121, 169.4886906899992)
(122, 177.09586192600182)
(123, 177.3295162320028)
(124, 187.97018890100026)
(125, 192.28106469400154)
(126, 192.56084083599944)
(127, 193.0785367899989)
(128, 193.49684474999958)
(129, 194.10650605900082)
(130, 194.52289781999934)
(131, 194.73121820300003)
(132, 194.8853282620039)
(133, 195.36979491700185)
(134, 195.54578477700124)
(135, 195.61321514000156)
(136, 196.95374139101477)
(137, 262.3132185380018)
(138, 291.2080539090002)
(139, 377.62509118900016)
(140, 442.76140190899605)
(141, 443.3944931839942)
(142, 443.52651643100035)
(143, 444.1278608410066)
(144, 444.1606766900077)
(145, 449.558451263998)
(146, 463.0016638110028)
(147, 581.9945247510004)
};
\addlegendentry{CPAChecker};
\addplot+[] coordinates {
(0, 0.10915527199995267)
(1, 0.11109248899992963)
(2, 0.11137738700017508)
(3, 0.11160536199986382)
(4, 0.11204850500052999)
(5, 0.11242767099975026)
(6, 0.11299808300009317)
(7, 0.11374727400016127)
(8, 0.11396220200003881)
(9, 0.11404127700006939)
(10, 0.11412964600003761)
(11, 0.1142793540002458)
(12, 0.11428007199992862)
(13, 0.11446146600064822)
(14, 0.11457362700002705)
(15, 0.1151909540003544)
(16, 0.11524142899997969)
(17, 0.11573702900000171)
(18, 0.11588723399927403)
(19, 0.11594303399942874)
(20, 0.11634281399983593)
(21, 0.11635911899975326)
(22, 0.1167189399993731)
(23, 0.11684953800022413)
(24, 0.11694277100014006)
(25, 0.11722273000032146)
(26, 0.11726847399995677)
(27, 0.11729966900020372)
(28, 0.11731704200064996)
(29, 0.1173857069998121)
(30, 0.11741040800006886)
(31, 0.1176578860004156)
(32, 0.11773158699998021)
(33, 0.11779192699941632)
(34, 0.11794800000006944)
(35, 0.11847874700015382)
(36, 0.11893762199997582)
(37, 0.11897404100000131)
(38, 0.11899584399998275)
(39, 0.11936975299977348)
(40, 0.11937182300061977)
(41, 0.11968812400004936)
(42, 0.1197027440000511)
(43, 0.12014059600005567)
(44, 0.12017629700039834)
(45, 0.12136217999977816)
(46, 0.12189978900005372)
(47, 0.12223334899999827)
(48, 0.12295304399958695)
(49, 0.12303510400033701)
(50, 0.12315083599969512)
(51, 0.12319157999991148)
(52, 0.12325740100004623)
(53, 0.12326340699974025)
(54, 0.12329281100028311)
(55, 0.12334569900008319)
(56, 0.1234033629998521)
(57, 0.12342904099932639)
(58, 0.12369448500066937)
(59, 0.12372162599990588)
(60, 0.12373186899981192)
(61, 0.12419216499984032)
(62, 0.12428279200048564)
(63, 0.12461346500003856)
(64, 0.12471729200024129)
(65, 0.12484687699998176)
(66, 0.12493767599971761)
(67, 0.12539577300003657)
(68, 0.12540897600047174)
(69, 0.1254322240001784)
(70, 0.12567072100046062)
(71, 0.1260226370000055)
(72, 0.12662471899966476)
(73, 0.1267741230003594)
(74, 0.12677813199934462)
(75, 0.1270394410003064)
(76, 0.1276129409998248)
(77, 0.12777774599999248)
(78, 0.12800868599970272)
(79, 0.1281598059999851)
(80, 0.12870342699989124)
(81, 0.1289371940001729)
(82, 0.128949159000058)
(83, 0.12991965699984576)
(84, 0.13006436500018026)
(85, 0.13057825999931083)
(86, 0.1308014380001623)
(87, 0.13115689100050076)
(88, 0.13140910800007077)
(89, 0.13163855600032548)
(90, 0.1316411729999345)
(91, 0.1327432480002244)
(92, 0.13281668800027546)
(93, 0.13346909600022627)
(94, 0.1335507819999293)
(95, 0.13361524099991584)
(96, 0.13369358700037992)
(97, 0.13369358999989345)
(98, 0.133774228999755)
(99, 0.13462772000002587)
(100, 0.13464683999973204)
(101, 0.13470892799978174)
(102, 0.13473453799997515)
(103, 0.13492787399991357)
(104, 0.13576975600062724)
(105, 0.1363985709995177)
(106, 0.13669546799974341)
(107, 0.1369331530004274)
(108, 0.13716287500028557)
(109, 0.13848296200001187)
(110, 0.13869657100076438)
(111, 0.1388852649997716)
(112, 0.13889706999998452)
(113, 0.14007712100010394)
(114, 0.1403816239999287)
(115, 0.14064574500025628)
(116, 0.1406709659995613)
(117, 0.14107807500022318)
(118, 0.14155481699981465)
(119, 0.14279944900044939)
(120, 0.14338836200022342)
(121, 0.14346568199999865)
(122, 0.1436346689997663)
(123, 0.14484315200024866)
(124, 0.14521029500065197)
(125, 0.14524701699997422)
(126, 0.14565893500002858)
(127, 0.14616902699981438)
(128, 0.14625234800041653)
(129, 0.14634927200040693)
(130, 0.14973748000011255)
(131, 0.15015121899978112)
(132, 0.150190439000653)
(133, 0.15029134499991414)
(134, 0.1513834699999279)
(135, 0.15177304900043964)
(136, 0.1523229310005263)
(137, 0.15309989999991558)
(138, 0.15322220500002004)
(139, 0.1534368020002148)
(140, 0.15375768199919548)
(141, 0.15717583399998603)
(142, 0.15773679900030402)
(143, 0.1582918720005182)
(144, 0.1596995169998081)
(145, 0.162361945000157)
(146, 0.16444378799951664)
(147, 0.16473170100016432)
(148, 0.16527513999972143)
(149, 0.1666074649992879)
(150, 0.1672718290001285)
(151, 0.1677464049998889)
(152, 0.1679147599998032)
(153, 0.168184838000343)
(154, 0.16913540299992746)
(155, 0.17552953799986426)
(156, 0.17602354799964814)
(157, 0.17822420299944497)
(158, 0.18376525400071841)
(159, 0.18486693700015167)
(160, 0.188841176999631)
(161, 0.1908042940001451)
(162, 0.19200917199941614)
(163, 0.19641679800042766)
(164, 0.19719582500010802)
(165, 0.2005583440004557)
(166, 0.20164210199982335)
(167, 0.20715709400019477)
(168, 0.20742839700051263)
(169, 0.2079027390000192)
(170, 0.20939113699932932)
(171, 0.21050012800003515)
(172, 0.21054284799993184)
(173, 0.21381490300063888)
(174, 0.21936014800030534)
(175, 0.22205560600013996)
(176, 0.22277980600028968)
(177, 0.22295361099986621)
(178, 0.2291221210000458)
(179, 0.23692949799988128)
(180, 0.24705031099983898)
(181, 0.24971239499973308)
(182, 0.2504746009999508)
(183, 0.25337113499972475)
(184, 0.2551738619995376)
(185, 0.2558859549999397)
(186, 0.25970718000007764)
(187, 0.27665038499981165)
(188, 0.2810839990000318)
(189, 0.2860693310000215)
(190, 0.3080661780004448)
(191, 0.30979482699996197)
(192, 0.3485320990002947)
(193, 0.3609832189995359)
(194, 0.37077617999966606)
(195, 0.37214105500061123)
(196, 0.3806685289999905)
(197, 0.38802462999956333)
(198, 0.41330802899938135)
(199, 0.5178575199997795)
(200, 0.5454625639999904)
(201, 0.5468602129994906)
(202, 0.588560563000101)
(203, 0.6954002890001902)
(204, 0.7956775650000054)
(205, 0.854748261000168)
(206, 0.8774424779999208)
(207, 1.2233252680000533)
(208, 1.3204300830002467)
(209, 1.5452619760001198)
(210, 1.6267229499999303)
(211, 1.640968466999766)
(212, 2.1739809089995106)
(213, 2.478241731000253)
(214, 2.529429340999741)
(215, 2.679949676999968)
(216, 3.814238400999784)
(217, 4.355365660000189)
(218, 4.398053317999711)
(219, 5.683562993000123)
(220, 9.188449637999838)
(221, 10.742884717999914)
(222, 502.588431913)
};
\addlegendentry{GPS};
\addplot+[] coordinates {
(0, 0.2587477290071547)
(1, 0.26318039699981455)
(2, 0.26559491899388377)
(3, 0.2685734009864973)
(4, 0.2697577539947815)
(5, 0.2718409929948393)
(6, 0.2722366630041506)
(7, 0.27482441600295715)
(8, 0.2886458500142908)
(9, 0.30112380599894095)
(10, 0.3015593500022078)
(11, 0.30237090900482144)
(12, 0.31255737600440625)
(13, 0.31941005699627567)
(14, 0.3194990729971323)
(15, 0.32141917199624004)
(16, 0.3222534399974393)
(17, 0.32583932900161017)
(18, 0.3263328650064068)
(19, 0.32635038499574875)
(20, 0.33008182200137526)
(21, 0.3335629279972636)
(22, 0.33424072500201873)
(23, 0.33466409600077895)
(24, 0.33665992399619427)
(25, 0.34000327099784045)
(26, 0.34656978600105504)
(27, 0.34680999699776294)
(28, 0.3482799440025701)
(29, 0.35823327799880644)
(30, 0.36206282100101816)
(31, 0.36939284300024156)
(32, 0.3741845940021449)
(33, 0.3822257020001416)
(34, 0.3834679920000781)
(35, 0.38616675599769223)
(36, 0.3874252519890433)
(37, 0.38879905000067083)
(38, 0.39163450200430816)
(39, 0.3939185280032689)
(40, 0.39490861700323876)
(41, 0.4003490689974569)
(42, 0.40568966300270404)
(43, 0.40716143299869145)
(44, 0.4121033709961921)
(45, 0.41521284299960826)
(46, 0.4167642819957109)
(47, 0.4250941939972108)
(48, 0.4277566319942707)
(49, 0.4496082630066667)
(50, 0.4680222099996172)
(51, 0.47463410799537087)
(52, 0.4848355059948517)
(53, 0.4943358710006578)
(54, 0.5030625979998149)
(55, 0.5225050699991698)
(56, 0.5452527079978609)
(57, 0.5673685709989513)
(58, 0.5834646099974634)
(59, 0.5872292640015075)
(60, 0.5920739089997369)
(61, 0.6455406169989146)
(62, 0.6650337749997561)
(63, 0.7528398819995346)
(64, 0.774071121999441)
(65, 0.8170255449986144)
(66, 0.8273003739886917)
(67, 0.8408431240022765)
(68, 0.9762088899988157)
(69, 1.3813250360035454)
(70, 2.0178904550120933)
(71, 2.3503550439927494)
(72, 2.4561629519957933)
(73, 3.4319423679989995)
(74, 4.031778096992639)
(75, 4.445307744012098)
(76, 6.188127897010418)
(77, 6.202661781004281)
(78, 6.23746412199398)
(79, 6.312680328002898)
(80, 6.95990902199992)
(81, 7.772332617001666)
(82, 8.502283409994561)
(83, 8.624021695002739)
(84, 8.777378566999687)
(85, 19.885777358002088)
(86, 27.993831682997552)
(87, 30.918190472002607)
(88, 42.9589230499987)
(89, 43.13913690499612)
(90, 43.54412237799261)
(91, 86.42170583000552)
(92, 86.55679803000385)
(93, 105.7259394390021)
(94, 172.03438567799458)
(95, 172.30932735999522)
(96, 202.91382706600052)
(97, 285.2722109739989)
(98, 285.85472899900196)
(99, 333.5244787630072)
(100, 333.53415089800546)
(101, 333.536513982006)
(102, 333.5476078500069)
(103, 333.55516283700126)
(104, 333.57804496500466)
(105, 333.59048187400913)
(106, 333.604433581997)
(107, 333.6083857049962)
(108, 333.62801625000066)
(109, 333.6298777149932)
(110, 333.63215240200225)
(111, 333.64487039198866)
(112, 333.65983630999835)
(113, 333.68091592499695)
(114, 333.7048669979995)
(115, 333.7123407110048)
(116, 333.7460259209911)
(117, 333.7721251250041)
(118, 333.77793276500597)
(119, 333.78605357799825)
(120, 333.8161567599891)
(121, 333.82960161899973)
(122, 333.93356926699926)
(123, 333.9770318140072)
(124, 349.24586900899885)
(125, 393.49725493998267)
(126, 393.64984352399915)
(127, 393.67752833204577)
(128, 393.76254021099885)
(129, 393.7704478039959)
(130, 393.8584650140001)
(131, 393.96368264099874)
(132, 393.99248405499384)
(133, 394.00484570200206)
(134, 394.00732214099844)
(135, 394.20639986999595)
(136, 394.215778728998)
(137, 394.2363156939973)
(138, 394.29124046200013)
(139, 394.6684711140042)
(140, 394.7566548540053)
(141, 394.9726063209964)
(142, 395.1774393550004)
(143, 395.8625529810088)
(144, 396.0756383150001)
(145, 396.07707413799653)
(146, 396.37554308400286)
(147, 396.43875814700004)
(148, 401.7595750709952)
(149, 401.7933246809989)
(150, 405.1151171640013)
(151, 405.78472404400236)
(152, 407.0380983609939)
(153, 417.9779747949942)
(154, 423.32946534499933)
(155, 427.6472275790002)
(156, 432.5555502539937)
(157, 432.6161534000021)
(158, 571.0390769169971)
(159, 572.4900918850035)
};
\addlegendentry{Symbiotic};
\addplot+[] coordinates {
(0, 2.9183095299522392)
(1, 2.9516937480075285)
(2, 2.989772849000474)
(3, 3.0035863160010194)
(4, 3.0081610749998617)
(5, 3.0137610799974937)
(6, 3.014775636000195)
(7, 3.015669699998398)
(8, 3.022917395002878)
(9, 3.022981721999713)
(10, 3.0233263580012135)
(11, 3.0255567390004217)
(12, 3.0287567719933577)
(13, 3.045037728999887)
(14, 3.045287644000382)
(15, 3.045568992999506)
(16, 3.050403372999881)
(17, 3.0507860529996833)
(18, 3.0538647639987175)
(19, 3.0558557769982144)
(20, 3.0674202959999093)
(21, 3.06759294999938)
(22, 3.0720833529994707)
(23, 3.0749385960007203)
(24, 3.079575998999644)
(25, 3.080073239000285)
(26, 3.0850625590001073)
(27, 3.08632851099901)
(28, 3.092536341999221)
(29, 3.092638808000629)
(30, 3.0928209159997095)
(31, 3.0955485019985645)
(32, 3.0974108990003515)
(33, 3.103573679999954)
(34, 3.1062128999983543)
(35, 3.106377741999495)
(36, 3.111160487998859)
(37, 3.112989851997554)
(38, 3.1171085409987427)
(39, 3.1173709069989854)
(40, 3.1198556419999477)
(41, 3.126580059000048)
(42, 3.1396518250003282)
(43, 3.148816413999157)
(44, 3.149720809000428)
(45, 3.1517914830001246)
(46, 3.1558938449998095)
(47, 3.1593931020006494)
(48, 3.162779092999699)
(49, 3.1637134939992393)
(50, 3.1660608339989267)
(51, 3.168850828998984)
(52, 3.1773992420003196)
(53, 3.178138958999625)
(54, 3.1795958369998516)
(55, 3.180435596999814)
(56, 3.1809294370000316)
(57, 3.181970061999891)
(58, 3.184047015998658)
(59, 3.1845989409994218)
(60, 3.1853995809979097)
(61, 3.1864863490009157)
(62, 3.1874010729989095)
(63, 3.196690756998578)
(64, 3.2101741579999725)
(65, 3.2158302960015135)
(66, 3.2173031019992777)
(67, 3.2176496639986)
(68, 3.217768613998487)
(69, 3.2180959330016776)
(70, 3.219418050999593)
(71, 3.2201157759991474)
(72, 3.221414160000222)
(73, 3.2216347789999418)
(74, 3.2266376959996705)
(75, 3.2272645510020084)
(76, 3.231138754999847)
(77, 3.2315883209994354)
(78, 3.2322592359996634)
(79, 3.2327084300000024)
(80, 3.233506382000087)
(81, 3.235711819000244)
(82, 3.2364404830004787)
(83, 3.239597760999459)
(84, 3.2405329479997818)
(85, 3.2422711889998936)
(86, 3.24534068200046)
(87, 3.2457077390008635)
(88, 3.2465876159994878)
(89, 3.24917096399804)
(90, 3.251033685999573)
(91, 3.2510968759997922)
(92, 3.2571078649998526)
(93, 3.262732874998619)
(94, 3.2653347739997116)
(95, 3.266986819000522)
(96, 3.2709902410001632)
(97, 3.272266167001362)
(98, 3.275422275000892)
(99, 3.2756184019999637)
(100, 3.2781244939978933)
(101, 3.293812770000045)
(102, 3.295937317001517)
(103, 3.2995457320012065)
(104, 3.302080184002989)
(105, 3.3071599730001253)
(106, 3.312738365999394)
(107, 3.3170613669972226)
(108, 3.3249974099999235)
(109, 3.326897568998902)
(110, 3.3281080320011824)
(111, 3.329603322999901)
(112, 3.3413852179983223)
(113, 3.3415206120007497)
(114, 3.345884547001333)
(115, 3.3473544209991815)
(116, 3.357565168998917)
(117, 3.362692718001199)
(118, 3.3679398089989263)
(119, 3.368929214000673)
(120, 3.376127441999415)
(121, 3.383821966999676)
(122, 3.3989387770016037)
(123, 3.4031906360014545)
(124, 3.4188921939994543)
(125, 3.4268662489994313)
(126, 3.4436596399991686)
(127, 3.464409906000583)
(128, 3.486260323999886)
(129, 3.5633690200011188)
(130, 3.577168867999717)
(131, 3.5802854650009976)
(132, 3.8623364929999298)
(133, 3.8651743730001726)
(134, 3.9231428719995165)
(135, 3.9637599180000507)
(136, 4.060714550999819)
(137, 4.091391654999825)
(138, 4.1130660670005454)
(139, 4.117723049999768)
(140, 4.1235292889996344)
(141, 4.129331299000114)
(142, 4.1599244749995705)
(143, 4.16173566299949)
(144, 4.173324563998904)
(145, 4.180919096999787)
(146, 4.181444531001034)
(147, 4.191818432000218)
(148, 4.207587780998438)
(149, 4.21045394600003)
(150, 4.225185859000703)
(151, 4.2439953840003)
(152, 4.24682359200051)
(153, 4.279460389999258)
(154, 4.285543910000342)
(155, 4.318315680999149)
(156, 4.403274443000555)
(157, 4.406252312999641)
(158, 4.586835304000488)
(159, 4.619177773998672)
(160, 4.685777013000916)
(161, 4.709680721000041)
(162, 4.7478811989999485)
(163, 4.752245204001156)
(164, 4.765705695001088)
(165, 4.830140138001298)
(166, 4.935556110000107)
(167, 4.972202597999058)
(168, 5.0334981759988295)
(169, 5.061228345000018)
(170, 5.129466690000299)
(171, 5.183059695999873)
(172, 5.238049554000099)
(173, 5.31064220799999)
(174, 5.733801013997436)
(175, 6.1119328910026525)
(176, 6.127887467999244)
(177, 6.1582679160001135)
(178, 6.179684261000148)
(179, 6.204066730999784)
(180, 6.445771839000372)
(181, 7.116033714999503)
(182, 7.517123837002146)
(183, 7.580775094000273)
(184, 9.73951088800095)
(185, 9.842695703002391)
(186, 10.260585326999717)
(187, 11.385877157001232)
(188, 11.508763743999225)
(189, 19.153048230000422)
(190, 19.792506145000516)
(191, 28.331578924000496)
(192, 35.02799677599978)
(193, 39.45760324100047)
(194, 97.13610598599917)
(195, 108.02009412200096)
(196, 119.36328144599975)
(197, 138.0265162570031)
(198, 155.1300569340001)
(199, 155.42697380000027)
(200, 157.11782145599955)
(201, 203.372423797)
(202, 203.42366265200053)
(203, 203.47548266000013)
(204, 211.5676722370008)
(205, 252.15604196500135)
(206, 355.4204406890003)
(207, 355.63702718199966)
};
\addlegendentry{VeriAbs};
\end{axis}
\end{tikzpicture}

%% file: related-work.tex
\section{Related Work} \label{sec:relwork}

{GPS builds upon three different lines of work: directed testing, static analysis, and software model checking---we discuss each of the three below.}

\bfpara{Directed test generation and execution.} \emph{Directed testing} sometimes refers to a technique that aims to maximize code coverage (e.g., DART \cite{DART}); in contrast, GPS directs testing toward a particular error location.
More in line with the goals of GPS is \emph{directed greybox fuzzing} \cite{AFLGo,Hawkeye,Beacon,MC2}, which
aims to generate test inputs that reach a designated error location.
Perhaps most similar to our technique is Beacon \cite{Beacon}, which uses a backwards static analysis (propagating information from the error location back to the entry of the program) to over-approximate the set of states that \textit{may} reach the error.  GPS uses a dualized variant of Tarjan's algorithm to compute path-to-error summaries for each location to achieve a similar result. \emph{Directed symbolic execution} \cite{directedSymbExec} uses heuristics for guiding a symbolic-execution algorithm toward a particular error location, instead of maximizing coverage. The main difference between this work and GPS is that GPS employs summarization and program instrumentation to guide the search toward particular paths to error, whereas directed symbolic execution leverages a distance metric over the structure of an interprocedural control-flow graph. 
The salient difference between GPS and
the aforementioned
techniques is that GPS uses a synergistic interaction of static-analysis and software-model-checking techniques that make it capable of \textit{verifying} safety properties while achieving a completeness result for property refutation, even for programs with an unbounded state space.

Saxena et al.\ \cite{ISSTA:SPMS09} identified programs like EX-1 
(\Cref{fig:ex1}) as being challenging
for concolic-execution engines.
EX-1 is an L\&K problem in which there is no chain of flow dependences between variable $\N$ and the branch-condition ``$\minVar < 1000$,'' but only \emph{implicit flow} \cite{cacm:DD77} between them.
In compiler terminology, $\N$ influences $\minVar$ only via the \emph{control dependence} \cite{kn:FOW87} from $\I < \N$ to $\minVar\texttt{++}$.

{
\bfpara{Invariant checking and synthesis.} The invariant-finding features of GPS consist of two components: (1) invariant synthesis from static-analysis-generated summaries and Craig interpolation, and (2) an {\sc Impact}-based \cite{cav:mcmillan06} invariant-checking procedure to check whether the synthesized invariants are inductive. We discuss prior work related to each component below.

\begin{itemize}
    \item \emph{Static analysis and invariant generation.}
A recent line of work on algebraic program analysis \cite{cav:apa} demonstrates that it is possible to design summary-based static analyses that are competitive with state-of-the-art software model checkers in terms of verification ability.  The design of GPS was inspired by the question of whether this precision could be harnessed for property \textit{refutation} as well.  In principle, GPS can be instantiated
with any backward static-analysis algorithm,
and the dualization of Tarjan's method presented in Section~\ref{Se:SingleTargetSummaries} can be seen as a way of implementing a backward analysis using algebraic program analysis.   From this perspective, the crucial feature for GPS's success is the fact that algebraic program analyses can be very precise, whereas historically backward analyses are imprecise (relative to forward analyses).
    \item \emph{Invariant checking.}
    GPS uses the covering procedure in {\sc Impact} to test whether ART labels constitute an inductive invariant. The induction principle underlying the covering procedure is as follows: In the ART, we add a covering edge from path $\pi\tau$ to its ancestor $\pi$ when the label at $\pi$ is an inductive invariant for $\tau$, and $\tau$ is an $m$-step unfolding of a loop (for some $m$). Thus, the covering procedure (in GPS and {\sc Impact}) tests whether an ART label is an invariant that is inductive with respect to an $m$-step unfolding of some loop.
    
    \hspace{1.5ex}
    A different, but related, approach is to test whether invariants are $k$-inductive using the principle of $k$-induction \cite{FMCAD:SSS2002}; however, the $k$-induction principle is incomparable to the induction principle behind {\sc Impact} and GPS. \IfCameraReady{See the extended version of our paper for a more thorough discussion of this topic.}{We illustrate the incomparability through the following examples:

\begin{example}[Inductive w.r.t. a 2-step unfolding but not 2-inductive]
 \label{ex:impact_induction_better}
 The property $2\cdot y = x$ is not 2-inductive for the loop inside the program 
\[
x = 0; y=0; \textbf{while}(*) \{ x\text{++}; y \text{+=} x\text{ mod }2; \}
\]
but is inductive for a 2-step unfolding of the loop. In particular, the verification condition for 2-induction requires that:
\[
(x = 0 \wedge y = 0 \wedge x’ = x + 1 \wedge y’ = x’ \mod 2 ) \implies 2\cdot y’ = x’
\]
which fails to hold
\twrchanged{because}
the property $2\cdot y=x$ 
\twrchanged{
only holds
}
on even iterations
\twrchanged{
of the loop body.
}
\end{example}

\begin{example}[2-Inductive but not inductive w.r.t. a 2-step unfolding]
\label{ex:k_induction_better}
As another example, the property $a > 0$ is 2-inductive for the loop inside the program
\[
a = 1, b = 2; tmp = *; \textbf{while}(*) \{ tmp = a; a = b; b = tmp + b; \} 
\]
but not inductive for a 2-step unfolding of the loop.
In particular, the verification condition for a 2-step unfolding of the loop requires
\twrchanged{the following property to hold:}
\[
\left( 
\begin{array}{cc}
    a > 0 &\wedge \\    
    (tmp' = a \wedge a' = b \wedge b' = tmp' + b) &\wedge \\
    (tmp'' = a' \wedge a'' = b' \wedge b'' = tmp'' + b')
\end{array}
 \right) \implies (a'' > 0)
\]
\twrchanged{
However, this property fails to hold
}
because the information that $b>0$ holds, which is needed to prove $a>0$ w.r.t. a 2-step unfolding of the loop,
\twrchanged{is not available in an inductive argument about the loop body.}
In contrast, in 2-induction, the verification condition becomes 
\[
\left( 
\begin{array}{cc}
    a > 0 &\wedge \\
    (tmp' = a \wedge a' = b \wedge b' = tmp' + b) &\wedge \\ 
    {\color{blue}a' > 0} & \wedge \\ 
    (tmp'' = a' \wedge a'' = b’ \wedge b'' = tmp'' + b')
\end{array}
 \right) \implies (a'' > 0)
\]
where the intermediate assumption $\color{blue}a'>0$ is added in the premise, in turn implying that $b>0$.
\end{example}

\hspace*{1.5ex}
To demystify this incomparability between $k$-induction and {\sc Impact}-style induction, 
\twrchanged{
consider their respective definitions,
}
presented in \Cref{tab:induction_defs}. In this table, each induction principle is defined in terms of an \emph{initialization} rule---checked at the initial state of a loop---and a \emph{consecution} rule---which requires that an invariant is inductively preserved across loop iterations.

\begin{table}[]
\small 
    \centering
    \begin{tabular}{c|c|c}
         & $k$-Induction & {\sc Impact}-style Induction \\
         \hline 
       Initialization  & $\left\{\begin{array}{c}P(X),\\ P(X)\wedge T(X,X') \wedge P(X'),\\ ..., \\P(X) \wedge T^k(X,X') \wedge P(X')\end{array}\right\}$ & $P(X)$ \\ 
       Consecution & $\left(\begin{array}{c}P(X) \wedge T(X,X') \wedge \\ P(X') \wedge T(X', X'')\wedge \\ ... \wedge \\ P(X^{(k-1)}) \wedge T(X^{(k-1)}, X^{(k)}) \end{array}\right) \implies P(X^{(k)})$ & 
       $P(X) \wedge T^k(X, X') \implies P(X')$
    \end{tabular}
    \caption{A comparison between $k$-induction and {\sc Impact}-style induction. Let $X$ denote the program vocabulary, $P(X)$ denote a property, and let $T(X, X')$ be the transition formula representing a loop body. Let $X^{(k)}$ denote $X$ primed $k$ times, and let $T^k$ denote the $k$-fold composition of transition formula $T$.}
    \label{tab:induction_defs}
\end{table}

\hspace*{1.5ex}
It is evident that the consecution rule for $k$-induction is more powerful than the consecution rule for {\sc Impact}-style induction.
As shown in the second row of \Cref{tab:induction_defs}, the $k$-induction consecution condition assumes more premises compared to {\sc Impact}-style induction---specifically $P(X'), ..., P(X^{k-1})$. This difference is the root cause behind \Cref{ex:k_induction_better}.
By contrast, in the initialization rule for $k$-induction, we need to assert that $P$ holds across each of the first $k$ iterations of the loop.
This requirement is much stronger than the initialization rule of {\sc Impact}-style induction, which just requires $P$ to hold in the pre-state of the loop. This difference is the root cause behind \Cref{ex:impact_induction_better}. 

\hspace*{1.5ex}}
There is also work that leverages the $k$-induction principle as the underlying invariant checker in an invariant-synthesis engine \cite{FMCAD:JD2016, FMCAD:GI2016, CAV:VKVGG2019}; incorporating the $k$-induction principle into the GPS framework is an interesting avenue for future work.

\end{itemize}

}

\input{related-works-feature-table}

\bfpara{Software model checking.}
There has been much work on software model checking; we focus on the most relevant works below, and give a comparison between GPS and related algorithms in \Cref{tab:feature_table}.

GPS builds on \emph{lazy abstraction with interpolants} ({\sc Impact}) \cite{cav:mcmillan06}, and similarly uses an abstract reachability tree to store explored states, and interpolation to perform refinement.  
There are two key innovations that GPS provides.
First, {\sc Impact} expands the ART by a depth-first search, which is in fact \emph{not} refutation-complete---as the algorithm can be forced to always extend the ART in an unbalanced manner (e.g., by always extending the \emph{leftmost} leaf). In contrast, GPS builds the tree by simulating execution from summary-directed tests and formally guarantees refutation-completeness, as the frontier paths in GPS are explored in a breadth-first manner.
Second, while {\sc Impact} proves safety one source-to-target path at a time, GPS uses path-to-target summaries to prove safety of regular sets of such paths. 

Perhaps most related to GPS's idea of combining test execution with interpolation-based refinement is the line of work on 
\textsc{Synergy} \cite{synergy}
and its successors 
\textsc{Dash} \cite{Dash}, 
\textsc{Smash} \cite{Smash}, 
and \textsc{McVeTo} \cite{CAV:TLLBDEAR10}. 
\textsc{Synergy}'s test-generation strategy might be thought of as an intermediate point between generating tests to maximize coverage and to reach an error location.  It finds a path to error through a finite-state abstraction of the program, and then uses the frontier edge of that path (which crosses from the part of the state space that has been explored to the part that has not) to generate a new test.  Using 
path summaries, GPS is able take the entire error path into account for test generation.  For instance,
in EX-1, GPS discovers the bug using a single test, whereas the \textsc{Synergy} family of algorithms requires 1000 tests, each test
unrolling the loop one more step.
\emph{Lazy Annotation} \cite{lazyAnnot} features an instrumentation step similar to GPS’s instrumentation-by-gas procedure to aid the algorithm’s convergence in the face of unbounded loops; however, lazy annotation does not explicitly guarantee refutation-completeness, nor does it investigate the impact of such an instrumentation procedure on algorithm performance.

\bfpara{Refutation-completeness.} 
 GPS benefits from the gas-instrumentation procedure (\Cref{ssec:gas}) for being refutation-complete; however, such a procedure only works due to how GPS performs refinement in \emph{breadth-first order} but eagerly explores generated tests in \emph{depth-first order}. This unique{, two-layered search strategy} 
 allows GPS to {efficiently} explore long paths-to-errors, while still retaining refutation-completeness due to the breadth-first refinement order. In contrast, {\sc Impact}-like algorithms \cite{cav:mcmillan06, CAV:CG12, fmcad:bkw10, ufo, Whale}{---many of which are described as algorithms implementing depth-first refinement---suffer from an inherent tension between achieving refutation-completeness and the issue of {efficiently} handling deep counterexamples. In particular, these algorithms} can either {be modified to} explore program states (and refine) in {breadth-first} order, and thus achieve refutation-completeness at the expense of performance, or perform both exploration and refinement in depth-first order, thus sacrificing refutation-completeness. {When restricted to intraprocedural programs, the {\sc Whale} algorithm of Albarghouthi et al.\ \cite{Whale} 
implements {\sc Impact} with breadth-first refinement and is thus refutation-complete, but (due to the tension described above) does not address the issue of handling deep counterexamples.}

\bfpara{Reasoning about sets of paths.} This paper introduces \textit{dead-end interpolation} as a means for generating candidate invariants that are high-quality in the sense that they are sufficient for proving that a (possibly infinite) set of paths cannot reach the error location.  Large-block encoding \cite{fmcad:bcgks09} shares a similar goal of enabling software model checkers to reason about sets of paths; in contrast to our work these sets are always finite. {In a similar spirit, \textsc{Whale} \cite{Whale} and \textsc{Ufo} \cite{ufo} extend \textsc{Impact} to operate over \emph{abstract reachability graphs} (ARGs), and perform refinement along subgraphs of the ARG. This approach also allows these algorithms to refine multiple paths at a time, but again each subgraph only represents a finite number of paths.}  Another related technique is acceleration \cite{ATVA:HIKKR2012}, which can be used to reason about infinitely many paths, {and} which is similar in spirit to summarization.  They differ in that acceleration is \textit{precise}, and applies only to loops of a particular form, whereas summarization is over-approximate and broadly applicable.

%% file: related-works-feature-table.tex
\begin{table}[]
\small
\begin{tabular}{c|c|c|c}
\toprule
& Only explores feasible paths & Invariant checking/synthesis & Refutation complete\\
\midrule
{\sc GPS (Our work)} & \checkmark & Set of paths & \checkmark \\
{\sc Dash} \cite{Dash} & \xmark & Set of paths & \xmark \\
{\sc Impact} \cite{cav:mcmillan06} & \xmark & Single path & \xmark \\
{\sc Dart} \cite{DART} & \checkmark & N/A & \xmark \\
\bottomrule
\end{tabular}
    \caption{A comparison between GPS and related algorithms. The ``Invariant Synthesis'' column refers to whether the algorithm considers a single path or multiple paths when proposing candidate invariants.}
    \label{tab:feature_table}
    \vspace*{-0.2in}
\end{table}

%% file: conclusion.tex
\section{Conclusion}

We presented a novel algorithm, GPS, for checking control-state reachability of intraprocedural programs. 
Our experiments confirmed that GPS is a particularly performant model-checking algorithm that can handle both challenging verification tasks and challenging refutation tasks. GPS opens up several exciting avenues for future work. First, the way GPS employs static analysis-generated summaries to help both test generation and refinement means that \emph{any improvement} in the precision of the (summary-generating) static analysis leads to an improvement in both the quality of the a directed test and an interpolant sequence generated by GPS. Thus, any future work on more precise path summary-generation methods would immediately benefit GPS. It is also interesting to consider using GPS to answer intra-procedural reachability queries in an inter-procedural model checking algorithm such as Spacer~\cite{spacer}. Finally, it would be interesting to consider whether GPS can be extended to handle richer classes of programs, such as those manipulating arrays, pointers, or heaps. 

%% file: appendix.tex
\clearpage
\section{Additional Statistics about Benchmarks and Evaluation}
\label{appendix:info_about_l_and_k_suite}

\begin{figure}
    \centering

    \begin{subfigure}{0.45\textwidth}
    \includegraphics[width=0.8\linewidth]{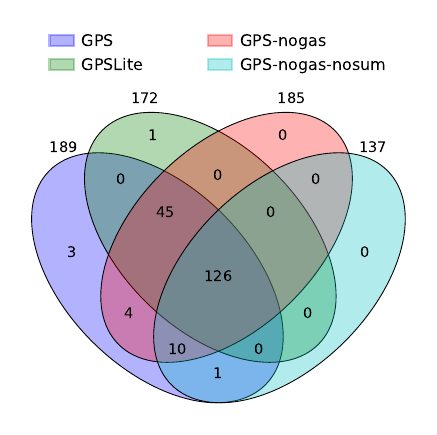}
    \caption{Venn diagram of safe benchmark results, across all three benchmark suites, for RQ \#1 (200 benchmarks total)}
        \label{fig:diagram-ablation-safe}
    \end{subfigure}
    \hfill 
    \begin{subfigure}{0.45\textwidth}
    \includegraphics[width=0.8\linewidth]{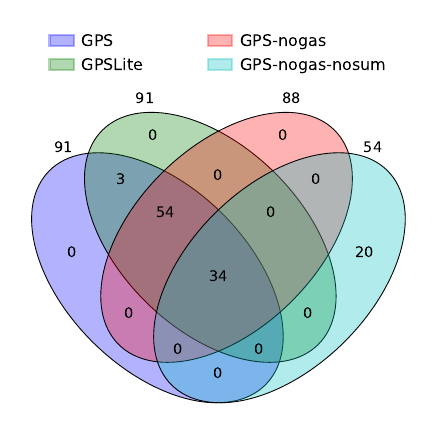}
    \caption{Venn diagram of unsafe benchmark results, across all three benchmark suites, for RQ \#1 (96 benchmarks total)}
        \label{fig:diagram-ablation-unsafe}
    \end{subfigure}
    \\

    \begin{subfigure}{\textwidth}\centering
    \resizebox{0.7\textwidth}{!}{\centering
        \input{figures/loops-ablation/new-scatter.cameraready}
    }
    \caption{Scatterplot of results for RQ \#1.}
    \label{fig:rq1-scatter-compare}
    \end{subfigure}    
    \caption{Detailed benchmark results for RQ \#1. 
    }
    \label{fig:rq1-benchmark-results}
\end{figure}

\begin{figure}[t]
    \centering
    \includegraphics[width=\linewidth]{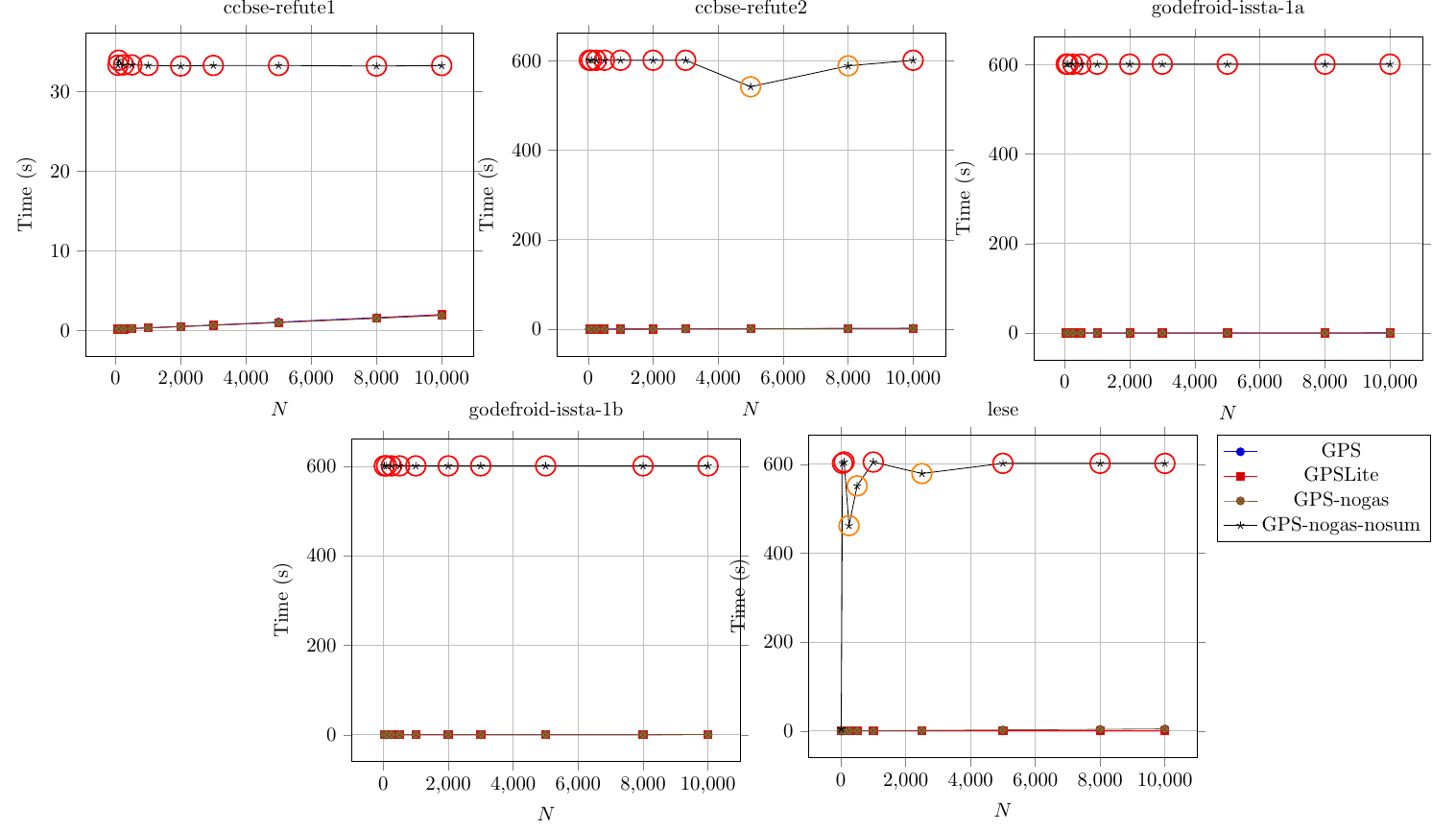}
    \caption{Parameterized Lock\& Key benchmarks for RQ \#1. Timeout threshold is 600s. Points shown in red circle denote a timeout, points shown in orange circle denote tool reporting out-of-memory. Note that GPS results are shown in red line near the bottom of each figure. }
    \label{fig:parametric_benchmarks_for_rq1}
\end{figure}

\begin{figure}[t]
    \centering
    \includegraphics[width=\linewidth]{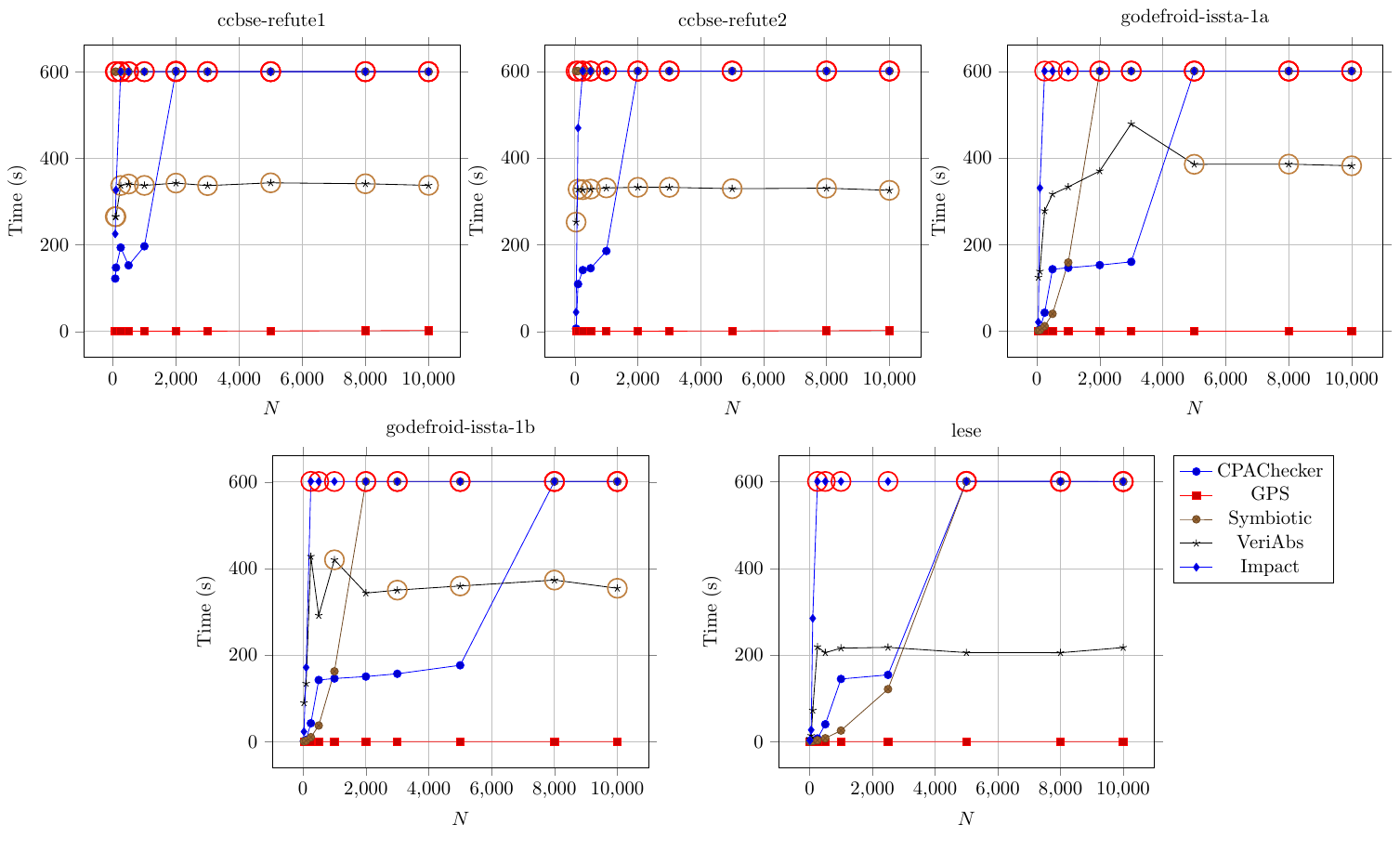}
    \caption{Parameterized Lock\& Key benchmarks for RQs \#2 and \#3. Timeout threshold is 600s. Points shown in red circle denote a timeout, points shown in brown circle denote tool reporting unknown result. Note that GPS results are shown in red line near the bottom of each figure. }
    \label{fig:parametric_benchmarks_for_rq2_3}
\end{figure}

Venn diagrams and scatterplots about the ablations done for RQ \#1 may be found in ~\Cref{fig:rq1-benchmark-results}. 
Detailed line plots for the parametrized L\&K benchmarks for RQs \#1 through \#3 may be found in~\Cref{fig:parametric_benchmarks_for_rq1} and~\Cref{fig:parametric_benchmarks_for_rq2_3}.


\section{Proofs}
\label{appendix:proofs}
{For a weighted graph $G$,} let $\mathit{AllPaths}_G = \bigcup_{u \in G}\textit{Paths}(s, u)$ denote all paths originating from the source vertex of $G$. For any path $\tau \in \mathit{AllPaths}_G$ and an ART $\Lambda$, let $\TreePrefixTwo{\Lambda}{\tau}$ denote the longest path in $\Lambda$ that is a prefix of $\tau$. Let $\TreeSuffixTwo{\Lambda}{\tau}$ be the path $\tau'$ such that $\tau = \TreePrefixTwo{\Lambda}{\tau}\tau'$. Note that $\tau'$ is a path from $\pdst{\TreePrefixTwo{\Lambda}{\tau}}$ that is not contained in $\Lambda$. 
A \textbf{cycle} is a path $\pi$ such that $\pdst{\pi} = \psrc{\pi}$. A path $\pi$ \textbf{contains a cycle} if there is a sub-path $\pi'$ of $\pi$ that is a cycle. 
We say that a path $\tau$ in a weighted graph $G$ is a \textbf{contraction} of another path $\tau'$, if $\tau' = \tau_0 \tau_1 \tau_2$ for some $\tau_0,\tau_1,\tau_2$ with $\tau_2$ being a cycle and $\tau=\tau_0\tau_2$ (i.e., $\tau$ is obtained from $\tau'$ by removing a cycle). $\tau_1$ may be the empty word, so every path is a contraction of itself.
A \textbf{proper contraction} of $\pi$ is a contraction of $\pi$ that isn't itself.
We define the \textbf{contraction order} $\tau \preceq \tau'$ to mean when $\tau$ is a contraction of $\tau'$, and note that $\preceq$ is a well-founded partial order, whose minimal elements are simple paths (i.e., paths that do not visit the same vertex twice).


\label{sec:Proofs}

\subsection{Proof of \Cref{thm:art_cfg_safe}}

First, we state three technical lemmas.

\begin{lemma}[Trichotomy of Complete ART] \label{lem:general_trichotomy}
Given a weighted graph $G=(V, E, w, s, t)$, a complete and well-labeled ART $\Lambda$ of $G$, and any path $\tau$ of $G$, either
\begin{enumerate}
  \item
    $\TreePrefixTwo{\Lambda}{\tau}= \tau$, and $\tau$ is an internal node 
    of
    $\Lambda$;
  \item
    $\TreePrefixTwo{\Lambda}{\tau}$ is pruned in $\Lambda$;
    or
  \item
    $\TreePrefixTwo{\Lambda}{\tau}$ is covered in $\Lambda$ by some other $\pi \in \TreeNodes{\Lambda}$ that is a tree ancestor of $\tau$ in $\Lambda$.
\end{enumerate}
\end{lemma}
\begin{proof}
    Assume, for the sake of producing a contradiction,
    that $\TreePrefixTwo{\Lambda}{\tau}$ does not satisfy any 
    of the situations
    above.
    Then, $\TreePrefixTwo{\Lambda}{\tau}$
    must be a leaf of $\Lambda$ that is neither covered nor pruned. This contradicts the fact that $\Lambda$ is complete. $\qed$
\end{proof}

\begin{lemma}
    \label{lem:pruned_implies_unsat}
    Given a weighted graph $G = (V, E, w, s, t)$ and well-labeled ART $\Lambda$ for $G$, for any $s$-$t$ path $\tau$ in $G$, if $\TreePrefixTwo{\Lambda}{\tau}$ is pruned in $\Lambda$, then $\TreeLbl{\Lambda}{\TreePrefixTwo{\Lambda}{\tau}} \wedge w(\TreeSuffixTwo{\Lambda}{\tau})$ is UNSAT.
\end{lemma}
\begin{proof}
Let $\tau_1 = \TreePrefixTwo{\Lambda}{\tau}$ and $\tau_2 = \TreeSuffixTwo{\Lambda}{\tau}$. Because  $\tau_1$ is pruned, we have that $\TreeLbl{\Lambda}{\tau_1} \wedge \mathsf{Sum}(G, \pdst{\tau_1}, t)$ is UNSAT.  By well-labeledness
of $\Lambda$, we have $\spost{\top}{w(\tau_1)} \models \TreeLbl{\Lambda}{\tau_1}$. Because  $\textsf{Sum}(G,\pdst{\tau_1},t)$ over-approximates all paths from $\pdst{\tau_1}$ to $t$, we have $w(\tau_2) \models \textsf{Sum}(G,\pdst{\tau_1},t)$ and so $\TreeLbl{\Lambda}{\tau_1} \land w(\tau_2)$ is UNSAT. $\qed$
\end{proof}
\begin{lemma}[Contraction]
    \label{lem:error_contraction}
  Given a weighted graph $G$ with a minimal feasible $s$-$t$ path, and assuming that a well-labeled, complete ART \rfchanged{$\Lambda = (T, L, \mathsf{Cov})$} exists for $G$, for any contraction $\tau$ of the minimal feasible $s$-$t$ path in $G$, it holds that 
  $\TreeLbl{\Lambda}{\TreePrefixTwo{\Lambda}{\tau}} \wedge w(\TreeSuffixTwo{\Lambda}{\tau})$ is UNSAT.
\end{lemma}

\begin{figure}[bt!]
    \centering
    \includegraphics[width=0.6\linewidth]{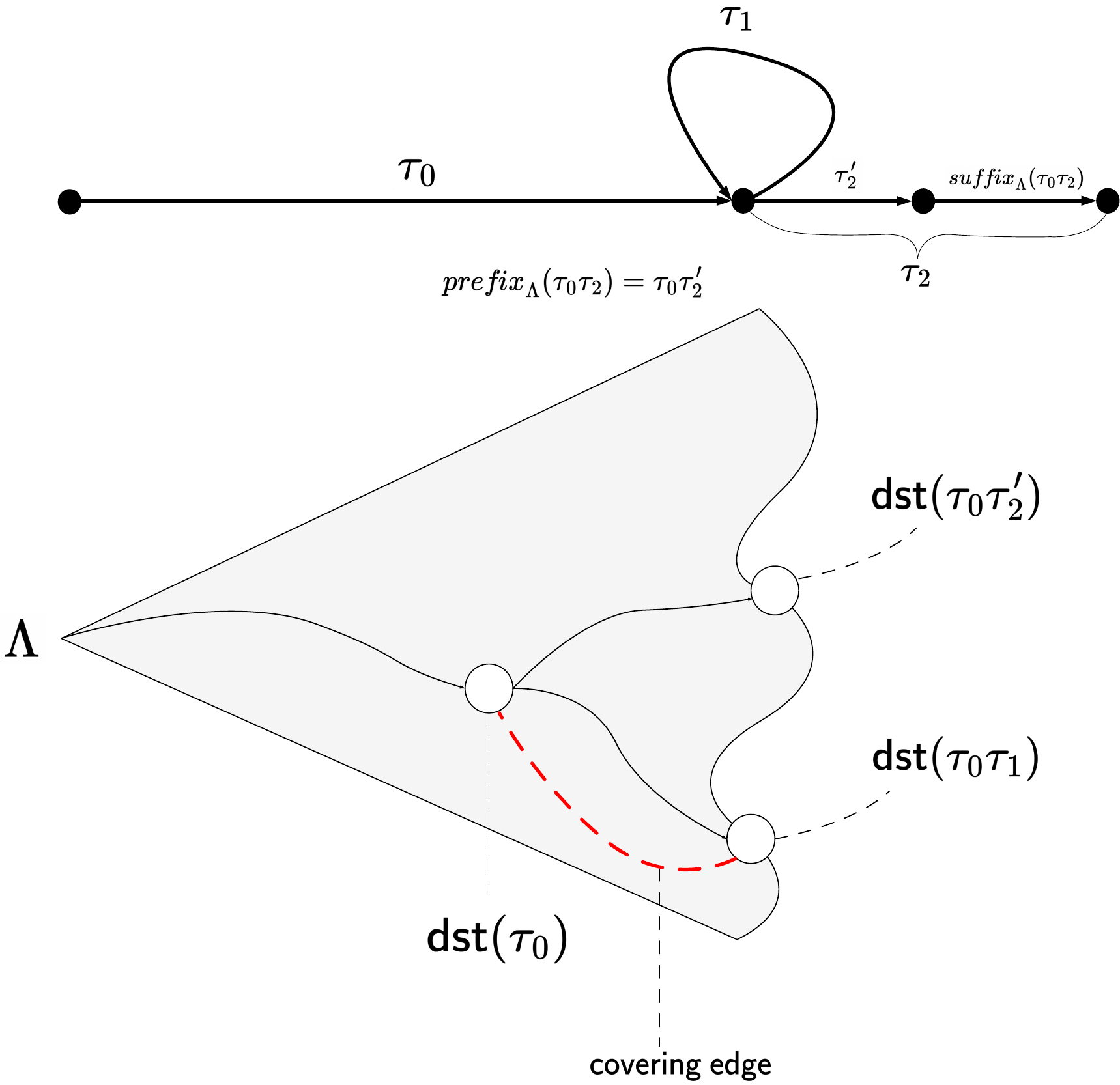}
    \caption{Illustration of applying case (2) of \Cref{lem:general_trichotomy} during the induction step in the proof of \Cref{lem:error_contraction}.}
    \label{fig:contraction_lem_illustration}
\end{figure}

\begin{proof}
 By strong induction on contraction order. 

 \bfpara{Base case.} Suppose that $\tau$ is a contraction of $\pi$ and minimal with respect to contraction order. Only case (2) of \Cref{lem:general_trichotomy} applies, because
 \begin{itemize}
     \item For case (1), we would have $\TreePrefixTwo{\Lambda}{\tau} = \tau$, and $\tau$ is internal node for $\Lambda$, which is impossible, because $\pdst{\tau}$ is the
     terminal
     vertex of $G$, which has no successors.
     \item For case (3), we would have $\TreePrefixTwo{\Lambda}{\tau}$ is covered in $\Lambda$ by some prefix path $\tau'$ of $\TreePrefixTwo{\Lambda}{\tau}$, but $\TreePrefixTwo{\Lambda}{\tau}$ is a simple path, so for all
     ancestors
     $\tau' \in \TreeNodes{\Lambda}$, $\pdst{\tau'}\neq \pdst{\TreePrefixTwo{\Lambda}{\tau}}$, and thus $\TreePrefixTwo{\Lambda}{\tau}$ cannot be covered.
 \end{itemize}

Therefore, only case (2) of \Cref{lem:general_trichotomy} is possible and the proof goal holds as a corollary of \Cref{lem:pruned_implies_unsat}.

\bfpara{Induction hypothesis.} Let $\tau$ be a contraction of $\pi$, and suppose that the above lemma holds for all proper contractions of $\tau$.

\bfpara{Induction step.} Here, we show that case (1) in \Cref{lem:general_trichotomy} is not possible, whereas both cases (2) and (3) of \Cref{lem:general_trichotomy} 
imply
the above proof goal:
\begin{itemize}
    \item For case (1), we would have $\TreePrefixTwo{\Lambda}{\tau} = \tau$, and $\tau$ is an internal node of $\Lambda$, which is impossible, because $\pdst{\tau}$ maps to the
     terminal
    vertex of $G$ which has no successors.
    \item For case (2), the proof goal holds as a corollary of \Cref{lem:pruned_implies_unsat}.
    \item For case (3), WLOG, write
    $\tau = \tau_0\tau_1\tau_2$,
    where $\TreePrefixTwo{\Lambda}{\tau} = \tau_0\tau_1$,
    $\TreeSuffixTwo{\Lambda}{\tau} = \tau_2$, and 
    $\tau_0$ covers $\tau_0\tau_1$ in $\Lambda$.
    Consider $\TreePrefixTwo{\Lambda}{\tau_0\tau_2}$.
    By the induction hypothesis, $\TreeLbl{\Lambda}{\TreePrefixTwo{\Lambda}{\tau_0\tau_2}} \wedge \TreeSuffixTwo{\Lambda}{\tau_0\tau_2}$ is UNSAT. 
    Let $\tau_2'$ be such that $\TreePrefixTwo{\Lambda}{\tau_0\tau_2} = \tau_0\tau_2'$.
    We have:
    \begin{enumerate}[label=(\alph*)]
        \item $\TreeLbl{\Lambda}{\tau_0\tau_1} \vDash \TreeLbl{\Lambda}{\tau_0}$ by
        the well-coveredness
        condition of $\Lambda_i$;
        \item $\spost{\TreeLbl{\Lambda}{\tau_0}}{w(\tau_2')} \vDash \TreeLbl{\Lambda_i}{\tau_0\tau_2'}$ by the consecution condition of $\Lambda$;
        \item $\TreeLbl{\Lambda} {\tau_0\tau_2'} \wedge \TreeSuffixTwo{\Lambda}{\tau_0\tau_2}$ is UNSAT, by the induction hypothesis; 
    \end{enumerate}    
    Altogether, (b) and (c) imply that $\TreeLbl{\Lambda}{\tau_0} \wedge w(\tau_2)$ is UNSAT, by the following reasoning:
    \begin{enumerate}
        \item Rewrite $\TreeLbl{\Lambda}{\tau_0} \wedge w(\tau_2)$ as $\TreeLbl{\Lambda}{\tau_0} \wedge (w(\tau_2') \circ w(\TreeSuffixTwo{\Lambda}{\tau_0\tau_2}))$
        \item Rewrite the above further, as $\spost{\TreeLbl{\Lambda}{\tau_0}}{w(\tau_2')} \wedge w(\TreeSuffixTwo{\Lambda}{\tau_0\tau_2})$
        \item Substitute in (b), getting the above implies $\TreeLbl{\Lambda}{\tau_0\tau_2'} \wedge w(\TreeSuffixTwo{\Lambda}{\tau_0\tau_2})$
        \item By (c), the above expression is UNSAT.
    \end{enumerate}
    By (a), $\TreeLbl{\Lambda_i}{\tau_0\tau_1} \wedge w(\tau_2)$ is UNSAT, so $\tau$ satisfies the proof goal above. Figure~\ref{fig:contraction_lem_illustration} illustrates the proof of this case.
\end{itemize}
 $\qed$
\end{proof}

We are now ready to prove \Cref{thm:art_cfg_safe}:

\bfpara{\Cref{thm:art_cfg_safe}.} Let $G = \tuple{V,E,w,s,t}$ be a weighted graph and let $\Lambda$ be a well-labeled ART for $G$.  If $\Lambda$ is complete and well-labeled, then $G$ is safe.

\bfpara{Proof.} 
Assume  that $G$ is not safe, but $\Lambda$ is a complete and well-labeled ART for $G$. By \Cref{lem:error_contraction}, we have a contradiction, because $\TreeLbl{\Lambda}{\TreePrefixTwo{\Lambda}{\tau}} \wedge w(\TreeSuffixTwo{\Lambda}{\tau})$ implies that $\tau$ is not a feasible $s$-$t$ path.
$\qed$

\subsection{Proof of Soundness (\Cref{thm:gps_sound})}

We first prove a series of Lemmas, which state that all sub-routines of GPS that modify an ART preserve well-labeledness of an ART.

\begin{lemma}[$\textsc{Explore}(-)$ Preserves Well-labeledness]
\label{lem:explore_well_labeled}
Given a weighted graph $G= (V,E,w,s,t)$, a well-labeled ART $\Lambda$ of $G$, a leaf node $v$ of $\Lambda$ and a model $M$ such that $M \vDash \TreeLbl{\Lambda}{v} \wedge \mathsf{Sum}(\TreeVtx{\Lambda}{v}, t)$, $\textsc{Explore}(G, \Lambda, v, M)$ (\Cref{fig:SummaryGuidedTesting}) produces an ART that is well-labeled.
\end{lemma}
\begin{proof}
At any step $k$ through the main loop of $\textsc{Explore}(-)$, the only modification to $\Lambda$ consists of path additions to $\TreeNodes{\Lambda}$; such additions preserve the consecution condition of labels in $\Lambda$ by labeling all newly added paths in $\Lambda$ with label $\top$. Hence, $\textsc{Explore}(G, \Lambda, v, M)$ preserves the well-labeledness of $\Lambda$. $\qed$
\end{proof}

Given an ART $\Lambda$, a node $\tau$ in $\Lambda$ corresponding to a weighted-graph path $\tau = \tau_1 \tau_2\tau_3$, where $\tau_3 = e_1e_2...e_k$ ($e_1,...,e_k$ are edges) such that $\tau \in \Lambda$, and a sequence $\sigma = [S_1, ..., S_k]$ 
of state formulas, we say that $\sigma$ is a \textbf{valid refinement of $(\tau_2, \tau_3)$} if (1) $\TreeLbl{\Lambda}{\tau_2} \vDash S_1$, and (2) $\spost{S_i}{F_i} \vDash S_{i+1}$ for $1 \leq i < k$ where $F_i = w(e_i)$. A length-$k$
sequence $\sigma$ of state formulas is a \textbf{valid refinement of $\pi \in \TreeNodes{\Lambda}$} if it is a valid refinement for some $(\pi', \pi)$ such that $\pi = \pi' e_1,...,e_k$.

\begin{lemma}[$\textsc{Refine}(-)$ Preserves Well-labeledness]
\label{lem:refine_well_labeld}
Given a well-labeled ART $\Lambda$, a weighted-graph path $\pi \in \TreeNodes{\Lambda}$, and a sequence of state formulas $\sigma = \phi_1,...,\phi_k$ such that $\sigma$ is a valid refinement for $\pi$, $\textsc{Refine}(\Lambda, \pi, \sigma)$ preserves the well-labeledness of $\Lambda$.
\end{lemma}
\begin{proof}
Write $\pi = \pi'e_1,...,e_k$ where $e_1,...,e_k$ are weighted-graph edges and $\pi'$ is a path that is a prefix of $\pi$. Let $l_i = \TreeLbl{\Lambda}{\pi'e_1,...,e_i}$ for $1 \leq i \leq k$. After the relabeling operations in the main loop of $\textsc{Refine}(-)$, each label from $\pi'e_1$ to  $\pi=\pi'e_1,...,e_k$ becomes $l_i \wedge \phi_i$.

The newly modified labels preserve the consecution condition because (1) the original labels $\{l_i\}$ preserve the consecution condition (by well-labeledness of input ART $\Lambda$), thus $\spost{l_{i-1}}{w_\Lambda(u_{i-1}, u_{i})} \vDash l_{i}$ for $0 < i \leq k$; (2) because $\sigma$ is a valid refinement for $v$ in $\Lambda$, we have $\spost{\phi_i}{w(e_i))} \vDash \phi_{i + 1}$ for $0 < i < k$. 

The newly modified labels do not destroy the well-covered condition in the input ART $\Lambda$ because any covering relations
that
the label modifications might violate are removed by the main loop in $\textsc{Refine}(-)$ (\Cref{fig:Explorerefine}). $\qed$
\end{proof}

\begin{lemma}[$\textsc{TryCover}(-)$ Preserves Well-labeledness]
\label{lem:try_cover_well_labeled}
Given a well-labeled ART $\Lambda$, the $\textsc{TryCover}(\Lambda, v)$ procedure preserves the well-labeledness of $\Lambda$.
\end{lemma}
\begin{proof}

Let $U = \{\tau \sqsubset \pi; \pdst{\tau} = \pdst{\pi}\}$ be the set of prefixes of $\pi$ that \emph{might} cover $\pi$. Notice that any 
modification
to $\Lambda$ in $\textsc{TryCover}(-)$ is guarded by a query to $\mathsf{Check}(\TreeLbl{\Lambda}{\tau}, \tau^{-1}\pi :: \neg \TreeLbl{\Lambda}{\tau})$ that yields UNSAT (\Cref{fig:Explorerefine}), for some $\tau \in U$. If this query is indeed UNSAT,
then we recover a sequence $\sigma$ of valid refinements for $(\tau, \pi)$ such that after refining
the labels from $\tau$ to $\pi$, $\TreeLbl{\Lambda}{\tau} \vDash \TreeLbl{\Lambda}{\pi}$. Because refinement preserves well-labeledness (\Cref{lem:refine_well_labeld}) and the labels at $(\tau, \pi)$ satisfy the
well-coveredness
condition after refinement, the modifications to $\Lambda$ preserve well-labeledness. $\qed$
\end{proof}

\begin{lemma}[Main Loop Preserves Well-labeledness]
\label{lem:main_loop_well_labeled}
Given a weighted graph $G = (V, E, w, s, t)$
and
a well-labeled ART $\Lambda$ of $G$, each iteration of the main loop in the GPS procedure (\Cref{fig:gps-intraproc}) preserves the well-labeledness of $\Lambda$.
\end{lemma}
\begin{proof}
We prove
the property
by induction on the number of iterations of the main loop of GPS (\Cref{fig:gps-intraproc}). 

\textbf{Base case.} Upon entering the main loop, GPS creates an ART $\Lambda$ containing an empty covering only $\TreeRoot{\Lambda}$---which is labeled by $\top$ and maps to the entry vertex of the weighted graph $G$. In this case, all three well-labeledness conditions---initiation, consecution, well-coveredness---are vacuously preserved.

\textbf{Inductive hypothesis.} Assume 
that
after $k$ iterations of the main loop, the ART $\Lambda$ is well-labeled. 

We show that $\Lambda$ remains well-labeled after the 
$k$+$1^{\textit{st}}$
iteration. Suppose that $\TreeFrontiers{\Lambda}$ is non-empty; for if it is empty, then $\Lambda$ is a well-labeled and complete ART after step $k$ and the algorithm terminates. As described in \Cref{sec:algorithm}, at this iteration, GPS retrieves a node $u$ from $\TreeFrontiers{\Lambda}$ and performs one of three operations:
\begin{itemize}
    \item First, GPS tries to cover $u$ by an ancestor node $v$ in $\Lambda$; if GPS succeeds, then by \Cref{lem:try_cover_well_labeled} above, iteration $k+1$ preserves well-labeledness of $\Lambda$.
    \item If
    the previous step fails, 
    GPS tries to refine the label from $\TreeRoot{\Lambda}$ to $u$; if GPS succeeds, then by \Cref{lem:refine_well_labeld} above, iteration $k+1$ preserves well-labeledness of $\Lambda$.
    \item If refinement fails, then GPS retrieves a test input $M$ and uses $\textsc{Explore}(G, \Lambda, u, M)$ to expand $\Lambda$ at $u$. In this case, then by \Cref{lem:explore_well_labeled} above, iteration $k+1$ preserves well-labeledness of $\Lambda$.
\end{itemize} $\qed$
\end{proof}
 We are now ready to prove the soundness theorem, restated as follows:

\bfpara{\Cref{thm:gps_sound}.} {Let $G$ be a weighted graph. If $\textsc{GPS}(G)$ returns $\mathsf{Safe}$, then $G$ is safe; if it returns $(\mathsf{Unsafe}, \pi)$, then $\pi$ is a feasible path from the source of $G$ to its target.}

\bfpara{Proof.}  First, observe that \emph{the $\textsc{GPS(-)}$ procedure produces either a complete ART $\Lambda$ or a feasible $s$-$t$ path in $G$:} If $\Lambda$ is not complete, then there is still some node $n$ left in $\TreeFrontiers{\Lambda}$ and the main loop (line~\ref{line:GPS:mainloop} of Figure~\ref{fig:gps-intraproc}) would not terminate. Next, by \Cref{lem:main_loop_well_labeled} above, each iteration of the GPS procedure preserves well-labeledness of ART $\Lambda$. Thus, GPS terminates either when an ART $\Lambda$ is complete or a feasible $s$-$t$ path is found. Hence, by \Cref{thm:art_cfg_safe} (proven above), GPS is sound. $\qed$

\subsection{Proof of Refutation Completeness (\Cref{thm:intraproc_refutation_complete})}

Our proof strategy for refutation completeness resembles a real analysis-style proof that an infinite sequence always converges to an upper bound: we first assume,
for the sake of producing a contradiction,
that GPS does not terminate for a weighted-graph $G$ that has a feasible $s$-$t$ path, deriving an infinite sequence of ARTs $\Lambda_0, ..., \Lambda_i, ...$ from this assumption, and then we show that this sequence \emph{must converge} to an ART $\Lambda$ that exhibits the same properties as a well-labeled and complete ART {(informally speaking)}, hence deriving a contradiction. As such, the overall refutation-completeness proof mimics the proof of \Cref{thm:art_cfg_safe}, but works over an infinite sequence of ARTs: in particular, the two main technical lemmas below---\Cref{lem:trichotomy} and \Cref{lem:error_contraction_omega}---are analogues of technical lemmas \Cref{lem:general_trichotomy} and \Cref{lem:error_contraction} used to prove \Cref{thm:art_cfg_safe}, correspondingly.

We first restate the refutation-completeness theorem:

\bfpara{\Cref{thm:intraproc_refutation_complete}.}  
Let  $P$ be a (call-free) procedure. If $G$ is unsafe, then $\textsc{GPS}(G)$  terminates with the result ``Unsafe''.

\bfpara{Proof.} Suppose,
for the sake of producing a contradiction,
that there is a minimal error path $\pi$, but GPS does not terminate.  Let $\Lambda_0,\Lambda_1,\dots$ be the infinite sequence of ARTs obtained by running GPS.  For any path $\tau$ in $G$, there is a maximal prefix $\tau_1$ of $\tau$ such that $\tau_1 \in \TreeNodes{\Lambda_i}$ for some $i$.  Writing $\tau = \tau_1\tau_2$, we define $\textit{prefix}(\tau) = \TreePrefixTwo{\Lambda_i}{\tau} = \tau_1$ and $\textit{suffix}(\tau) = \TreeSuffixTwo{\Lambda_i}{\tau} = \tau_2$.
We say that a path $\tau$ is \textbf{maximal} if 
$\textit{prefix}(\tau) = \tau$. Next, we state two technical lemmas that resemble lemmas \Cref{lem:general_trichotomy} and \Cref{lem:error_contraction} used in the proof of \Cref{thm:art_cfg_safe}:

\begin{lemma}[Trichotomy of Sequence of ARTs] \label{lem:trichotomy}
Consider an infinite sequence of ARTs
$\Lambda_0, ..., \Lambda_i, ...$ obtained from a non-terminating run of GPS.
For any path $\tau$ of $G$, either
\begin{enumerate}
\item $\textit{prefix}(\tau) = \tau$, and $\tau$ is an internal node for some
$\Lambda_i$ (and thus for $\Lambda_j$ as well,
for all $j > i$);
\item There is some $i$ such that $\textit{prefix}(\tau)$ is pruned in $\Lambda_i$ (and thus
in $\Lambda_j$ as well, for all $j > i$); or
\item $\textit{prefix}(\tau)$ is covered in $\Lambda_i$ for infinitely many $i$.
\end{enumerate}
\end{lemma}
\begin{proof}
Let $\tau$ be a path.  Let $i$ be the least integer such that $\textit{prefix}(\tau) \in \TreeNodes{\Lambda_i}$.  Because $\textit{prefix}(\tau)$ was added to the ART by {\sc Explore}, it must be either internal or in the frontier of $\Lambda_i$.   If it is internal, then because $\textit{prefix}(\tau)$ is the longest prefix of $\tau$ in $\TreeNodes{\Lambda_i}$, we must have
$\textit{prefix}(\tau) = \tau$ and we are in case 1.

Suppose that $\textit{prefix}(\tau)$ is in the frontier of $\Lambda_i$.  Suppose that there is no iteration $j$ such that $\textit{prefix}(\tau)$ is internal or pruned in $\Lambda_j$---we must prove that it is covered infinitely often.  So we let $j$ be arbitrary, and show that there is some $k \geq j$ such that $\textit{prefix}(\tau)$ is covered in $\Lambda_k$.  Supposing that $\textit{prefix}(\tau)$ is not covered in $\Lambda_j$ itself, it must belong to the frontier of $\Lambda_j$.  There must be some $k \geq j$ such that GPS selects $\textit{prefix}(\tau)$ at the $k^{\textit{th}}$ iteration of its main loop, because the body of the main loop is terminating and the frontier queue is visited in FIFO order.  By assumption, $\textit{prefix}(\tau)$ is neither expanded nor pruned in $\Lambda_{k+1}$, so it must be covered.
$\qed$
\end{proof}


\begin{lemma}[Contraction for Sequence of ARTs]
    \label{lem:error_contraction_omega}
  For any contraction $\tau$ of the minimal feasible error path $\pi$, there is some $i$ \rfchanged{and ART $\Lambda_i = (T, L, \mathsf{Cov})$} such that
  $\textit{prefix}(\tau) \in \TreeNodes{\Lambda_i}$
and $\TreeLbl{\Lambda_i}{\textit{prefix}(\tau)} \land w(\textit{suffix}(\tau))$ is unsatisfiable.
\end{lemma}
\begin{proof}
 By strong induction on contraction order. 

 \bfpara{Base case.} Suppose that $\tau$ is a contraction of $\pi$ and minimal with respect to contraction order. Only case (2)
 of
 \Cref{lem:trichotomy} applies, because
 \begin{itemize}
     \item For case (1), we would have $\textit{prefix}(\tau) = \tau$, and $\tau$ is internal node for some $\Lambda_i$, which is impossible, because $\tau$ maps to the
     terminal
     vertex of $G$, which has no successors.
     \item For case (3), we would have $\textit{prefix}(\tau)$ is covered in $\Lambda_i$ for infinitely many $i$.
     However,
     $\tau$ is a simple path, so no ancestor of $\textit{prefix}(\tau)$ in $\Lambda$ corresponds to the same vertex as $\textit{prefix}(\tau)$, and thus $\textit{prefix}(\tau)$ cannot be covered.
 \end{itemize}

Thus, the lemma holds as a corollary of case 2 of \Cref{lem:trichotomy} in the base case.

 \bfpara{Induction hypothesis.} Let $\tau$ be a contraction of $\pi$, and suppose that the lemma holds for all proper contractions of $\tau$.

\bfpara{Induction step.} Here, we show that case (1) in \Cref{lem:trichotomy} is not possible, whereas both cases (2) and (3) of \Cref{lem:trichotomy} 
imply
the proof goal:
\begin{itemize}
    \item For case (1), we would have $\textit{prefix}(\tau) = \tau$, and $\tau$ is an internal node for some $\Lambda_i$, which is impossible, because $\tau$ maps to the
     terminal
    vertex of $G$, which has no successors.
    \item For case (2), because
    $\textit{prefix}(\tau)$ is pruned, $\TreeLbl{\Lambda_i}{\textit{prefix}(\tau)} \land w(\textit{suffix}(\tau))$ must be unsatisfiable by \Cref{lem:pruned_implies_unsat}.
    \item For case (3),
    WLOG, write
    $\tau = \tau_0\tau_1\tau_2$,
    where $\textit{prefix}(\tau) = \tau_0\tau_1$,
    $\textit{suffix}(\tau) = \tau_2$, and 
    $\tau_0$ covers $\tau_0\tau_1$ in $\Lambda_i$ for infinitely many $i$.
    Consider $\textit{prefix}(\tau_0\tau_2)$.
    By the induction hypothesis, there is some $j$ such that
    $\TreeLbl{\Lambda_j}{\textit{prefix}(\tau_0\tau_2)} \wedge \textit{suffix}(\tau_0\tau_2)$ is UNSAT. Fix an index $i$ such that $i > j$, and $\tau_0\tau_1$ appears covered by $\tau_0$ in $\Lambda_i$. 
    Let $\tau_2'$ be such that
    $\textit{prefix}(\tau_0\tau_2) = \tau_0\tau_2'$.
%
    %
   %
    We have:
    \begin{itemize}
        \item  (a) $\TreeLbl{\Lambda_i}{\tau_0\tau_1} \vDash \TreeLbl{\Lambda_i}{\tau_0}$ by
        the well-coveredness
        condition of $\Lambda_i$;
        \item (b) $\spost{\TreeLbl{\Lambda_i}{\tau_0}}{w(\tau_2')} \vDash \TreeLbl{\Lambda_i}{\tau_0\tau_2'}$ by the consecution condition of $\Lambda_i$;
        \item (c) $\TreeLbl{\Lambda_i} {\tau_0\tau_2'} \wedge \textit{suffix}(\tau_0\tau_2)$ is UNSAT, by the induction hypothesis; 
        
    \end{itemize}
    Altogether, (b) and (c) imply that $\TreeLbl{\Lambda_i}{\tau_0} \wedge w(\tau_2)$ is UNSAT, by the following reasoning:
    \begin{itemize}
        \item Rewrite $\TreeLbl{\Lambda_i}{\tau_0} \wedge w(\tau_2)$ as $\TreeLbl{\Lambda_i}{\tau_0} \wedge (w(\tau_2') \circ w(\textit{suffix}(\tau_0\tau_2)))$
        \item Rewrite the above further, as $\spost{\TreeLbl{\Lambda_i}{\tau_0}}{w(\tau_2')} \wedge w(\textit{suffix}(\tau_0\tau_2))$
        \item Substitute in (b), getting the above implies $\TreeLbl{\Lambda_i}{\tau_0\tau_2'} \wedge w(\textit{suffix}(\tau_0\tau_2))$
        \item By (c), the above expression is UNSAT.
    \end{itemize}
    
    By (a), $\TreeLbl{\Lambda_i}{\tau_0\tau_1} \wedge w(\tau_2)$ is UNSAT, so $\tau$ satisfies the proof goal of Lemma 3.
    $\qed$ 
\end{itemize}
\end{proof}

\noindent
\textit{Resumption of the proof of \Cref{thm:intraproc_refutation_complete}:}
The property that $\textsc{GPS}(G)$  terminates with the result ``Unsafe'' follows from \Cref{lem:error_contraction_omega},
because $\pi$ is a contraction of itself.  $\TreeLbl{\Lambda_i}{\textit{prefix}(\pi)} \land w(\textit{suffix}(\pi))$ contradicts the assumption that $\pi$ is feasible. $\qed$




%% file: figures/loops-ablation/new-scatter.cameraready.tex
\begin{tikzpicture}
            \begin{axis}[
            xlabel={Benchmarks Solved},
            ylabel={Running time (s)},
            title=Benchmarks from Suites \#1 and \#2,
            grid=both,
            tick align = outside,
            yticklabel style={/pgf/number format/fixed},     
            scaled x ticks = false,
            xticklabel style={/pgf/number format/fixed},
            legend style={at={(1.05,1)}, anchor=north west}
            ]
            
\addplot+[] coordinates {
(0, 0.10915527199995267)
(1, 0.11109248899992963)
(2, 0.11137738700017508)
(3, 0.11160536199986382)
(4, 0.11204850500052999)
(5, 0.11242767099975026)
(6, 0.11299808300009317)
(7, 0.11374727400016127)
(8, 0.11396220200003881)
(9, 0.11404127700006939)
(10, 0.11412964600003761)
(11, 0.1142793540002458)
(12, 0.11428007199992862)
(13, 0.11446146600064822)
(14, 0.11457362700002705)
(15, 0.1151909540003544)
(16, 0.11524142899997969)
(17, 0.11573702900000171)
(18, 0.11588723399927403)
(19, 0.11594303399942874)
(20, 0.11634281399983593)
(21, 0.11635911899975326)
(22, 0.1167189399993731)
(23, 0.11684953800022413)
(24, 0.11694277100014006)
(25, 0.11722273000032146)
(26, 0.11726847399995677)
(27, 0.11729966900020372)
(28, 0.11731704200064996)
(29, 0.1173857069998121)
(30, 0.11741040800006886)
(31, 0.1176578860004156)
(32, 0.11773158699998021)
(33, 0.11779192699941632)
(34, 0.11794800000006944)
(35, 0.11847874700015382)
(36, 0.11893762199997582)
(37, 0.11897404100000131)
(38, 0.11899584399998275)
(39, 0.11936975299977348)
(40, 0.11937182300061977)
(41, 0.11968812400004936)
(42, 0.1197027440000511)
(43, 0.12014059600005567)
(44, 0.12017629700039834)
(45, 0.12136217999977816)
(46, 0.12189978900005372)
(47, 0.12223334899999827)
(48, 0.12295304399958695)
(49, 0.12303510400033701)
(50, 0.12315083599969512)
(51, 0.12319157999991148)
(52, 0.12325740100004623)
(53, 0.12326340699974025)
(54, 0.12329281100028311)
(55, 0.12334569900008319)
(56, 0.1234033629998521)
(57, 0.12342904099932639)
(58, 0.12369448500066937)
(59, 0.12372162599990588)
(60, 0.12373186899981192)
(61, 0.12419216499984032)
(62, 0.12428279200048564)
(63, 0.12461346500003856)
(64, 0.12471729200024129)
(65, 0.12484687699998176)
(66, 0.12493767599971761)
(67, 0.12539577300003657)
(68, 0.12540897600047174)
(69, 0.1254322240001784)
(70, 0.12567072100046062)
(71, 0.1260226370000055)
(72, 0.12662471899966476)
(73, 0.1267741230003594)
(74, 0.12677813199934462)
(75, 0.1270394410003064)
(76, 0.1276129409998248)
(77, 0.12777774599999248)
(78, 0.12800868599970272)
(79, 0.1281598059999851)
(80, 0.12870342699989124)
(81, 0.1289371940001729)
(82, 0.128949159000058)
(83, 0.12991965699984576)
(84, 0.13006436500018026)
(85, 0.13057825999931083)
(86, 0.1308014380001623)
(87, 0.13115689100050076)
(88, 0.13140910800007077)
(89, 0.13163855600032548)
(90, 0.1316411729999345)
(91, 0.1327432480002244)
(92, 0.13281668800027546)
(93, 0.13346909600022627)
(94, 0.1335507819999293)
(95, 0.13361524099991584)
(96, 0.13369358700037992)
(97, 0.13369358999989345)
(98, 0.133774228999755)
(99, 0.13462772000002587)
(100, 0.13464683999973204)
(101, 0.13470892799978174)
(102, 0.13473453799997515)
(103, 0.13492787399991357)
(104, 0.13576975600062724)
(105, 0.1363985709995177)
(106, 0.13669546799974341)
(107, 0.1369331530004274)
(108, 0.13716287500028557)
(109, 0.13848296200001187)
(110, 0.13869657100076438)
(111, 0.1388852649997716)
(112, 0.13889706999998452)
(113, 0.14007712100010394)
(114, 0.1403816239999287)
(115, 0.14064574500025628)
(116, 0.1406709659995613)
(117, 0.14107807500022318)
(118, 0.14155481699981465)
(119, 0.14279944900044939)
(120, 0.14338836200022342)
(121, 0.14346568199999865)
(122, 0.1436346689997663)
(123, 0.14484315200024866)
(124, 0.14521029500065197)
(125, 0.14524701699997422)
(126, 0.14565893500002858)
(127, 0.14616902699981438)
(128, 0.14625234800041653)
(129, 0.14634927200040693)
(130, 0.14973748000011255)
(131, 0.15015121899978112)
(132, 0.150190439000653)
(133, 0.15029134499991414)
(134, 0.1513834699999279)
(135, 0.15177304900043964)
(136, 0.1523229310005263)
(137, 0.15309989999991558)
(138, 0.15322220500002004)
(139, 0.1534368020002148)
(140, 0.15375768199919548)
(141, 0.15717583399998603)
(142, 0.15773679900030402)
(143, 0.1582918720005182)
(144, 0.1596995169998081)
(145, 0.162361945000157)
(146, 0.16444378799951664)
(147, 0.16473170100016432)
(148, 0.16527513999972143)
(149, 0.1666074649992879)
(150, 0.1672718290001285)
(151, 0.1677464049998889)
(152, 0.1679147599998032)
(153, 0.168184838000343)
(154, 0.16913540299992746)
(155, 0.17552953799986426)
(156, 0.17602354799964814)
(157, 0.17822420299944497)
(158, 0.18376525400071841)
(159, 0.18486693700015167)
(160, 0.188841176999631)
(161, 0.1908042940001451)
(162, 0.19200917199941614)
(163, 0.19641679800042766)
(164, 0.19719582500010802)
(165, 0.2005583440004557)
(166, 0.20164210199982335)
(167, 0.20715709400019477)
(168, 0.20742839700051263)
(169, 0.2079027390000192)
(170, 0.20939113699932932)
(171, 0.21050012800003515)
(172, 0.21054284799993184)
(173, 0.21381490300063888)
(174, 0.21936014800030534)
(175, 0.22205560600013996)
(176, 0.22277980600028968)
(177, 0.22295361099986621)
(178, 0.2291221210000458)
(179, 0.23692949799988128)
(180, 0.24705031099983898)
(181, 0.24971239499973308)
(182, 0.2504746009999508)
(183, 0.25337113499972475)
(184, 0.2551738619995376)
(185, 0.2558859549999397)
(186, 0.25970718000007764)
(187, 0.27665038499981165)
(188, 0.2810839990000318)
(189, 0.2860693310000215)
(190, 0.3080661780004448)
(191, 0.30979482699996197)
(192, 0.3485320990002947)
(193, 0.3609832189995359)
(194, 0.37077617999966606)
(195, 0.37214105500061123)
(196, 0.3806685289999905)
(197, 0.38802462999956333)
(198, 0.41330802899938135)
(199, 0.5178575199997795)
(200, 0.5454625639999904)
(201, 0.5468602129994906)
(202, 0.588560563000101)
(203, 0.6954002890001902)
(204, 0.7956775650000054)
(205, 0.854748261000168)
(206, 0.8774424779999208)
(207, 1.2233252680000533)
(208, 1.3204300830002467)
(209, 1.5452619760001198)
(210, 1.6267229499999303)
(211, 1.640968466999766)
(212, 2.1739809089995106)
(213, 2.478241731000253)
(214, 2.529429340999741)
(215, 2.679949676999968)
(216, 3.814238400999784)
(217, 4.355365660000189)
(218, 4.398053317999711)
(219, 5.683562993000123)
(220, 9.188449637999838)
(221, 10.742884717999914)
(222, 502.588431913)
};
\addlegendentry{GPS};
\addplot+[] coordinates {
(0, 0.1097170159991947)
(1, 0.11090476499884971)
(2, 0.11110649799957173)
(3, 0.11135800400006701)
(4, 0.11222641600033967)
(5, 0.11261421800008975)
(6, 0.11323654399893712)
(7, 0.11345179000272765)
(8, 0.11363788599919644)
(9, 0.11422201999812387)
(10, 0.11452293100046518)
(11, 0.11472489399966435)
(12, 0.11496061900106724)
(13, 0.11527754800044931)
(14, 0.11554965800314676)
(15, 0.11594811100076186)
(16, 0.11669245200027945)
(17, 0.11680366799919284)
(18, 0.11685965500146267)
(19, 0.11693042399929254)
(20, 0.11737745800201083)
(21, 0.11753179999868735)
(22, 0.11756719699769747)
(23, 0.11779148800269468)
(24, 0.11824245599927963)
(25, 0.1186999339988688)
(26, 0.11885167000218644)
(27, 0.11888083700250718)
(28, 0.1189718499990704)
(29, 0.11906559899944114)
(30, 0.11933301699900767)
(31, 0.11938456799907726)
(32, 0.1194667509989813)
(33, 0.11951735600086977)
(34, 0.11960201100009726)
(35, 0.11981562599976314)
(36, 0.11994727899946156)
(37, 0.1201502209987666)
(38, 0.12020452700016904)
(39, 0.12058510999850114)
(40, 0.12062304699975357)
(41, 0.1207583280011022)
(42, 0.1208800610002072)
(43, 0.12098005100051523)
(44, 0.12120805000085966)
(45, 0.1213609219994396)
(46, 0.12244253699827823)
(47, 0.12247184799707611)
(48, 0.12282728500213125)
(49, 0.12293045100159361)
(50, 0.12306725500093307)
(51, 0.12319969799864339)
(52, 0.1233702040008211)
(53, 0.123377686000822)
(54, 0.12338983600056963)
(55, 0.12339942499966128)
(56, 0.123595592998754)
(57, 0.12424069100052293)
(58, 0.12442386700058705)
(59, 0.12456104199918627)
(60, 0.1247711479991267)
(61, 0.1249258469997585)
(62, 0.12502631800089148)
(63, 0.12512360400069156)
(64, 0.12531667800067225)
(65, 0.1254144200029259)
(66, 0.1261597820011957)
(67, 0.12617287400280475)
(68, 0.12627659300051164)
(69, 0.12644008200004464)
(70, 0.12647441499939305)
(71, 0.1266337230008503)
(72, 0.12766250500135357)
(73, 0.12777002599978005)
(74, 0.1277876650019607)
(75, 0.12810141099907923)
(76, 0.1283252339999308)
(77, 0.12837773199862568)
(78, 0.12842293200083077)
(79, 0.1284885570021288)
(80, 0.12861595299909823)
(81, 0.12872851899919624)
(82, 0.12878667599943583)
(83, 0.1295536829966295)
(84, 0.12963384900103847)
(85, 0.13028158499946585)
(86, 0.13040840600297088)
(87, 0.13078219499948318)
(88, 0.1314027789994725)
(89, 0.13183210100032738)
(90, 0.13227250999989337)
(91, 0.13265326300097513)
(92, 0.13271649499802152)
(93, 0.13346713700047985)
(94, 0.13357819299926632)
(95, 0.13379214799897454)
(96, 0.13427585099998396)
(97, 0.13529463200029568)
(98, 0.13533364499744494)
(99, 0.13542301999768824)
(100, 0.1355890160011768)
(101, 0.13566667300074187)
(102, 0.13627075100157526)
(103, 0.1364913389988942)
(104, 0.13657745599994087)
(105, 0.136856278997584)
(106, 0.1370076409984904)
(107, 0.13737153800138913)
(108, 0.13792744900274556)
(109, 0.13902298199900542)
(110, 0.13921937600025558)
(111, 0.13949861899891403)
(112, 0.1395543539983919)
(113, 0.14001009999992675)
(114, 0.1400999259967648)
(115, 0.14015892799943686)
(116, 0.14073192099931475)
(117, 0.14231128900064505)
(118, 0.14428617199882865)
(119, 0.14490321899938863)
(120, 0.14582626100309426)
(121, 0.14590338099878863)
(122, 0.1463698760016996)
(123, 0.1464533440012019)
(124, 0.14740496599915787)
(125, 0.14765288199851057)
(126, 0.1485842069996579)
(127, 0.14905855400138535)
(128, 0.1493915270002617)
(129, 0.14970995800103992)
(130, 0.1499096700008522)
(131, 0.1512654019988986)
(132, 0.15275690099952044)
(133, 0.15279199500037066)
(134, 0.15305055399949197)
(135, 0.15330803100005141)
(136, 0.15403925600185175)
(137, 0.15513821199783706)
(138, 0.15620076499908464)
(139, 0.15713908699945023)
(140, 0.1576614099994913)
(141, 0.15993090000119992)
(142, 0.163319039000271)
(143, 0.1645072909996088)
(144, 0.164568762998897)
(145, 0.16821913099920494)
(146, 0.16826657299679937)
(147, 0.16865063500154065)
(148, 0.16997504299797583)
(149, 0.17160704399793758)
(150, 0.17467982300149743)
(151, 0.18076568800097448)
(152, 0.18153536099998746)
(153, 0.1856988609979453)
(154, 0.18953223700009403)
(155, 0.18998269599978812)
(156, 0.19013548200018704)
(157, 0.1958631580000656)
(158, 0.19713993000186747)
(159, 0.19952708500204608)
(160, 0.20370092400116846)
(161, 0.2042509950006206)
(162, 0.20499335199929192)
(163, 0.20642446900092182)
(164, 0.21077317700110143)
(165, 0.21138512200013793)
(166, 0.21490919600182679)
(167, 0.2186202900011267)
(168, 0.22453789399878588)
(169, 0.22653726799944707)
(170, 0.23341603099834174)
(171, 0.2336214819988527)
(172, 0.23369824399924255)
(173, 0.24213446900103008)
(174, 0.24808188499991957)
(175, 0.24829608599975472)
(176, 0.2508568039993406)
(177, 0.25109581899960176)
(178, 0.25197122899953683)
(179, 0.2520157479993941)
(180, 0.2535667749980348)
(181, 0.26032142999974894)
(182, 0.2612765240010049)
(183, 0.26411494499916444)
(184, 0.2679399669996201)
(185, 0.2779041979993053)
(186, 0.29972698699930334)
(187, 0.3011040099991078)
(188, 0.30596229000002495)
(189, 0.3260067010014609)
(190, 0.34277132699935464)
(191, 0.37686027200106764)
(192, 0.39948349699989194)
(193, 0.42179375599880586)
(194, 0.43272249400251894)
(195, 0.48729205499694217)
(196, 0.547681496002042)
(197, 0.5749376309977379)
(198, 0.6172519120009383)
(199, 0.6915455260004819)
(200, 0.8469699839988607)
(201, 4.370133933000034)
(202, 4.393745620000118)
(203, 4.728980316000161)
(204, 5.636586676999286)
(205, 10.771712776999266)
};
\addlegendentry{GPSLite};
\addplot+[] coordinates {
(0, 0.10538192999956664)
(1, 0.1068209110017051)
(2, 0.10712048599816626)
(3, 0.10882555799616966)
(4, 0.10918188000505324)
(5, 0.10919300500245299)
(6, 0.10940678499900969)
(7, 0.10947457799920812)
(8, 0.10950977599713951)
(9, 0.11005652599851601)
(10, 0.11005708199809305)
(11, 0.11058989999582991)
(12, 0.11067101899971021)
(13, 0.11095888399722753)
(14, 0.11135384400404291)
(15, 0.11139375699713128)
(16, 0.11156353000114905)
(17, 0.11288331799732987)
(18, 0.11316483299742686)
(19, 0.11324192999745719)
(20, 0.1135853230007342)
(21, 0.11401236399979098)
(22, 0.11449677099881228)
(23, 0.11467532599635888)
(24, 0.11484908199781785)
(25, 0.11494905799918342)
(26, 0.11507875900133513)
(27, 0.11543444100243505)
(28, 0.11551437099842587)
(29, 0.11554183800035389)
(30, 0.11564207699848339)
(31, 0.11583968700142577)
(32, 0.11642465600016294)
(33, 0.11645635899913032)
(34, 0.116458884003805)
(35, 0.11654895600076998)
(36, 0.11665418300253805)
(37, 0.11675314400054049)
(38, 0.1167946670029778)
(39, 0.11699668299843324)
(40, 0.11700635699526174)
(41, 0.11705854599858867)
(42, 0.1171200399985537)
(43, 0.11721507399488473)
(44, 0.11746106099599274)
(45, 0.11779504799778806)
(46, 0.11780967200320447)
(47, 0.11826311900222208)
(48, 0.11830132999602938)
(49, 0.1185427320015151)
(50, 0.11857504600629909)
(51, 0.11866497599839931)
(52, 0.11869967999518849)
(53, 0.11874729300325271)
(54, 0.11882246500317706)
(55, 0.11888602300314233)
(56, 0.11906016399734654)
(57, 0.11916220000421163)
(58, 0.11935694199928548)
(59, 0.11979196999891428)
(60, 0.11981512100464897)
(61, 0.11991059599677101)
(62, 0.12001226800202858)
(63, 0.12019663999672048)
(64, 0.12031043299793964)
(65, 0.1204763119967538)
(66, 0.12056968600518303)
(67, 0.12086547599756159)
(68, 0.12125167699559825)
(69, 0.12156035299994983)
(70, 0.12161193700012518)
(71, 0.1216891090007266)
(72, 0.12174702599440934)
(73, 0.12184209399856627)
(74, 0.12184521400195081)
(75, 0.12243352599762147)
(76, 0.12247568699967815)
(77, 0.12252107900712872)
(78, 0.12291965600161348)
(79, 0.12336529699678067)
(80, 0.1236877880000975)
(81, 0.12414533799892524)
(82, 0.12516248400061158)
(83, 0.1252424609992886)
(84, 0.12560733199643437)
(85, 0.1260592720063869)
(86, 0.12626709500182187)
(87, 0.12685125400457764)
(88, 0.12700551099987933)
(89, 0.12703929800045444)
(90, 0.12709901399648516)
(91, 0.1272834609990241)
(92, 0.12762908199511003)
(93, 0.12780208099866286)
(94, 0.12805630399816437)
(95, 0.1286024040018674)
(96, 0.12865608499851078)
(97, 0.12978427200141596)
(98, 0.12978819400450448)
(99, 0.13008725899999263)
(100, 0.13025011999707203)
(101, 0.13030006600456545)
(102, 0.13031563199911034)
(103, 0.130339625000488)
(104, 0.1304633969994029)
(105, 0.1305786140001146)
(106, 0.1309367379944888)
(107, 0.13136657000723062)
(108, 0.13148690999514656)
(109, 0.1315955380050582)
(110, 0.1319412180018844)
(111, 0.1322262640023837)
(112, 0.13256442800047807)
(113, 0.1327425060007954)
(114, 0.13350852100120392)
(115, 0.13559709300170653)
(116, 0.13565757300239056)
(117, 0.13658580999617698)
(118, 0.13687356599984923)
(119, 0.1371228490024805)
(120, 0.137789824999345)
(121, 0.13806367899815086)
(122, 0.1385013980034273)
(123, 0.1387648550007725)
(124, 0.13886064899998019)
(125, 0.13912972300022375)
(126, 0.13914336500602076)
(127, 0.13948743500077398)
(128, 0.1395659350018832)
(129, 0.14014230400061933)
(130, 0.14034070399793563)
(131, 0.14040345400280785)
(132, 0.14074616899597459)
(133, 0.14170446699426975)
(134, 0.14217345300130546)
(135, 0.14239265799551504)
(136, 0.14252248899720144)
(137, 0.14354445799835958)
(138, 0.14465243599988753)
(139, 0.14519360599661013)
(140, 0.14544080100313295)
(141, 0.14569977700011805)
(142, 0.14609941699745832)
(143, 0.1476507400002447)
(144, 0.14834317500208272)
(145, 0.1495272850006586)
(146, 0.15031157399789663)
(147, 0.15034290000039618)
(148, 0.15173087299626786)
(149, 0.15204296999581857)
(150, 0.15322594600002049)
(151, 0.15472227099962765)
(152, 0.1547510960008367)
(153, 0.1582854629959911)
(154, 0.15906621500471374)
(155, 0.16123401799995918)
(156, 0.16332993900141446)
(157, 0.16527932299504755)
(158, 0.16611745199770667)
(159, 0.1663845439979923)
(160, 0.1729293600001256)
(161, 0.18250093800452305)
(162, 0.1861394610023126)
(163, 0.1864573470011237)
(164, 0.19095330399431987)
(165, 0.1925258830015082)
(166, 0.19290529600402806)
(167, 0.1981576580001274)
(168, 0.19896001600136515)
(169, 0.2011017519980669)
(170, 0.20139499299693853)
(171, 0.20228854499873705)
(172, 0.20391485199797899)
(173, 0.2056539210025221)
(174, 0.20756275999883655)
(175, 0.20814446499571204)
(176, 0.20995208300155355)
(177, 0.2193491450016154)
(178, 0.23117362100310856)
(179, 0.23474316500505665)
(180, 0.2418561430022237)
(181, 0.243270789993403)
(182, 0.24667033000150695)
(183, 0.24683847800042713)
(184, 0.25106498500099406)
(185, 0.2806862809957238)
(186, 0.28129475899913814)
(187, 0.28541715200117324)
(188, 0.2879795689950697)
(189, 0.30399268199835205)
(190, 0.31639117899612756)
(191, 0.35166595799819333)
(192, 0.3620963529974688)
(193, 0.37381905199436005)
(194, 0.38257576400064863)
(195, 0.38353402300708694)
(196, 0.4157862770007341)
(197, 0.4574725159982336)
(198, 0.4639131099975202)
(199, 0.5515088359970832)
(200, 0.5556065939963446)
(201, 0.6009310149966041)
(202, 0.6138111519976519)
(203, 0.8748339130033855)
(204, 0.8845122630009428)
(205, 1.1385300839974661)
(206, 1.5327974749961868)
(207, 1.9799262520027696)
(208, 2.424150186001498)
(209, 3.8047971990017686)
(210, 4.038245429001108)
(211, 4.566370399996231)
(212, 4.627103997001541)
(213, 4.63166315700073)
(214, 5.354096301998652)
(215, 6.230225808998512)
(216, 7.149136758001987)
(217, 15.32275610200304)
(218, 589.5048688210009)
};
\addlegendentry{GPS-nogas};
\addplot+[] coordinates {
(0, 0.1350452140031848)
(1, 0.15538565500173718)
(2, 0.15641428000526503)
(3, 0.15874825700302608)
(4, 0.16319084097631276)
(5, 0.1639139829785563)
(6, 0.16435251801158302)
(7, 0.16458559897728264)
(8, 0.16565314299077727)
(9, 0.16609113299637102)
(10, 0.16696910600876436)
(11, 0.16697044999455102)
(12, 0.1673251820029691)
(13, 0.16734584199730307)
(14, 0.16870893599116243)
(15, 0.17058608902152628)
(16, 0.1705992039933335)
(17, 0.1711576520174276)
(18, 0.17135464400053024)
(19, 0.17202678599278443)
(20, 0.17280231599579565)
(21, 0.17292637898935936)
(22, 0.17323244200088084)
(23, 0.1739001199894119)
(24, 0.17411980699398555)
(25, 0.17427444399800152)
(26, 0.17432014000951312)
(27, 0.17539309998392127)
(28, 0.17584509498556145)
(29, 0.17606145999161527)
(30, 0.17608095199102536)
(31, 0.17638020002050325)
(32, 0.17687404199386947)
(33, 0.17691391499829479)
(34, 0.17763921999721788)
(35, 0.17776918300660327)
(36, 0.1778475670143962)
(37, 0.17829090499435551)
(38, 0.1784105509868823)
(39, 0.1784120410156902)
(40, 0.17844895698362961)
(41, 0.1784530610020738)
(42, 0.17870349501026794)
(43, 0.17892819800181314)
(44, 0.17895003999001347)
(45, 0.17946177400881425)
(46, 0.1807763859978877)
(47, 0.18083642798592336)
(48, 0.1812857009936124)
(49, 0.18180399099946953)
(50, 0.18182876298669726)
(51, 0.18202378999558277)
(52, 0.18260879497393034)
(53, 0.18272537298616953)
(54, 0.1827612929919269)
(55, 0.18286826499388553)
(56, 0.18357126897899434)
(57, 0.18481552400044166)
(58, 0.18528883199905977)
(59, 0.18561875700834207)
(60, 0.18864751199726015)
(61, 0.18890659901080653)
(62, 0.1891426790098194)
(63, 0.18930750599247403)
(64, 0.18947559199295938)
(65, 0.19059489699429832)
(66, 0.1916215629898943)
(67, 0.19194300399976783)
(68, 0.19245313899591565)
(69, 0.19253767700865865)
(70, 0.19289324100827798)
(71, 0.19363075599540025)
(72, 0.19382210102048703)
(73, 0.19432304799556732)
(74, 0.19487490301253274)
(75, 0.19490875999326818)
(76, 0.19530556799145415)
(77, 0.1962541679968126)
(78, 0.19666671301820315)
(79, 0.19669191099819727)
(80, 0.19704658098635264)
(81, 0.19840423201094382)
(82, 0.19874694201280363)
(83, 0.20093720400473103)
(84, 0.20101455799886025)
(85, 0.20125129801454023)
(86, 0.20128323999233544)
(87, 0.20197577599901706)
(88, 0.20224007999058813)
(89, 0.20236871499218978)
(90, 0.2035239620017819)
(91, 0.205464827013202)
(92, 0.20678643201244995)
(93, 0.20769406302133575)
(94, 0.20854583699838258)
(95, 0.20982644698233344)
(96, 0.21492942899931222)
(97, 0.216900486004306)
(98, 0.21724552699015476)
(99, 0.22111825598403811)
(100, 0.2212580629857257)
(101, 0.2238526600121986)
(102, 0.2321227380016353)
(103, 0.23358632400049828)
(104, 0.23479550302727148)
(105, 0.24788177400478162)
(106, 0.24858956900425255)
(107, 0.2565029239922296)
(108, 0.2566797859908547)
(109, 0.2582622950139921)
(110, 0.2590112049947493)
(111, 0.263152490981156)
(112, 0.2728752459806856)
(113, 0.27484863900463097)
(114, 0.2749345180054661)
(115, 0.27670977599336766)
(116, 0.2819755809905473)
(117, 0.28721406700788066)
(118, 0.2879357899946626)
(119, 0.2918214220262598)
(120, 0.2918365660007112)
(121, 0.29247764899628237)
(122, 0.297204698988935)
(123, 0.2976523670076858)
(124, 0.2993135610013269)
(125, 0.30012080099550076)
(126, 0.31853971199598163)
(127, 0.32785929201054387)
(128, 0.32991416499135084)
(129, 0.33083731099031866)
(130, 0.3310037129849661)
(131, 0.33613859800971113)
(132, 0.3418678309826646)
(133, 0.36038009502226487)
(134, 0.36664795997785404)
(135, 0.3909638390119653)
(136, 0.396891048992984)
(137, 0.40105089597636834)
(138, 0.42143308799131773)
(139, 0.4323256330098957)
(140, 0.484129529009806)
(141, 0.5577180170221254)
(142, 0.5943303359963465)
(143, 0.6539812150003854)
(144, 0.6892640839796513)
(145, 0.7298712229821831)
(146, 0.7457494930131361)
(147, 0.776614743983373)
(148, 0.8535495759861078)
(149, 1.1328239819849841)
(150, 1.175390865013469)
(151, 2.14825772299082)
(152, 2.6089763109921478)
(153, 3.260625941999024)
(154, 3.5749219940043986)
(155, 4.222498107003048)
(156, 4.789195716002723)
(157, 10.870408310001949)
(158, 12.070777460001409)
(159, 13.866069672018057)
(160, 32.998622932995204)
(161, 41.75210785100353)
(162, 42.48964202500065)
(163, 102.80068129100255)
(164, 166.0975438959722)
(165, 242.03598242500448)
(166, 340.4998726689955)
(167, 381.25358184502693)
(168, 403.6307786499965)
};
\addlegendentry{GPS-nogas-nosum};
\end{axis}
\end{tikzpicture}

%% file: main-OOPSLA25-ExtendedVersion.bbl

\begin{thebibliography}{46}


\ifx \showCODEN    \undefined \def \showCODEN     #1{\unskip}     \fi
\ifx \showDOI      \undefined \def \showDOI       #1{#1}\fi
\ifx \showISBNx    \undefined \def \showISBNx     #1{\unskip}     \fi
\ifx \showISBNxiii \undefined \def \showISBNxiii  #1{\unskip}     \fi
\ifx \showISSN     \undefined \def \showISSN      #1{\unskip}     \fi
\ifx \showLCCN     \undefined \def \showLCCN      #1{\unskip}     \fi
\ifx \shownote     \undefined \def \shownote      #1{#1}          \fi
\ifx \showarticletitle \undefined \def \showarticletitle #1{#1}   \fi
\ifx \showURL      \undefined \def \showURL       {\relax}        \fi
\providecommand\bibfield[2]{#2}
\providecommand\bibinfo[2]{#2}
\providecommand\natexlab[1]{#1}
\providecommand\showeprint[2][]{arXiv:#2}

\bibitem[Afzal et~al\mbox{.}(2019)]%
        {tool-veriabs}
\bibfield{author}{\bibinfo{person}{Mohammad Afzal}, \bibinfo{person}{A. Asia}, \bibinfo{person}{Avriti Chauhan}, \bibinfo{person}{Bharti Chimdyalwar}, \bibinfo{person}{Priyanka Darke}, \bibinfo{person}{Advaita Datar}, \bibinfo{person}{Shrawan Kumar}, {and} \bibinfo{person}{R. Venkatesh}.} \bibinfo{year}{2019}\natexlab{}.
\newblock \showarticletitle{VeriAbs : Verification by Abstraction and Test Generation}. In \bibinfo{booktitle}{\emph{2019 34th IEEE/ACM International Conference on Automated Software Engineering (ASE)}}. \bibinfo{pages}{1138--1141}.
\newblock
\urldef\tempurl%
\url{https://doi.org/10.1109/ASE.2019.00121}
\showDOI{\tempurl}


\bibitem[Albarghouthi et~al\mbox{.}(2012a)]%
        {Whale}
\bibfield{author}{\bibinfo{person}{Aws Albarghouthi}, \bibinfo{person}{Arie Gurfinkel}, {and} \bibinfo{person}{Marsha Chechik}.} \bibinfo{year}{2012}\natexlab{a}.
\newblock \showarticletitle{Whale: An Interpolation-Based Algorithm for Inter-procedural Verification}. In \bibinfo{booktitle}{\emph{Verification, Model Checking, and Abstract Interpretation}}, \bibfield{editor}{\bibinfo{person}{Viktor Kuncak} {and} \bibinfo{person}{Andrey Rybalchenko}} (Eds.). \bibinfo{publisher}{Springer Berlin Heidelberg}, \bibinfo{address}{Berlin, Heidelberg}, \bibinfo{pages}{39--55}.
\newblock
\showISBNx{978-3-642-27940-9}
\urldef\tempurl%
\url{https://doi.org/10.1007/978-3-642-27940-9_4}
\showURL{%
\tempurl}


\bibitem[Albarghouthi et~al\mbox{.}(2012b)]%
        {ufo}
\bibfield{author}{\bibinfo{person}{Aws Albarghouthi}, \bibinfo{person}{Yi Li}, \bibinfo{person}{Arie Gurfinkel}, {and} \bibinfo{person}{Marsha Chechik}.} \bibinfo{year}{2012}\natexlab{b}.
\newblock \showarticletitle{Ufo: A Framework for Abstraction- and Interpolation-Based Software Verification}. In \bibinfo{booktitle}{\emph{Computer Aided Verification}}, \bibfield{editor}{\bibinfo{person}{P.~Madhusudan} {and} \bibinfo{person}{Sanjit~A. Seshia}} (Eds.). \bibinfo{publisher}{Springer Berlin Heidelberg}, \bibinfo{address}{Berlin, Heidelberg}, \bibinfo{pages}{672--678}.
\newblock
\showISBNx{978-3-642-31424-7}
\urldef\tempurl%
\url{https://doi.org/10.1007/978-3-642-31424-7_48}
\showURL{%
\tempurl}


\bibitem[Beckman et~al\mbox{.}(2008)]%
        {Dash}
\bibfield{author}{\bibinfo{person}{Nels~E. Beckman}, \bibinfo{person}{Aditya~V. Nori}, \bibinfo{person}{Sriram~K. Rajamani}, {and} \bibinfo{person}{Robert~J. Simmons}.} \bibinfo{year}{2008}\natexlab{}.
\newblock \showarticletitle{Proofs from Tests}. In \bibinfo{booktitle}{\emph{Proceedings of the 2008 International Symposium on Software Testing and Analysis}} (Seattle, WA, USA) \emph{(\bibinfo{series}{ISSTA '08})}. \bibinfo{publisher}{Association for Computing Machinery}, \bibinfo{address}{New York, NY, USA}, \bibinfo{pages}{3–14}.
\newblock
\showISBNx{9781605580500}
\urldef\tempurl%
\url{https://doi.org/10.1145/1390630.1390634}
\showDOI{\tempurl}


\bibitem[Beyer(2024)]%
        {svcomp2024}
\bibfield{author}{\bibinfo{person}{Dirk Beyer}.} \bibinfo{year}{2024}\natexlab{}.
\newblock \showarticletitle{State of the Art in Software Verification and Witness Validation: SV-COMP 2024}. In \bibinfo{booktitle}{\emph{Tools and Algorithms for the Construction and Analysis of Systems}}, \bibfield{editor}{\bibinfo{person}{Bernd Finkbeiner} {and} \bibinfo{person}{Laura Kov{\'a}cs}} (Eds.). \bibinfo{publisher}{Springer Nature Switzerland}, \bibinfo{address}{Cham}, \bibinfo{pages}{299--329}.
\newblock
\showISBNx{978-3-031-57256-2}
\urldef\tempurl%
\url{https://doi.org/10.1007/978-3-031-57256-2_15}
\showURL{%
\tempurl}


\bibitem[Beyer et~al\mbox{.}(2009)]%
        {fmcad:bcgks09}
\bibfield{author}{\bibinfo{person}{Dirk Beyer}, \bibinfo{person}{Alessandro Cimatti}, \bibinfo{person}{Alberto Griggio}, \bibinfo{person}{M.~Erkan Keremoglu}, \bibinfo{person}{Simon~Fraser University}, {and} \bibinfo{person}{Roberto Sebastiani}.} \bibinfo{year}{2009}\natexlab{}.
\newblock \showarticletitle{Software model checking via large-block encoding}. In \bibinfo{booktitle}{\emph{2009 Formal Methods in Computer-Aided Design}}. \bibinfo{pages}{25--32}.
\newblock
\urldef\tempurl%
\url{https://doi.org/10.1109/FMCAD.2009.5351147}
\showDOI{\tempurl}


\bibitem[Beyer and Keremoglu(2011)]%
        {tool-cpachecker}
\bibfield{author}{\bibinfo{person}{Dirk Beyer} {and} \bibinfo{person}{M.~Erkan Keremoglu}.} \bibinfo{year}{2011}\natexlab{}.
\newblock \showarticletitle{CPAchecker: A Tool for Configurable Software Verification}. In \bibinfo{booktitle}{\emph{Computer Aided Verification}}, \bibfield{editor}{\bibinfo{person}{Ganesh Gopalakrishnan} {and} \bibinfo{person}{Shaz Qadeer}} (Eds.). \bibinfo{publisher}{Springer Berlin Heidelberg}, \bibinfo{address}{Berlin, Heidelberg}, \bibinfo{pages}{184--190}.
\newblock
\showISBNx{978-3-642-22110-1}
\urldef\tempurl%
\url{https://doi.org/10.1007/978-3-642-22110-1_16}
\showURL{%
\tempurl}


\bibitem[Beyer et~al\mbox{.}(2010)]%
        {fmcad:bkw10}
\bibfield{author}{\bibinfo{person}{Dirk Beyer}, \bibinfo{person}{M.~Erkan Keremoglu}, {and} \bibinfo{person}{Philipp Wendler}.} \bibinfo{year}{2010}\natexlab{}.
\newblock \showarticletitle{Predicate abstraction with adjustable-block encoding}. In \bibinfo{booktitle}{\emph{Formal Methods in Computer Aided Design}}. \bibinfo{pages}{189--197}.
\newblock


\bibitem[Beyer et~al\mbox{.}(2019)]%
        {benchexec}
\bibfield{author}{\bibinfo{person}{Dirk Beyer}, \bibinfo{person}{Stefan L{\"o}we}, {and} \bibinfo{person}{Philipp Wendler}.} \bibinfo{year}{2019}\natexlab{}.
\newblock \showarticletitle{Reliable benchmarking: requirements and solutions}.
\newblock \bibinfo{journal}{\emph{International Journal on Software Tools for Technology Transfer}} \bibinfo{volume}{21}, \bibinfo{number}{1} (\bibinfo{year}{2019}), \bibinfo{pages}{1--29}.
\newblock
\showISBNx{1433-2787}
\urldef\tempurl%
\url{https://doi.org/10.1007/s10009-017-0469-y}
\showDOI{\tempurl}


\bibitem[B\"{o}hme et~al\mbox{.}(2017)]%
        {AFLGo}
\bibfield{author}{\bibinfo{person}{Marcel B\"{o}hme}, \bibinfo{person}{Van-Thuan Pham}, \bibinfo{person}{Manh-Dung Nguyen}, {and} \bibinfo{person}{Abhik Roychoudhury}.} \bibinfo{year}{2017}\natexlab{}.
\newblock \showarticletitle{Directed Greybox Fuzzing}. In \bibinfo{booktitle}{\emph{Proceedings of the 2017 ACM SIGSAC Conference on Computer and Communications Security}} (Dallas, Texas, USA) \emph{(\bibinfo{series}{CCS '17})}. \bibinfo{publisher}{Association for Computing Machinery}, \bibinfo{address}{New York, NY, USA}, \bibinfo{pages}{2329–2344}.
\newblock
\showISBNx{9781450349468}
\urldef\tempurl%
\url{https://doi.org/10.1145/3133956.3134020}
\showDOI{\tempurl}


\bibitem[Bradley(2012)]%
        {ICThree}
\bibfield{author}{\bibinfo{person}{Aaron~R. Bradley}.} \bibinfo{year}{2012}\natexlab{}.
\newblock \showarticletitle{Understanding IC3}. In \bibinfo{booktitle}{\emph{Theory and Applications of Satisfiability Testing -- SAT 2012}}, \bibfield{editor}{\bibinfo{person}{Alessandro Cimatti} {and} \bibinfo{person}{Roberto Sebastiani}} (Eds.). \bibinfo{publisher}{Springer Berlin Heidelberg}, \bibinfo{address}{Berlin, Heidelberg}, \bibinfo{pages}{1--14}.
\newblock
\showISBNx{978-3-642-31612-8}
\urldef\tempurl%
\url{https://doi.org/10.1007/978-3-642-31612-8_1}
\showURL{%
\tempurl}


\bibitem[Chen et~al\mbox{.}(2018)]%
        {Hawkeye}
\bibfield{author}{\bibinfo{person}{Hongxu Chen}, \bibinfo{person}{Yinxing Xue}, \bibinfo{person}{Yuekang Li}, \bibinfo{person}{Bihuan Chen}, \bibinfo{person}{Xiaofei Xie}, \bibinfo{person}{Xiuheng Wu}, {and} \bibinfo{person}{Yang Liu}.} \bibinfo{year}{2018}\natexlab{}.
\newblock \showarticletitle{Hawkeye: Towards a Desired Directed Grey-box Fuzzer}. In \bibinfo{booktitle}{\emph{Proceedings of the 2018 ACM SIGSAC Conference on Computer and Communications Security}} (Toronto, Canada) \emph{(\bibinfo{series}{CCS '18})}. \bibinfo{publisher}{Association for Computing Machinery}, \bibinfo{address}{New York, NY, USA}, \bibinfo{pages}{2095–2108}.
\newblock
\showISBNx{9781450356930}
\urldef\tempurl%
\url{https://doi.org/10.1145/3243734.3243849}
\showDOI{\tempurl}


\bibitem[Cimatti and Griggio(2012)]%
        {CAV:CG12}
\bibfield{author}{\bibinfo{person}{A. Cimatti} {and} \bibinfo{person}{A. Griggio}.} \bibinfo{year}{2012}\natexlab{}.
\newblock \showarticletitle{Software Model Checking via {IC3}}. In \bibinfo{booktitle}{\emph{CAV}}.
\newblock


\bibitem[Cyphert and Kincaid(2024)]%
        {POPL:CK2024}
\bibfield{author}{\bibinfo{person}{John Cyphert} {and} \bibinfo{person}{Zachary Kincaid}.} \bibinfo{year}{2024}\natexlab{}.
\newblock \showarticletitle{Solvable Polynomial Ideals: The Ideal Reflection for Program Analysis}.
\newblock \bibinfo{journal}{\emph{Proc. ACM Program. Lang.}} \bibinfo{volume}{8}, \bibinfo{number}{POPL}, Article \bibinfo{articleno}{25} (\bibinfo{date}{Jan.} \bibinfo{year}{2024}), \bibinfo{numpages}{29}~pages.
\newblock
\urldef\tempurl%
\url{https://doi.org/10.1145/3632867}
\showDOI{\tempurl}


\bibitem[Darke et~al\mbox{.}(2021)]%
        {veriabs-svcomp24}
\bibfield{author}{\bibinfo{person}{Priyanka Darke}, \bibinfo{person}{Sakshi Agrawal}, {and} \bibinfo{person}{R. Venkatesh}.} \bibinfo{year}{2021}\natexlab{}.
\newblock \showarticletitle{VeriAbs: A Tool for Scalable Verification by Abstraction (Competition Contribution)}. In \bibinfo{booktitle}{\emph{Tools and Algorithms for the Construction and Analysis of Systems}}, \bibfield{editor}{\bibinfo{person}{Jan~Friso Groote} {and} \bibinfo{person}{Kim~Guldstrand Larsen}} (Eds.). \bibinfo{publisher}{Springer International Publishing}, \bibinfo{address}{Cham}, \bibinfo{pages}{458--462}.
\newblock
\showISBNx{978-3-030-72013-1}
\urldef\tempurl%
\url{https://doi.org/10.1007/978-3-030-72013-1_32}
\showURL{%
\tempurl}


\bibitem[Denning and Denning(1977)]%
        {cacm:DD77}
\bibfield{author}{\bibinfo{person}{D.E. Denning} {and} \bibinfo{person}{P.J. Denning}.} \bibinfo{year}{1977}\natexlab{}.
\newblock \showarticletitle{Certification of Programs for Secure Information Flow}.
\newblock \bibinfo{journal}{\emph{Commun.\ ACM}} \bibinfo{volume}{20}, \bibinfo{number}{7} (\bibinfo{year}{1977}), \bibinfo{pages}{504--513}.
\newblock
\urldef\tempurl%
\url{https://doi.org/10.1145/359636.359712}
\showDOI{\tempurl}


\bibitem[Dietsch et~al\mbox{.}(2017)]%
        {DBLP:conf/sigsoft/DietschHMNP17}
\bibfield{author}{\bibinfo{person}{Daniel Dietsch}, \bibinfo{person}{Matthias Heizmann}, \bibinfo{person}{Betim Musa}, \bibinfo{person}{Alexander Nutz}, {and} \bibinfo{person}{Andreas Podelski}.} \bibinfo{year}{2017}\natexlab{}.
\newblock \showarticletitle{{Craig} vs.\ {Newton} in software model checking}. In \bibinfo{booktitle}{\emph{Proceedings of the 2017 11th Joint Meeting on Foundations of Software Engineering, {ESEC/FSE} 2017, Paderborn, Germany, September 4-8, 2017}}, \bibfield{editor}{\bibinfo{person}{Eric Bodden}, \bibinfo{person}{Wilhelm Sch{\"{a}}fer}, \bibinfo{person}{Arie van Deursen}, {and} \bibinfo{person}{Andrea Zisman}} (Eds.). \bibinfo{publisher}{{ACM}}, \bibinfo{pages}{487--497}.
\newblock
\urldef\tempurl%
\url{https://doi.org/10.1145/3106237.3106307}
\showDOI{\tempurl}


\bibitem[Farzan and Kincaid(2015)]%
        {fmcad:fk15}
\bibfield{author}{\bibinfo{person}{Azadeh Farzan} {and} \bibinfo{person}{Zachary Kincaid}.} \bibinfo{year}{2015}\natexlab{}.
\newblock \showarticletitle{Compositional recurrence analysis}. In \bibinfo{booktitle}{\emph{2015 Formal Methods in Computer-Aided Design (FMCAD)}}. \bibinfo{pages}{57--64}.
\newblock
\urldef\tempurl%
\url{https://doi.org/10.1109/FMCAD.2015.7542253}
\showDOI{\tempurl}


\bibitem[Ferrante et~al\mbox{.}(1987)]%
        {kn:FOW87}
\bibfield{author}{\bibinfo{person}{Jeanne Ferrante}, \bibinfo{person}{Karl~J. Ottenstein}, {and} \bibinfo{person}{Joe~D. Warren}.} \bibinfo{year}{1987}\natexlab{}.
\newblock \showarticletitle{The program dependence graph and its use in optimization}.
\newblock \bibinfo{journal}{\emph{ACM Trans. Program. Lang. Syst.}} \bibinfo{volume}{9}, \bibinfo{number}{3} (\bibinfo{date}{July} \bibinfo{year}{1987}), \bibinfo{pages}{319–349}.
\newblock
\showISSN{0164-0925}
\urldef\tempurl%
\url{https://doi.org/10.1145/24039.24041}
\showDOI{\tempurl}


\bibitem[Godefroid et~al\mbox{.}(2005)]%
        {DART}
\bibfield{author}{\bibinfo{person}{Patrice Godefroid}, \bibinfo{person}{Nils Klarlund}, {and} \bibinfo{person}{Koushik Sen}.} \bibinfo{year}{2005}\natexlab{}.
\newblock \showarticletitle{DART: Directed Automated Random Testing}.
\newblock \bibinfo{journal}{\emph{SIGPLAN Not.}} \bibinfo{volume}{40}, \bibinfo{number}{6} (\bibinfo{date}{jun} \bibinfo{year}{2005}), \bibinfo{pages}{213–223}.
\newblock
\showISSN{0362-1340}
\urldef\tempurl%
\url{https://doi.org/10.1145/1064978.1065036}
\showDOI{\tempurl}


\bibitem[Godefroid and Luchaup(2011)]%
        {ISSTA:GL11}
\bibfield{author}{\bibinfo{person}{Patrice Godefroid} {and} \bibinfo{person}{Daniel Luchaup}.} \bibinfo{year}{2011}\natexlab{}.
\newblock \showarticletitle{Automatic partial loop summarization in dynamic test generation}. In \bibinfo{booktitle}{\emph{Proceedings of the 2011 International Symposium on Software Testing and Analysis}} (Toronto, Ontario, Canada) \emph{(\bibinfo{series}{ISSTA '11})}. \bibinfo{publisher}{Association for Computing Machinery}, \bibinfo{address}{New York, NY, USA}, \bibinfo{pages}{23–33}.
\newblock
\showISBNx{9781450305624}
\urldef\tempurl%
\url{https://doi.org/10.1145/2001420.2001424}
\showDOI{\tempurl}


\bibitem[Godefroid et~al\mbox{.}(2010)]%
        {Smash}
\bibfield{author}{\bibinfo{person}{Patrice Godefroid}, \bibinfo{person}{Aditya~V. Nori}, \bibinfo{person}{Sriram~K. Rajamani}, {and} \bibinfo{person}{Sai~Deep Tetali}.} \bibinfo{year}{2010}\natexlab{}.
\newblock \showarticletitle{Compositional May-Must Program Analysis: Unleashing the Power of Alternation}. In \bibinfo{booktitle}{\emph{Proceedings of the 37th Annual ACM SIGPLAN-SIGACT Symposium on Principles of Programming Languages}} (Madrid, Spain) \emph{(\bibinfo{series}{POPL '10})}. \bibinfo{publisher}{Association for Computing Machinery}, \bibinfo{address}{New York, NY, USA}, \bibinfo{pages}{43–56}.
\newblock
\showISBNx{9781605584799}
\urldef\tempurl%
\url{https://doi.org/10.1145/1706299.1706307}
\showDOI{\tempurl}


\bibitem[Gulavani et~al\mbox{.}(2006)]%
        {synergy}
\bibfield{author}{\bibinfo{person}{Bhargav~S. Gulavani}, \bibinfo{person}{Thomas~A. Henzinger}, \bibinfo{person}{Yamini Kannan}, \bibinfo{person}{Aditya~V. Nori}, {and} \bibinfo{person}{Sriram~K. Rajamani}.} \bibinfo{year}{2006}\natexlab{}.
\newblock \showarticletitle{SYNERGY: A New Algorithm for Property Checking}. In \bibinfo{booktitle}{\emph{Proceedings of the 14th ACM SIGSOFT International Symposium on Foundations of Software Engineering}} (Portland, Oregon, USA) \emph{(\bibinfo{series}{SIGSOFT '06/FSE-14})}. \bibinfo{publisher}{Association for Computing Machinery}, \bibinfo{address}{New York, NY, USA}, \bibinfo{pages}{117–127}.
\newblock
\showISBNx{1595934685}
\urldef\tempurl%
\url{https://doi.org/10.1145/1181775.1181790}
\showDOI{\tempurl}


\bibitem[Gurfinkel and Ivrii(2017)]%
        {FMCAD:GI2016}
\bibfield{author}{\bibinfo{person}{Arie Gurfinkel} {and} \bibinfo{person}{Alexander Ivrii}.} \bibinfo{year}{2017}\natexlab{}.
\newblock \showarticletitle{K-Induction Without Unrolling}. In \bibinfo{booktitle}{\emph{Formal Methods in Computer-Aided Design (FMCAD)}}. \bibinfo{publisher}{IEEE}, \bibinfo{pages}{1--9}.
\newblock
\urldef\tempurl%
\url{https://doi.org/10.23919/FMCAD.2017.8102243}
\showDOI{\tempurl}


\bibitem[Hojjat et~al\mbox{.}(2012)]%
        {ATVA:HIKKR2012}
\bibfield{author}{\bibinfo{person}{Hossein Hojjat}, \bibinfo{person}{Radu Iosif}, \bibinfo{person}{Filip Kone{\v{c}}n{\'y}}, \bibinfo{person}{Viktor Kuncak}, {and} \bibinfo{person}{Philipp R{\"u}mmer}.} \bibinfo{year}{2012}\natexlab{}.
\newblock \showarticletitle{Accelerating Interpolants}. In \bibinfo{booktitle}{\emph{Automated Technology for Verification and Analysis}}, \bibfield{editor}{\bibinfo{person}{Supratik Chakraborty} {and} \bibinfo{person}{Madhavan Mukund}} (Eds.). \bibinfo{publisher}{Springer Berlin Heidelberg}, \bibinfo{address}{Berlin, Heidelberg}, \bibinfo{pages}{187--202}.
\newblock
\showISBNx{978-3-642-33386-6}
\urldef\tempurl%
\url{https://doi.org/10.1007/978-3-642-33386-6_16}
\showURL{%
\tempurl}


\bibitem[Huang et~al\mbox{.}(2022)]%
        {Beacon}
\bibfield{author}{\bibinfo{person}{Heqing Huang}, \bibinfo{person}{Yiyuan Guo}, \bibinfo{person}{Qingkai Shi}, \bibinfo{person}{Peisen Yao}, \bibinfo{person}{Rongxin Wu}, {and} \bibinfo{person}{Charles Zhang}.} \bibinfo{year}{2022}\natexlab{}.
\newblock \showarticletitle{BEACON: Directed Grey-Box Fuzzing with Provable Path Pruning}. In \bibinfo{booktitle}{\emph{2022 IEEE Symposium on Security and Privacy (SP)}}. \bibinfo{pages}{36--50}.
\newblock
\urldef\tempurl%
\url{https://doi.org/10.1109/SP46214.2022.9833751}
\showDOI{\tempurl}


\bibitem[Jon{\'a}{\v{s}} et~al\mbox{.}(2024)]%
        {tool-symbiotic}
\bibfield{author}{\bibinfo{person}{Martin Jon{\'a}{\v{s}}}, \bibinfo{person}{Kristi{\'a}n Kumor}, \bibinfo{person}{Jakub Nov{\'a}k}, \bibinfo{person}{Jind{\v{r}}ich Sedl{\'a}{\v{c}}ek}, \bibinfo{person}{Marek Trt{\'i}k}, \bibinfo{person}{Luk{\'a}{\v{s}} Zaoral}, \bibinfo{person}{Paul{\'i}na Ayaziov{\'a}}, {and} \bibinfo{person}{Jan Strej{\v{c}}ek}.} \bibinfo{year}{2024}\natexlab{}.
\newblock \showarticletitle{Symbiotic 10: Lazy Memory Initialization and Compact Symbolic Execution}. In \bibinfo{booktitle}{\emph{Tools and Algorithms for the Construction and Analysis of Systems}}, \bibfield{editor}{\bibinfo{person}{Bernd Finkbeiner} {and} \bibinfo{person}{Laura Kov{\'a}cs}} (Eds.). \bibinfo{publisher}{Springer Nature Switzerland}, \bibinfo{address}{Cham}, \bibinfo{pages}{406--411}.
\newblock
\showISBNx{978-3-031-57256-2}
\urldef\tempurl%
\url{https://doi.org/10.1007/978-3-031-57256-2_29}
\showURL{%
\tempurl}


\bibitem[Jovanovic and Dutertre(2016)]%
        {FMCAD:JD2016}
\bibfield{author}{\bibinfo{person}{Dejan Jovanovic} {and} \bibinfo{person}{Bruno Dutertre}.} \bibinfo{year}{2016}\natexlab{}.
\newblock \showarticletitle{Property-Directed k-Induction}. In \bibinfo{booktitle}{\emph{Formal Methods in Computer-Aided Design (FMCAD)}}. \bibinfo{publisher}{IEEE}, \bibinfo{pages}{85--92}.
\newblock
\urldef\tempurl%
\url{https://doi.org/10.1109/FMCAD.2016.7886671}
\showDOI{\tempurl}


\bibitem[Kincaid et~al\mbox{.}(2017a)]%
        {DBLP:conf/pldi/KincaidBBR17}
\bibfield{author}{\bibinfo{person}{Zachary Kincaid}, \bibinfo{person}{Jason Breck}, \bibinfo{person}{Ashkan~Forouhi Boroujeni}, {and} \bibinfo{person}{Thomas~W. Reps}.} \bibinfo{year}{2017}\natexlab{a}.
\newblock \showarticletitle{Compositional recurrence analysis revisited}. In \bibinfo{booktitle}{\emph{Proceedings of the 38th {ACM} {SIGPLAN} Conference on Programming Language Design and Implementation, {PLDI} 2017, Barcelona, Spain, June 18-23, 2017}}. \bibinfo{pages}{248--262}.
\newblock
\urldef\tempurl%
\url{https://doi.org/10.1145/3062341.3062373}
\showDOI{\tempurl}


\bibitem[Kincaid et~al\mbox{.}(2017b)]%
        {POPL:KCBR2017}
\bibfield{author}{\bibinfo{person}{Zachary Kincaid}, \bibinfo{person}{John Cyphert}, \bibinfo{person}{Jason Breck}, {and} \bibinfo{person}{Thomas Reps}.} \bibinfo{year}{2017}\natexlab{b}.
\newblock \showarticletitle{Non-linear reasoning for invariant synthesis}.
\newblock \bibinfo{journal}{\emph{Proc. ACM Program. Lang.}} \bibinfo{volume}{2}, \bibinfo{number}{POPL}, Article \bibinfo{articleno}{54} (\bibinfo{date}{Dec.} \bibinfo{year}{2017}), \bibinfo{numpages}{33}~pages.
\newblock
\urldef\tempurl%
\url{https://doi.org/10.1145/3158142}
\showDOI{\tempurl}


\bibitem[Kincaid et~al\mbox{.}(2021)]%
        {cav:apa}
\bibfield{author}{\bibinfo{person}{Zachary Kincaid}, \bibinfo{person}{Thomas Reps}, {and} \bibinfo{person}{John Cyphert}.} \bibinfo{year}{2021}\natexlab{}.
\newblock \showarticletitle{Algebraic Program Analysis}. In \bibinfo{booktitle}{\emph{Computer Aided Verification}}, \bibfield{editor}{\bibinfo{person}{Alexandra Silva} {and} \bibinfo{person}{K.~Rustan~M. Leino}} (Eds.). \bibinfo{publisher}{Springer International Publishing}, \bibinfo{address}{Cham}, \bibinfo{pages}{46--83}.
\newblock
\showISBNx{978-3-030-81685-8}
\urldef\tempurl%
\url{https://doi.org/10.1007/978-3-030-81685-8_3}
\showURL{%
\tempurl}


\bibitem[Komuravelli et~al\mbox{.}(2014)]%
        {spacer}
\bibfield{author}{\bibinfo{person}{Anvesh Komuravelli}, \bibinfo{person}{Arie Gurfinkel}, {and} \bibinfo{person}{Sagar Chaki}.} \bibinfo{year}{2014}\natexlab{}.
\newblock \showarticletitle{SMT-Based Model Checking for Recursive Programs}. In \bibinfo{booktitle}{\emph{Computer Aided Verification}}, \bibfield{editor}{\bibinfo{person}{Armin Biere} {and} \bibinfo{person}{Roderick Bloem}} (Eds.). \bibinfo{publisher}{Springer International Publishing}, \bibinfo{address}{Cham}, \bibinfo{pages}{17--34}.
\newblock
\showISBNx{978-3-319-08867-9}
\urldef\tempurl%
\url{https://doi.org/10.1007/978-3-319-08867-9_2}
\showURL{%
\tempurl}


\bibitem[Ma et~al\mbox{.}(2011)]%
        {directedSymbExec}
\bibfield{author}{\bibinfo{person}{Kin-Keung Ma}, \bibinfo{person}{Khoo Yit~Phang}, \bibinfo{person}{Jeffrey~S. Foster}, {and} \bibinfo{person}{Michael Hicks}.} \bibinfo{year}{2011}\natexlab{}.
\newblock \showarticletitle{Directed Symbolic Execution}. In \bibinfo{booktitle}{\emph{Static Analysis}}, \bibfield{editor}{\bibinfo{person}{Eran Yahav}} (Ed.). \bibinfo{publisher}{Springer Berlin Heidelberg}, \bibinfo{address}{Berlin, Heidelberg}, \bibinfo{pages}{95--111}.
\newblock
\showISBNx{978-3-642-23702-7}
\urldef\tempurl%
\url{https://doi.org/10.1007/978-3-642-23702-7_11}
\showURL{%
\tempurl}


\bibitem[Majumdar and Sen(2007)]%
        {MajumdarSenICSE07}
\bibfield{author}{\bibinfo{person}{Rupak Majumdar} {and} \bibinfo{person}{Koushik Sen}.} \bibinfo{year}{2007}\natexlab{}.
\newblock \showarticletitle{Hybrid Concolic Testing}. In \bibinfo{booktitle}{\emph{29th International Conference on Software Engineering (ICSE'07)}}. \bibinfo{pages}{416--426}.
\newblock
\urldef\tempurl%
\url{https://doi.org/10.1109/ICSE.2007.41}
\showDOI{\tempurl}


\bibitem[McMillan(2006)]%
        {cav:mcmillan06}
\bibfield{author}{\bibinfo{person}{Kenneth~L. McMillan}.} \bibinfo{year}{2006}\natexlab{}.
\newblock \showarticletitle{Lazy Abstraction with Interpolants}. In \bibinfo{booktitle}{\emph{Computer Aided Verification}}, \bibfield{editor}{\bibinfo{person}{Thomas Ball} {and} \bibinfo{person}{Robert~B. Jones}} (Eds.). \bibinfo{publisher}{Springer Berlin Heidelberg}, \bibinfo{address}{Berlin, Heidelberg}, \bibinfo{pages}{123--136}.
\newblock
\showISBNx{978-3-540-37411-4}
\urldef\tempurl%
\url{https://doi.org/10.1007/11817963_14}
\showURL{%
\tempurl}


\bibitem[McMillan(2010)]%
        {lazyAnnot}
\bibfield{author}{\bibinfo{person}{Kenneth~L. McMillan}.} \bibinfo{year}{2010}\natexlab{}.
\newblock \showarticletitle{Lazy Annotation for Program Testing and Verification}. In \bibinfo{booktitle}{\emph{Computer Aided Verification}}, \bibfield{editor}{\bibinfo{person}{Tayssir Touili}, \bibinfo{person}{Byron Cook}, {and} \bibinfo{person}{Paul Jackson}} (Eds.). \bibinfo{publisher}{Springer Berlin Heidelberg}, \bibinfo{address}{Berlin, Heidelberg}, \bibinfo{pages}{104--118}.
\newblock
\showISBNx{978-3-642-14295-6}
\urldef\tempurl%
\url{https://doi.org/10.1007/978-3-642-14295-6_10}
\showURL{%
\tempurl}


\bibitem[McMillan(2018)]%
        {mcmillanInterpolantSummary}
\bibfield{author}{\bibinfo{person}{Kenneth~L. McMillan}.} \bibinfo{year}{2018}\natexlab{}.
\newblock \showarticletitle{Interpolation and Model Checking}.
\newblock In \bibinfo{booktitle}{\emph{Handbook of Model Checking}}, \bibfield{editor}{\bibinfo{person}{Edmund~M. Clarke}, \bibinfo{person}{Thomas~A. Henzinger}, \bibinfo{person}{Helmut Veith}, {and} \bibinfo{person}{Roderick Bloem}} (Eds.). \bibinfo{publisher}{Springer International Publishing}, \bibinfo{address}{Cham}, \bibinfo{pages}{421--446}.
\newblock
\showISBNx{978-3-319-10575-8}
\urldef\tempurl%
\url{https://doi.org/10.1007/978-3-319-10575-8_14}
\showDOI{\tempurl}


\bibitem[Molnar et~al\mbox{.}(2008)]%
        {ndss:glm08}
\bibfield{author}{\bibinfo{person}{David Molnar}, \bibinfo{person}{P Godefroid}, {and} \bibinfo{person}{M Levin}.} \bibinfo{year}{2008}\natexlab{}.
\newblock \showarticletitle{Automated whitebox fuzz testing}. In \bibinfo{booktitle}{\emph{Network and distributed system security symposium, NDSS}}. \bibinfo{pages}{416--426}.
\newblock


\bibitem[Saxena et~al\mbox{.}(2009)]%
        {ISSTA:SPMS09}
\bibfield{author}{\bibinfo{person}{Prateek Saxena}, \bibinfo{person}{Pongsin Poosankam}, \bibinfo{person}{Stephen McCamant}, {and} \bibinfo{person}{Dawn Song}.} \bibinfo{year}{2009}\natexlab{}.
\newblock \showarticletitle{Loop-extended symbolic execution on binary programs}. In \bibinfo{booktitle}{\emph{Proceedings of the Eighteenth International Symposium on Software Testing and Analysis}} (Chicago, IL, USA) \emph{(\bibinfo{series}{ISSTA '09})}. \bibinfo{publisher}{Association for Computing Machinery}, \bibinfo{address}{New York, NY, USA}, \bibinfo{pages}{225–236}.
\newblock
\showISBNx{9781605583389}
\urldef\tempurl%
\url{https://doi.org/10.1145/1572272.1572299}
\showDOI{\tempurl}


\bibitem[Sen et~al\mbox{.}(2005)]%
        {FSE:SMA05}
\bibfield{author}{\bibinfo{person}{Koushik Sen}, \bibinfo{person}{Darko Marinov}, {and} \bibinfo{person}{Gul Agha}.} \bibinfo{year}{2005}\natexlab{}.
\newblock \showarticletitle{CUTE: A Concolic Unit Testing Engine for C}.
\newblock \bibinfo{journal}{\emph{SIGSOFT Softw. Eng. Notes}} \bibinfo{volume}{30}, \bibinfo{number}{5}, \bibinfo{pages}{263–272}.
\newblock
\showISSN{0163-5948}
\urldef\tempurl%
\url{https://doi.org/10.1145/1095430.1081750}
\showDOI{\tempurl}


\bibitem[Shah et~al\mbox{.}(2022)]%
        {MC2}
\bibfield{author}{\bibinfo{person}{Abhishek Shah}, \bibinfo{person}{Dongdong She}, \bibinfo{person}{Samanway Sadhu}, \bibinfo{person}{Krish Singal}, \bibinfo{person}{Peter Coffman}, {and} \bibinfo{person}{Suman Jana}.} \bibinfo{year}{2022}\natexlab{}.
\newblock \showarticletitle{MC2: Rigorous and Efficient Directed Greybox Fuzzing}. In \bibinfo{booktitle}{\emph{Proceedings of the 2022 ACM SIGSAC Conference on Computer and Communications Security}} (Los Angeles, CA, USA) \emph{(\bibinfo{series}{CCS '22})}. \bibinfo{publisher}{Association for Computing Machinery}, \bibinfo{address}{New York, NY, USA}, \bibinfo{pages}{2595–2609}.
\newblock
\showISBNx{9781450394505}
\urldef\tempurl%
\url{https://doi.org/10.1145/3548606.3560648}
\showDOI{\tempurl}


\bibitem[Sheeran et~al\mbox{.}(2002)]%
        {FMCAD:SSS2002}
\bibfield{author}{\bibinfo{person}{Mary Sheeran}, \bibinfo{person}{Satnam Singh}, {and} \bibinfo{person}{Gunnar St{\aa}lmarck}.} \bibinfo{year}{2002}\natexlab{}.
\newblock \showarticletitle{Checking Safety Properties Using Induction and a SAT-Solver}. In \bibinfo{booktitle}{\emph{Formal Methods in Computer-Aided Design (FMCAD)}} \emph{(\bibinfo{series}{LNCS}, Vol.~\bibinfo{volume}{2517})}. \bibinfo{publisher}{Springer}, \bibinfo{pages}{108--125}.
\newblock


\bibitem[Silverman and Kincaid(2019)]%
        {CAV:SK2019}
\bibfield{author}{\bibinfo{person}{Jake Silverman} {and} \bibinfo{person}{Zachary Kincaid}.} \bibinfo{year}{2019}\natexlab{}.
\newblock \showarticletitle{Loop Summarization with Rational Vector Addition Systems}. In \bibinfo{booktitle}{\emph{Computer Aided Verification}}, \bibfield{editor}{\bibinfo{person}{Isil Dillig} {and} \bibinfo{person}{Serdar Tasiran}} (Eds.). \bibinfo{publisher}{Springer International Publishing}, \bibinfo{address}{Cham}, \bibinfo{pages}{97--115}.
\newblock
\showISBNx{978-3-030-25543-5}
\urldef\tempurl%
\url{https://doi.org/10.1007/978-3-030-25543-5_7}
\showURL{%
\tempurl}


\bibitem[Tarjan(1981)]%
        {tarjan}
\bibfield{author}{\bibinfo{person}{Robert~Endre Tarjan}.} \bibinfo{year}{1981}\natexlab{}.
\newblock \showarticletitle{Fast Algorithms for Solving Path Problems}.
\newblock \bibinfo{journal}{\emph{J. ACM}} \bibinfo{volume}{28}, \bibinfo{number}{3} (\bibinfo{date}{jul} \bibinfo{year}{1981}), \bibinfo{pages}{594–614}.
\newblock
\showISSN{0004-5411}
\urldef\tempurl%
\url{https://doi.org/10.1145/322261.322273}
\showDOI{\tempurl}


\bibitem[Thakur et~al\mbox{.}(2010)]%
        {CAV:TLLBDEAR10}
\bibfield{author}{\bibinfo{person}{Aditya Thakur}, \bibinfo{person}{Junghee Lim}, \bibinfo{person}{Akash Lal}, \bibinfo{person}{Amanda Burton}, \bibinfo{person}{Evan Driscoll}, \bibinfo{person}{Matt Elder}, \bibinfo{person}{Tycho Andersen}, {and} \bibinfo{person}{Thomas Reps}.} \bibinfo{year}{2010}\natexlab{}.
\newblock \showarticletitle{Directed Proof Generation for Machine Code}. In \bibinfo{booktitle}{\emph{Computer Aided Verification}}, \bibfield{editor}{\bibinfo{person}{Tayssir Touili}, \bibinfo{person}{Byron Cook}, {and} \bibinfo{person}{Paul Jackson}} (Eds.). \bibinfo{publisher}{Springer Berlin Heidelberg}, \bibinfo{address}{Berlin, Heidelberg}, \bibinfo{pages}{288--305}.
\newblock
\showISBNx{978-3-642-14295-6}
\urldef\tempurl%
\url{https://doi.org/10.1007/978-3-642-14295-6_27}
\showURL{%
\tempurl}


\bibitem[Vediramana~Krishnan et~al\mbox{.}(2019)]%
        {CAV:VKVGG2019}
\bibfield{author}{\bibinfo{person}{Hari~Govind Vediramana~Krishnan}, \bibinfo{person}{Yakir Vizel}, \bibinfo{person}{Vijay Ganesh}, {and} \bibinfo{person}{Arie Gurfinkel}.} \bibinfo{year}{2019}\natexlab{}.
\newblock \showarticletitle{Interpolating Strong Induction}. In \bibinfo{booktitle}{\emph{Computer Aided Verification}}, \bibfield{editor}{\bibinfo{person}{Isil Dillig} {and} \bibinfo{person}{Serdar Tasiran}} (Eds.). \bibinfo{publisher}{Springer International Publishing}, \bibinfo{address}{Cham}, \bibinfo{pages}{367--385}.
\newblock
\showISBNx{978-3-030-25543-5}
\urldef\tempurl%
\url{https://doi.org/10.1007/978-3-030-25543-5_21}
\showURL{%
\tempurl}


\end{thebibliography}
